\documentclass[sigconf]{acmart}

\AtBeginDocument{%
  \providecommand\BibTeX{{%
    \normalfont B\kern-0.5em{\scshape i\kern-0.25em b}\kern-0.8em\TeX}}}

\copyrightyear{2022}
\acmYear{2022}
\setcopyright{rightsretained}
\acmConference[CHI '22]{CHI Conference on Human Factors in Computing
Systems}{April 29-May 5, 2022}{New Orleans, LA, USA}
\acmBooktitle{CHI Conference on Human Factors in Computing Systems
(CHI '22), April 29-May 5, 2022, New Orleans, LA, USA}
\acmDOI{10.1145/3491102.3517673}
\acmISBN{978-1-4503-9157-3/22/04}

\usepackage{subfigure}
\usepackage{array}
\usepackage{booktabs, makecell, tabularx}
\usepackage{picins}

\usepackage[english]{babel}
\usepackage{hyperref}
\addto\extrasenglish{%
}

\newcommand{\rulesep}{\unskip\ \vrule\ }
\usepackage{review}

\usepackage{hyphenat}
\usepackage{xstring}
\usepackage{forloop}

\newsavebox\MyBreakChar%
\sbox\MyBreakChar{}%
\newsavebox\MySpaceBreakChar%
\sbox\MySpaceBreakChar{\hyp}%
\makeatletter%
\newcommand*{\BreakableChar}[1][\MyBreakChar]{%
  \leavevmode%
  \prw@zbreak%
  \discretionary{\usebox#1}{}{}%
  \prw@zbreak%
}%
\makeatother

\newcounter{index}%
\newcommand{\AddBreakableChars}[1]{%
  \StrLen{#1 }[\stringLength]%
  \forloop[1]{index}{1}{\value{index}<\stringLength}{%
    \StrChar{#1}{\value{index}}[\currentLetter]%
    \IfStrEqCase{\currentLetter}{%
        {*}{\currentLetter\BreakableChar[\MyBreakChar]}%
        {/}{\currentLetter\BreakableChar[\MyBreakChar]}%
        {+}{\currentLetter\BreakableChar[\MyBreakChar]}%
        {\&}{\currentLetter\BreakableChar[\MyBreakChar]}%
    }[\currentLetter]%
  }%
}%

\begin{document}

\title[The Pattern is in the Details]{The Pattern is in the Details: An Evaluation of Interaction Techniques for Locating, Searching, and Contextualizing Details in Multivariate Matrix Visualizations}

\author{Yalong Yang}
\orcid{0000-0001-9414-9911}
\affiliation{%
  \institution{Virginia Tech}
  \streetaddress{}
  \city{Blacksburg}
  \state{VA}
  \country{USA}
  \postcode{24060}}
\email{yalongyang@vt.edu}

\author{Wenyu Xia}
\affiliation{%
  \institution{Carnegie Mellon University}
  \streetaddress{}
  \city{Pittsburgh}
  \state{PA}
  \country{USA}
  \postcode{15213}}
\email{wenyux@andrew.cmu.edu}

\author{Fritz Lekschas}
\orcid{0000-0001-8432-4835}
\affiliation{%
  \institution{Harvard University}
  \streetaddress{}
  \city{Boston}
  \state{MA}
  \country{USA}
  \postcode{02134}}
\email{lekschas@seas.harvard.edu}

\author{Carolina Nobre}
\orcid{0000-0002-2892-0509}
\affiliation{%
  \institution{Harvard University}
  \streetaddress{}
  \city{Boston}
  \state{MA}
  \country{USA}
  \postcode{02134}}
\email{cnobre@seas.harvard.edu}

\author{Robert Krueger}
\orcid{0000-0002-6468-8356}
\affiliation{%
  \institution{Harvard University}
  \streetaddress{}
  \city{Boston}
  \state{MA}
  \country{USA}
  \postcode{02134}}
\email{krueger@seas.harvard.edu}

\author{Hanspeter Pfister}
\orcid{0000-0002-3620-2582}
\affiliation{%
  \institution{Harvard University}
  \streetaddress{}
  \city{Boston}
  \state{MA}
  \country{USA}
  \postcode{02134}}
\email{pfister@seas.harvard.edu}

\renewcommand{\shortauthors}{Yang and Xia, et al.}

\begin{abstract} %

Matrix visualizations are widely used to display large-scale network, tabular, set, or sequential data. They typically only encode a single value per cell, e.g., through color. However, this can greatly limit the visualizations’ utility when exploring multivariate data, where each cell represents a data point with multiple values (referred to as \emph{details}). Three well-established interaction approaches can be applicable in multivariate matrix visualizations (or MMV): \emph{focus+context}, \emph{pan\&zoom}, and \emph{overview+detail}. However, there is little empirical knowledge of how these approaches compare in exploring MMV. We report on two studies comparing them for locating, searching, and contextualizing details in MMV. We first compared four focus+context techniques and found that the \emph{fisheye lens} overall outperformed the others. We then compared the fisheye lens, to pan\&zoom and overview+detail. We found that \emph{pan\&zoom} was faster in locating and searching details, and as good as overview+detail in contextualizing details. 

\end{abstract}



\begin{teaserfigure}
  \centering
  \includegraphics[width=\linewidth]{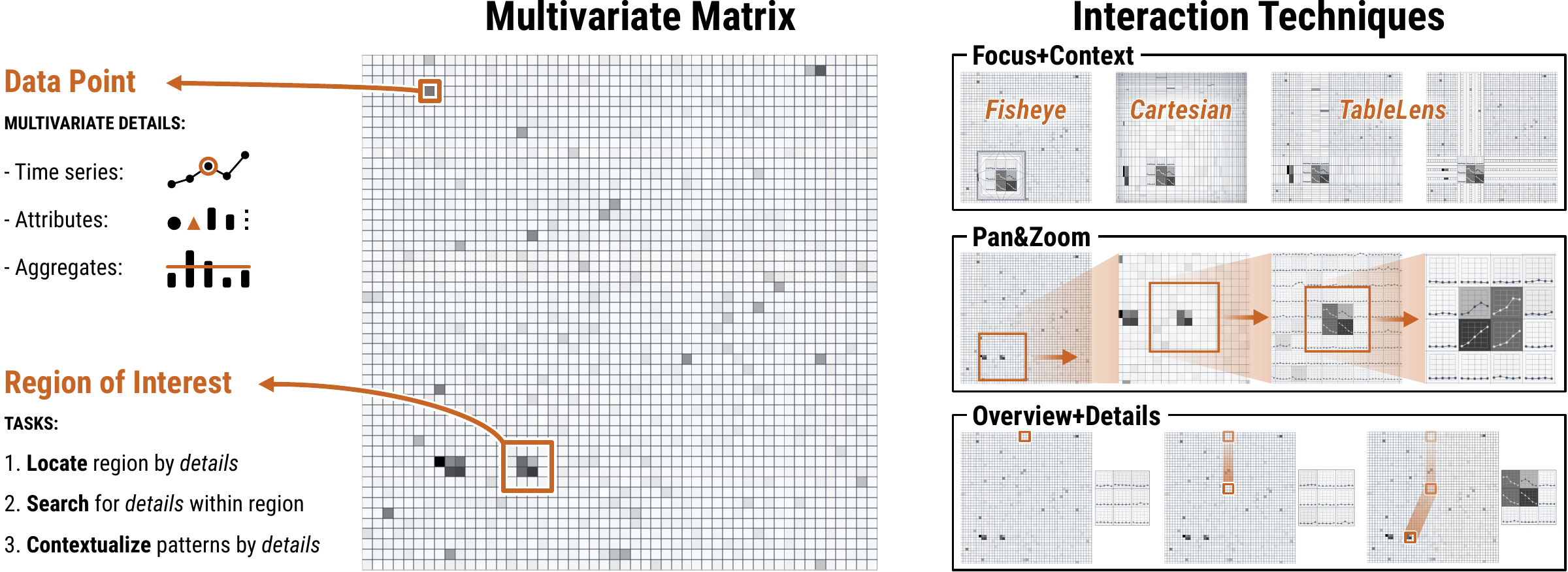}
  \vspace{0mm}
  \Description{Left sub-figure demonstrates different multivariate details (e.g., time series, attributes, aggregates) that can be visualized in a matrix and three representative tasks (i.e., locate, search, and contextualize) tested in the user studies. Right sub-figure shows three types of interaction techniques (i.e., focus+context, pan\&zoom, and overview+details) tested in the user study.}
  \caption{In a multivariate matrix visualization (MMV) each cell represents a \emph{data point} that is associated to multiple time points, attributes, or other data points (through aggregation). To study a \emph{region of interest}, an analyst typically needs to \emph{locate} the region, \emph{search} for details within the region, and identify \emph{contextual} patterns using details. Multiple interaction approaches can be used to conduct this exploration but which one is most effective for different tasks: focus+context, pan\&zoom, or overview+details?}
  \label{fig:teaser}
  \vspace{5mm}
\end{teaserfigure}

\def\flens{\textsc{Lens}}
\def\lcartesian{\textsc{Cartesian}}
\def\ltable{\textsc{Stretch}}
\def\ltabledc{\textsc{Step}}
\def\lfisheye{\textsc{Fisheye}}

\def\fsize{\textsc{Size}}
\def\llarge{\textsc{Large}}
\def\lsmall{\textsc{Small}}

\def\tlocate{\textsc{Locate}}
\def\tcompare{\textsc{Search}}
\def\tcontext{\textsc{Context}}

\def\ftechnique{\textsc{Technique}}
\def\lfc{\textsc{Focus+\BreakableChar{}Context}}
\def\lod{\textsc{Overview+\BreakableChar{}Detail}}
\def\lzoom{\textsc{Pan\&\BreakableChar{}Zoom}}

\maketitle

\section{Introduction}
\label{sec:introduction}

Plotting a series of data points in a regular two-dimensional grid---a \emph{matrix visualization}---is a space-efficient approach for visualizing large-scale and dense network~\cite{ghoniem2005readability,yalong_yang_many--many_2017}, tabular~\cite{niederer2017taco,neto2020explainable}, set~\cite{sadana2014onset}, or sequential data~\cite{anders2009visualization,kerpedjiev_higlass_2018,boix2021regulatory}.
In a matrix visualization, a cell typically only encodes a single value of a data point, e.g., through color.
However, for multivariate data, multiple attributes or values (called \emph{details} hereafter) are associated with each data point.
We refer to the matrix visualization of multivariate data as \emph{multivariate matrix visualization} (MMV).
MMV are widely used in various applications.
For example, analysts frequently use them to explore temporal data~\cite{behrisch2014visual,bach2014visualizing,bach_small_2015,wood_visualizing_2011,fischer_visual_2021,beck_state_2014,yi_timematrix_2010}, such as ecologists studied multi-year international food trade through MMV~\cite{kastner_rapid_2014}, and biologists studied dynamic Bayesian networks with MMVs to model probabilistic dependencies in gene regulation and brain connectivity across time~\cite{vogogias_visual_2020}.
Additionally, MMVs can also be used to show multiple attributes of a data point~\cite{sadana2014onset,yates2014visualizing,Pearce2020,horak_responsive_2021} or aggregated values from details~\cite{elmqvist_zame_2008,dang2016multilayermatrix,lekschas_hipiler_2018,lekschas_generic_2021}.
For instance, MMV can help pathologists interpret multiclass classifications by visualizing multiple class probabilities at once~\cite{Pearce2020} in histopathology~\cite{xu2017large}, and MMV can be used to help analyze complex multi-variate geographic data~\cite{goodwin_visualizing_2016}.

Exploring MMV requires people to investigate the \emph{details} in each single cell, which is usually challenging because each matrix cell’s display space is limited and often cannot show all data points in full detail.
To enable analysts to effectively explore MMV, a common strategy is to selectively visualize the details of a subset of data points. 
To this end, three general interaction approaches can be used for MMV: focus+context (or lens), pan\&zoom, and overview+details.
\yalong{In this work, we consider MMV where matrix cells change their representation from a single to a multi-value visualization (i.e., from a single color to a line chart) with these interaction techniques.}
However, adapting these interactions to MMV is not trivial as the MMV's special characteristics need to be taking into consideration.
Focus+context magnifies a selected region (referred to as the focus) within the context to show it in greater detail. To make space for the magnified region, the surrounding area (referred to as context) is compressed in size. 
Not all focus+context techniques are suitable for MMV. The distortion from many focus+context techniques, like a pixel-based fisheye lens, produce irregular shapes of cells that may prohibit effective exploration in MMV.
On the other hand, Responsive matrix cells~\cite{horak_responsive_2021}, Mélange~\cite{elmqvist_melange_2008,elmqvist_melange_2010}, LiveRAC~\cite{mclachlan_liverac_2008}, and TableLens~\cite{rao_table_1994} are representative focus+context techniques that are applicable to MMV. 
Overview+detail technique provides two spatially-separate views with different levels of detail. One view shows the details, and the other offers the context.
For example, Burch et al.~\cite{burch_matrix-based_2013} used overview+detail to facilitate the exploration of MMV.
Pan\&zoom presents the visualization at a certain detail level while enabling the user to zoom into the visualization and pan to other regions.
For instance, TimeMatrix~\cite{yi_timematrix_2010} provides pan\&zoom for the users to navigate a MMV in different levels of details.

Focus+context, pan\&zoom, and overview+details have been extensively compared in various applications~\cite{cockburn_review_2009,ronne_jakobsen_sizing_2011,yang_embodied_2021,baudisch_keeping_2002,burigat_map_2008,stefano_burigat_effectiveness_2013} (more details in \autoref{sec:related_work}).
However, mixed results were found about their effectiveness, indicating the applied scenarios might largely influence their performance.
Thus, it is not applicable and reliable to compile guidelines for MMV only from prior results.
Yet, to the best of our knowledge, there is no user study comparing them in the context of MMV.
To close this gap, we conducted two extensive user studies to compare the effectiveness of different interaction techniques for MMV.
Our goal is to better understand how people interactively explore multivariate details associated with data points in MMV.
Thus, in our evaluation, we did not vary the visual encoding and used a simple visualization within each matrix cell to reduce the complexity and potential confounding factors. 
To this end, we chose a line chart to visualize a time series in each matrix cell, as exploring temporal data is one of the most frequently reported applications for MMV~\cite{behrisch2014visual,bach2014visualizing,bach_small_2015,wood_visualizing_2011,fischer_visual_2021,beck_state_2014,yi_timematrix_2010}, and line charts are a widely-used technique for visualizing temporal data.
We are especially interested in the effectiveness of different interaction techniques for navigating MMV and retrieving details from matrix cells, as it is the unique aspect that distinguishes MMV from univariate matrix visualizations.
After analyzing the literature~\cite{andrienko_conceptual_2011,bach_descriptive_2017,beck_state_2014,goodwin_visualizing_2016,nobre_state_2019}, taxonomies~\cite{munzner2014visualization,yang_embodied_2021,laviola20173d,nilsson_natural_2018,lam_framework_2008,wang_baldonado_guidelines_2000} and real-world applications~\cite{kastner_rapid_2014,Pearce2020}, we derived and tested three fundamental \emph{interaction} tasks that cover a wide range of MMV use cases:
\emph{locating} a single cell and then inspecting the details inside; 
\emph{searching} a region of interest (ROI) of multiple cells to find the cells that match the target pattern; and 
\emph{contextualizing} patterns using details, which requires inspecting both the details and the context.  
These three tested tasks can act as ``primitive'' interactions to serve more sophisticated visual analytic scenarios.

Given the diversity of focus+context techniques~\cite{horak_responsive_2021,elmqvist_melange_2008,elmqvist_melange_2010,mclachlan_liverac_2008,rao_table_1994}, the many ways their distortions could impact the perception and task performance, and their overall good performance in some applications~\cite{baudisch_keeping_2002,gutwin_fisheyes_2003,shoemaker_supporting_2007}, we compared different lenses
in \textbf{our first study}. 
To identify representative lenses, we followed Carpendale's taxonomy for distortion~\cite{carpendale_extending_1997} and identified four lenses: 
Cartesian lens~\cite{sarkar_graphical_1992} applies non-linear orthogonal distortion; 
TableLens~\cite{rao_table_1994} with orthogonal distortion (step and stretch);
and a fisheye lens technique that is adapted to matrix visualizations~\cite{robertson_document_1993, carpendale_extending_1997}.
Overall, the results indicate that fisheye lens performed as well as or better than other techniques in the tested tasks. Participants also rated the fisheye lens as the easiest technique for locating matrix cells.

\textbf{Our second study} compared the fisheye lens~---~the overall best performing focus+context technique from the first study~---~against a pan\&zoom and an overview+detail technique. 
We found pan\&zoom was faster than focus+context and overview+detail techniques in locating and searching for details and as good as overview+detail in contextualizing details. Pan\&zoom was also rated with the highest usability and lowest mental demand in almost all tasks. 
Our results contribute empirical knowledge on the effectiveness of different interaction techniques for exploring MMV. 
We also discuss promising improvements over existing techniques and potential novel techniques inspired by our results.

\section{Related Work}
\label{sec:related_work}

A foundation of exploring MMV is enabling interactive inspection of multiple levels of detail.
Several interaction approaches have emerged for this purpose, such as focus+context, overview+detail, and pan\&zoom.
Cockburn et al. distill the issues with each approach~\cite{cockburn_review_2009}:
focus+context distorts the information space; overview+\BreakableChar{}detail requires extra effort for users to relate information between the overview and the detail view; pan\&zoom leads to higher working memory as the users can only see one view at a time.
Yet, it is still unclear to what extent these findings apply to MMV. 

\textbf{Focus+context (or lens) techniques.}
A common group of focus+context techniques is lenses, introduced by Bier et al.~\cite{bier_toolglass_1993,tominski_interactive_2017} as generic see-through interfaces between the application and the cursor. 
Lenses apply magnification to increase the detail in local areas. 
Lenses can further reveal hidden information, enhance data of interest~\cite{kruger_trajectorylensesset-based_2013}, or suppress distracting information~\cite{ellis_sampling_2005,hornbaek_reading_2001}.
While emphasizing details, matrix analysis tasks can necessitate all cells of the matrix to be concurrently visible. 
To achieve this, Carpendale et al.~\cite{carpendale_extending_1997} discuss various distortion possibilities with smooth transitions from focus to context in rectangular uniform grids (matrices). 
Depending on the data, different spatial mapping techniques can be advantageous. 
Bifocal Display~\cite{apperley_bifocal_1982} introduces a continuous one-dimensional distortion for 2D data by stretching a column in focus and pushing all other columns aside. 
The TableLens technique~\cite{rao_table_1994} distorts a 2D grid in two dimensions: stretching the columns and rows of the cell in focus only (non-continuous) and shifting the remaining non-magnified cells outward.
LiveRAC~\cite{mclachlan_liverac_2008} adapts the idea of TableLens in showing time-series data.
Document Lens~\cite{robertson_document_1993} offers 3D distortion of 2D fields.
Mélange~\cite{elmqvist_melange_2010} is a 3D distortion technique to ease comparison tasks. It folds the intervening space to guarantee the visibility of multiple focus
regions. Responsive matrix cells~\cite{horak_responsive_2021} combine focus+context with semantic zooming to allow analysts to go from the overview of the matrix to details in cells. 
Given the diversity of focus+context techniques, we tested the effectiveness of four representative lenses derived from Carpendale et al.'s taxonomy~\cite{carpendale_extending_1997}: Cartesian lens~\cite{sarkar_graphical_1992}, two TableLens variations~\cite{rao_table_1994}, and an adapted fisheye lens~\cite{robertson_document_1993}.

\textbf{Evaluating Focus+context techniques.}
Most previous studies on focus+context concentrate on parameter testing and different types of focus+context techniques have not be compared empirically.
McGuffin and Balakrshnan~\cite{mcguffin_acquisition_2002} investigated the acquisition of targets that dynamically grow in response to users' focus of attention. In their study with 12 participants, they found that performance is governed by the target's size and can be predicted with Fitts' law~\cite{fitts_information_1954}.
Gutwin et al.~\cite{gutwin_improving_2002} found that speed-coupled flattening improved focus-targeting when using fisheye distortion with 10 participants. 
However, fisheye techniques can also introduce reading difficulties. 
To alleviate this issue,
Zanella et al.~\cite{zanella_effects_2002} showed that grids aid readability in a larger study with 30 participants. Finally, Pietriga's study~\cite{pietriga_sigma_2008} with 10 participants compared different transitions between focus+context and found that gradually increasing translucence was the best choice.
Most previous studies were also with a small number of participants. With 48 participants, our study is less outlier-prone and potentially has a smaller margin of error.

\textbf{Overview+detail techniques.}
Prominent examples for 2D navigation are horizontal and vertical scrollbars with thumbnails~\cite{chimera_value_1998} and mini-maps~\cite{zammitto_visualization_2008}, as well as more distinct linked views~\cite{roberts_state_2007} with different perspectives 
for overview and details. 
MatLink~\cite{henry_matlink_2007} encodes links as curved edges to give detail at the border of the matrix for improving path-finding.
Lekschas et al.~\cite{lekschas_hipiler_2018} propose an overview+detail method to compare regions of interest at different scales through interactive small multiples. In their system, each small multiple provides a detailed view of a small local matrix pattern. They later show that this approach can be extended to support pattern-driven guidance in navigating in MMVs~\cite{lekschas_pattern-driven_2020}. 
CoCoNutTrix~\cite{isenberg_coconuttrix_2009} visualized network data using NodeTrix~\cite{henry_nodetrix_2007} on a high-resolution large display.
We used a standard overview+detail design in our second user study, where we placed the overview and detail view side-by-side, and the user can interactively select the ROI in the overview to update the detail view. 

\textbf{Pan\&Zoom techniques.} 
The literature distinguishes between geometric and semantic zooming. 
The former specifies the spatial scale of magnification. Van Wijk and Nuij summarize smooth and efficient geometric zooming and panning techniques and present a model to calculate a solution for optimal view animations ~\cite{wijk_smooth_2003}.
Semantic zooming, by contrast, changes the level of detail by varying the visual encoding, not only its physical size~\cite{boulos_use_2003}.
Lekschas et al.~\cite{lekschas_hipiler_2018} categorize interaction in matrices into content-agnostic and content-aware approaches. Content-agnostic approaches, such as geometric panning\&zooming, operate entirely on the view level, while the latter \textit{"incorporate the data to drive visualization."} 
ZAME~\cite{elmqvist_zame_2008} and TimeMatrix~\cite{yi_timematrix_2010} are content-aware technique that relies on semantic zoom. It first reorders~\cite{siirtola_interaction_1999} rows and columns to group related elements and then aggregates neighboring cells dependent on the current zoom level.
Horak et al.~\cite{horak_responsive_2021} provide both geometric and semantic zooming in matrices. However, their technique has not been empirically evaluated.
In our second user study, following a widely used design (e.g., Google Maps), we tested the pan\&zoom condition, which allows the user to scroll the mouse wheel to continuously zoom in and out a certain region of the matrix. 

\textbf{Evaluating focus+context, overview+detail, and pan\&zoom.}
These three interaction techniques have been extensively evaluated in various applications, but not in the context of MMV.
Baudisch et al.~\cite{baudisch_keeping_2002} found that focus+context had reduced error rates and time (up to 36\% faster) over pan\&zoom and overview+detail for finding connections on a circuit board and the closest hotels on a map.
Similarly, Gutwin et al.~\cite{gutwin_fisheyes_2003} concluded fisheye views to be advantageous over overview+details and zooming for large steering tasks. 
Shoemaker and Gutwin~\cite{shoemaker_supporting_2007} also found fisheye lens superior over standalone panning and over zooming for multi-point target acquisition on images.
On the other hand, Rønne and Hornbæk~\cite{ronne_jakobsen_sizing_2011} had opposite findings for locating, comparing, and tracking objects on geographic maps. They found fisheye had the worst performance, while overview+detail performed best.
These previous studies had mixed results for different applications, and none of them were conducted in the context of MMV.
Most similar to our second study is the study with 12 participants by Pietriga et al.~\cite{pietriga_pointing_2007} for multiscale search tasks. They found overview+details superior over the fisheye lens, while both techniques outperformed pan\&zoom.
They tested the conditions in a matrix-like application but with no multivariate details in the cells.
They also only tested one searching task, and the tested fisheye lens was the classical one with non-linear radial distortion, which breaks the regularity of the grid.

\begin{figure*}[t]
    \centering
        \subfigure[Cartesian lens (\lcartesian{}{})] {
        \includegraphics[height=4.15cm]{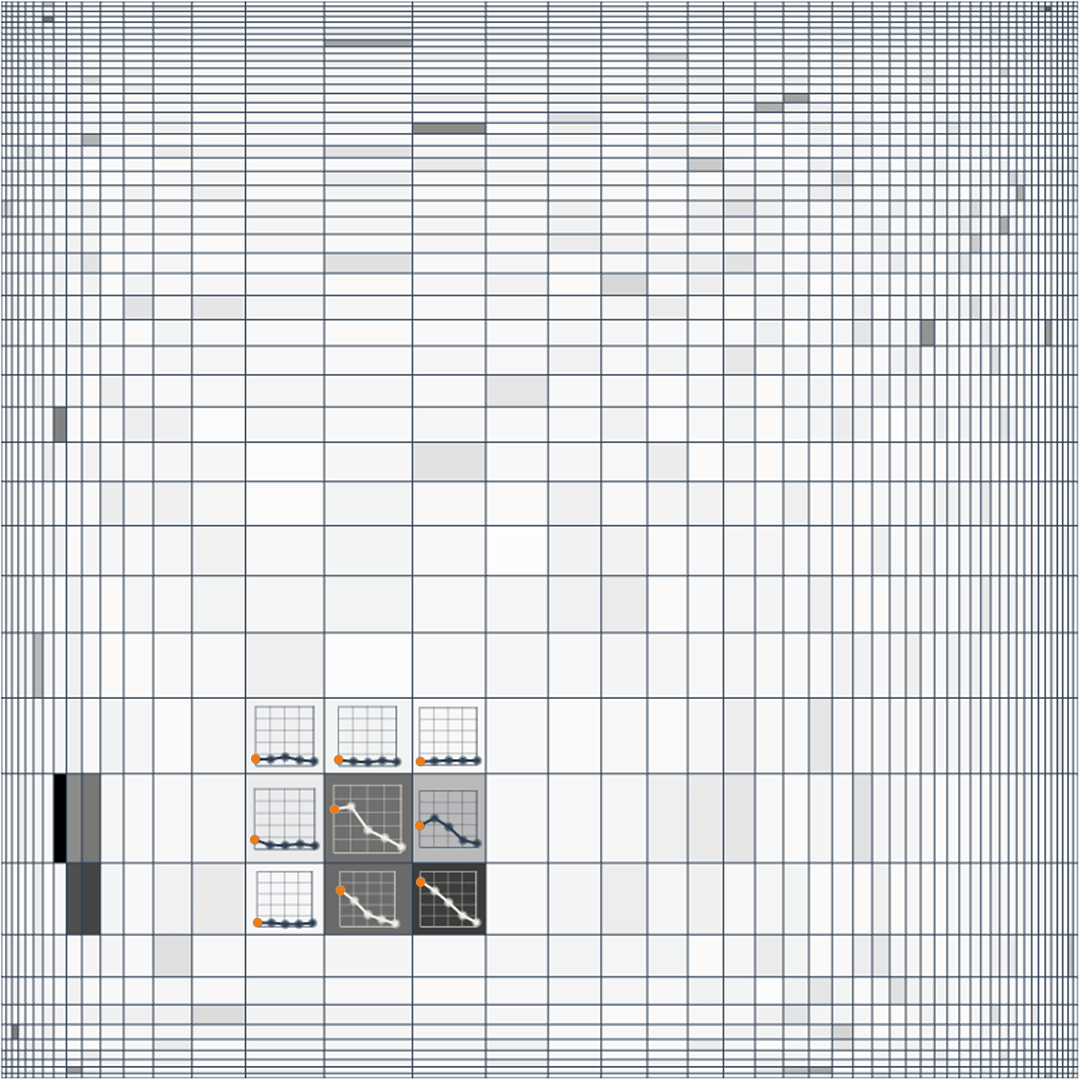}
            \label{fig:study-1-lenses-cartesian}
        }
        \subfigure[Fisheye lens (\lfisheye{})] {
        \includegraphics[height=4.15cm]{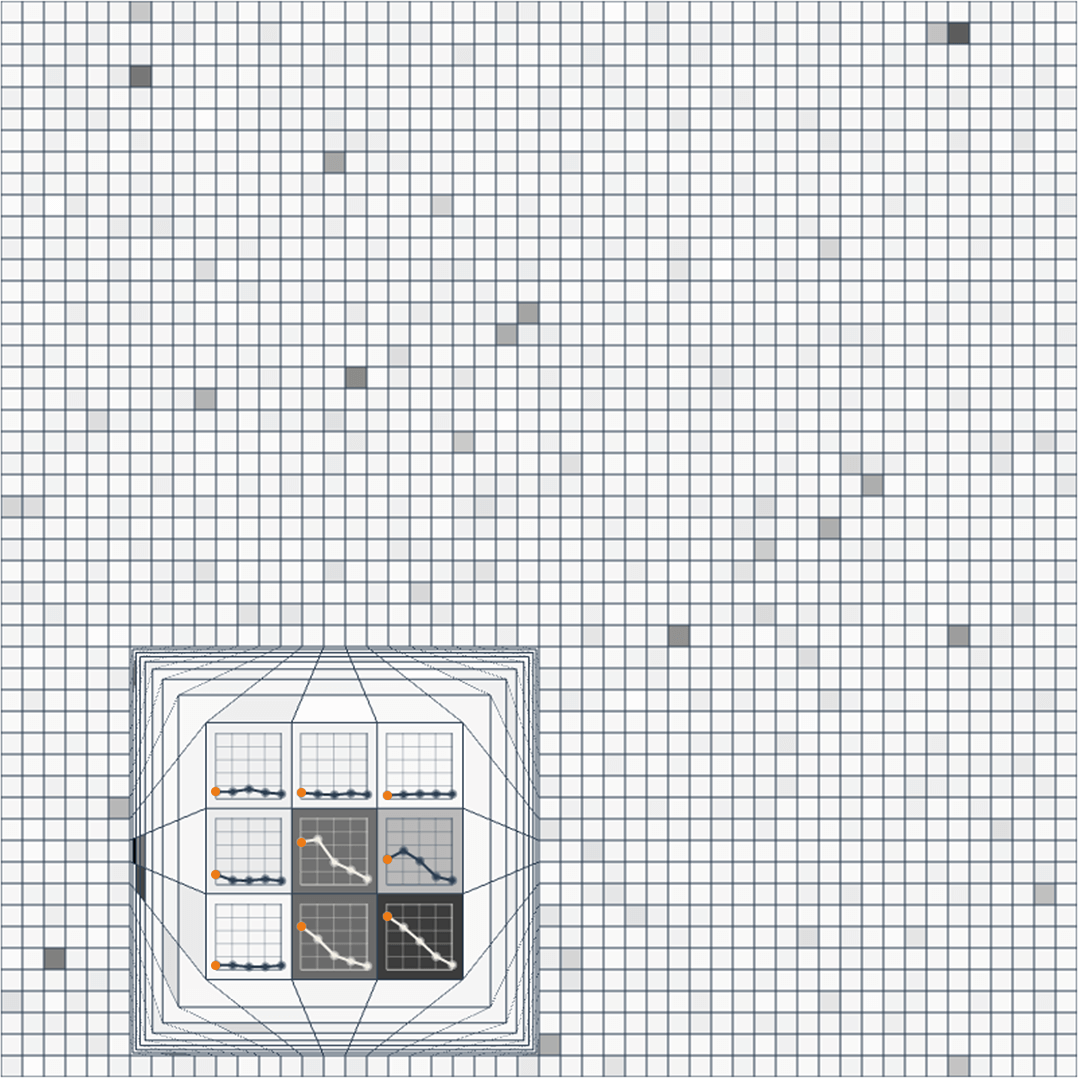}
            \label{fig:study-1-lenses-fisheye}
        }
        \subfigure[TableLens Stretch (\ltable{})] {
        \includegraphics[height=4.15cm]{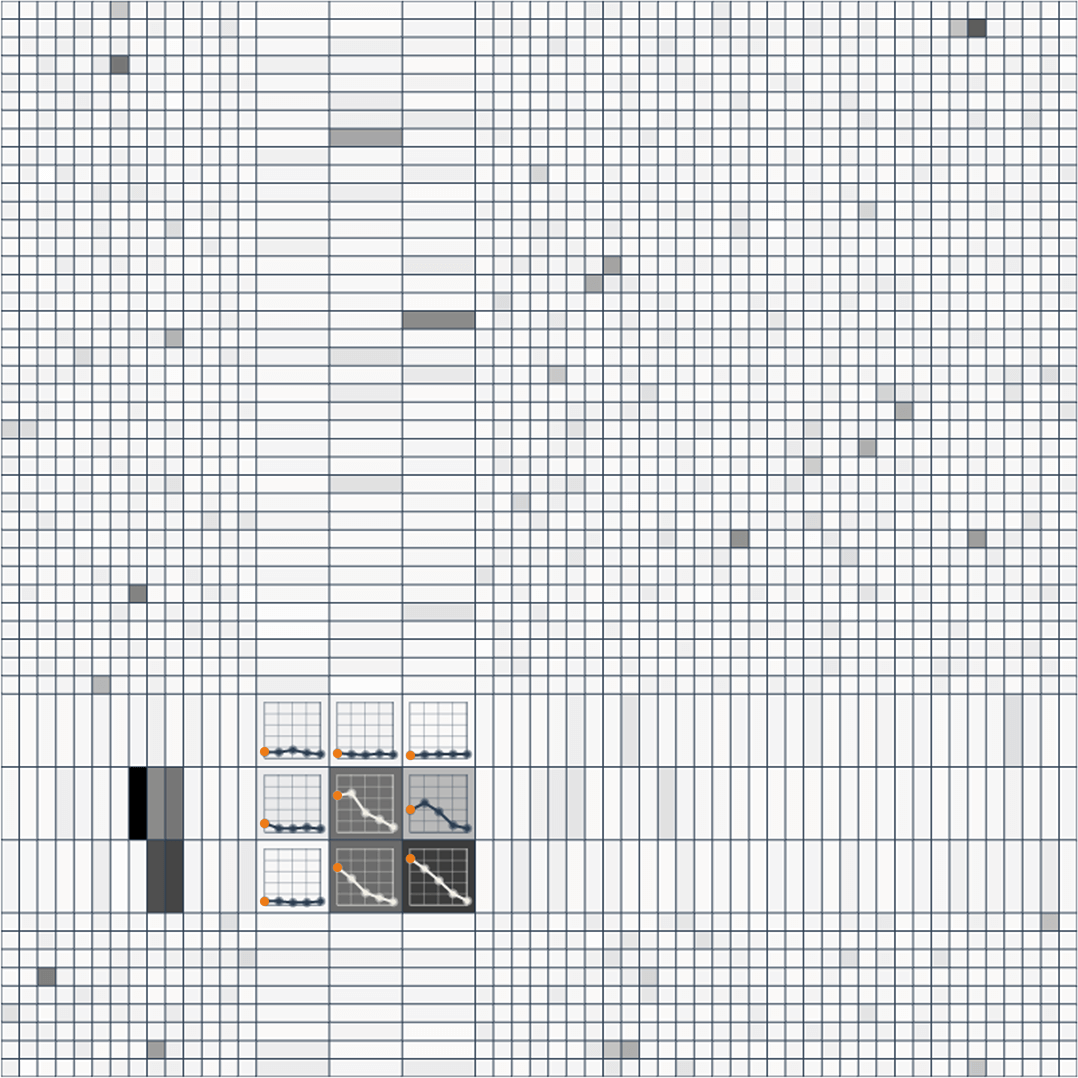}
            \label{fig:study-1-lenses-stretch}
        }
        \subfigure[TableLens Step (\ltabledc{})] {
        \includegraphics[height=4.63cm]{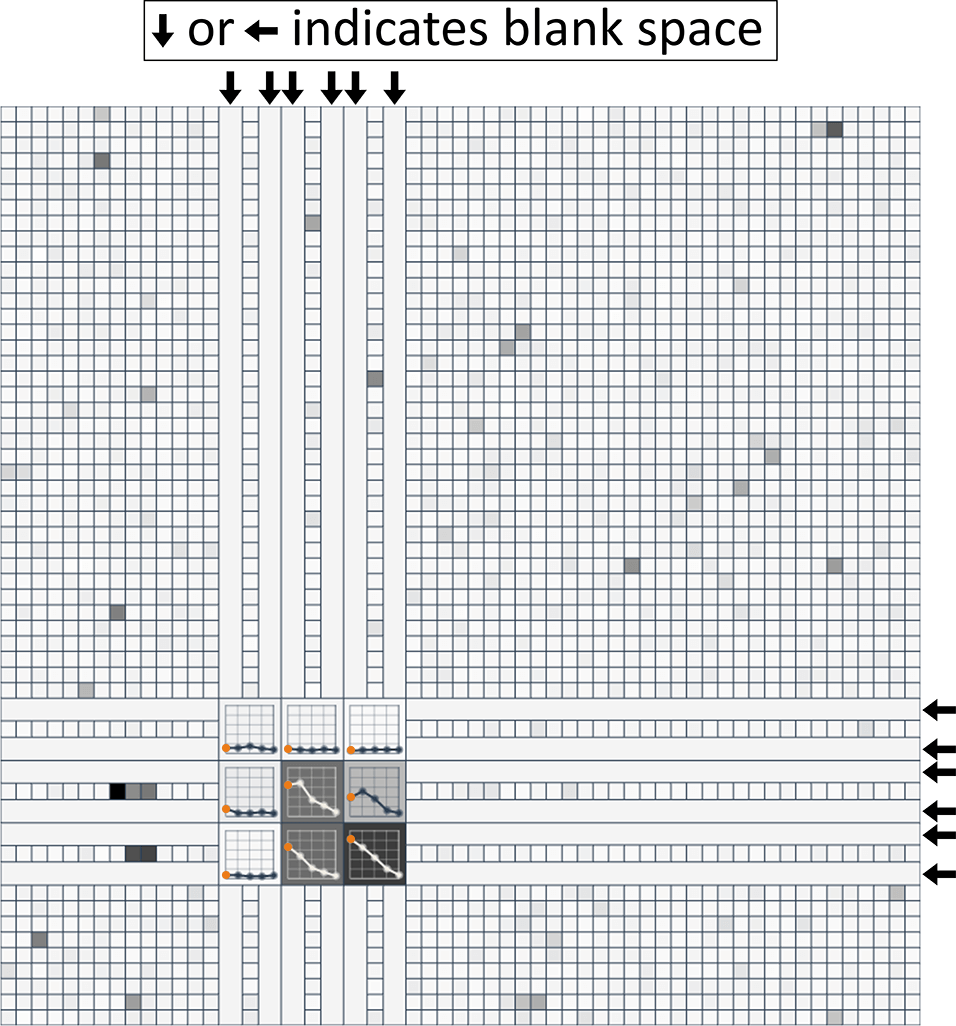}
            \label{fig:study-1-lenses-step}
        }
    \Description{Four difference lenses tested in the user study. Different characteristics of them are summarized in Table 1.}
    \caption{\textbf{Study 1 visualization conditions with 50$\times$50 matrices}: four interactive lenses tested in the user study. An interactive demo is available at
    \url{https://mmvdemo.github.io/},
    and has been tested with Chrome and Edge browsers.}
    \label{fig:study-1-lenses}
    \vspace{1mm}
\end{figure*}

\section{Study 1 --- Different Lenses in MMV}
This first study is intended to address the gaps in the literature described above in terms of how different distortions can impact perception and task performance in using focus+context (or lens) techniques for exploring MMV. 
As this study was the first to compare different lenses in MMV, there was little empirical knowledge about the user performance with different lenses. 
Thus, our first study is exploratory rather than confirmatory. 
We pre-registered the study at \url{https://osf.io/dxsr5}.
Meanwhile, test conditions are demonstrated in the supplementary video, and detailed results of statistical tests are provided in supplementary materials.

\subsection{Experimental Conditions}
\label{sec:study-1-exp_conditions}

Using lenses to explore MMV selectively enlarges an ROI of the matrix, so that the enlarged cells have enough space to show the detail.
These enlarged cells are also referred as focus cells.
To make space for the focus cells, lenses introduce two types of distortions: \emph{focal} and \emph{contextual} distortion. Focal distortion applies to cells at the inner border of the lens. Contextual distortion, on the other hand, applies to cells outside the lens.

Unlike lenses in a map or images, lenses in MMV have more constraints.
While, there are more flexibility with their elementary units in a map or images, the cells in MMV are all the same size and laid in a regular grid (i.e., rows and columns are orthogonal to each other).
Additionally, according to Carpendale et al.~\cite{carpendale_extending_1997}, gaps are also considered an important distortion characteristic.
In summary, we identified three characteristics for the distortions of lenses in MMV:
\textbf{\emph{regularity}}---whether the cells are rendered in a regular or orthogonal grid, 
\textbf{\emph{uniformity}}---whether the cells are sized uniformly, and 
\textbf{\emph{continuity}}---whether the cells are laid out continuously.
Those characteristics can be used to model both the \emph{focal} and \emph{contextual} distortions in the lenses.
We chose four lenses for our study, according to Carpendale et al.'s distortion taxonomy~\cite{carpendale_extending_1997}:

\begin{table}[t]
    \vspace{0.5em}
	\centering
    \includegraphics[width=0.98\columnwidth]{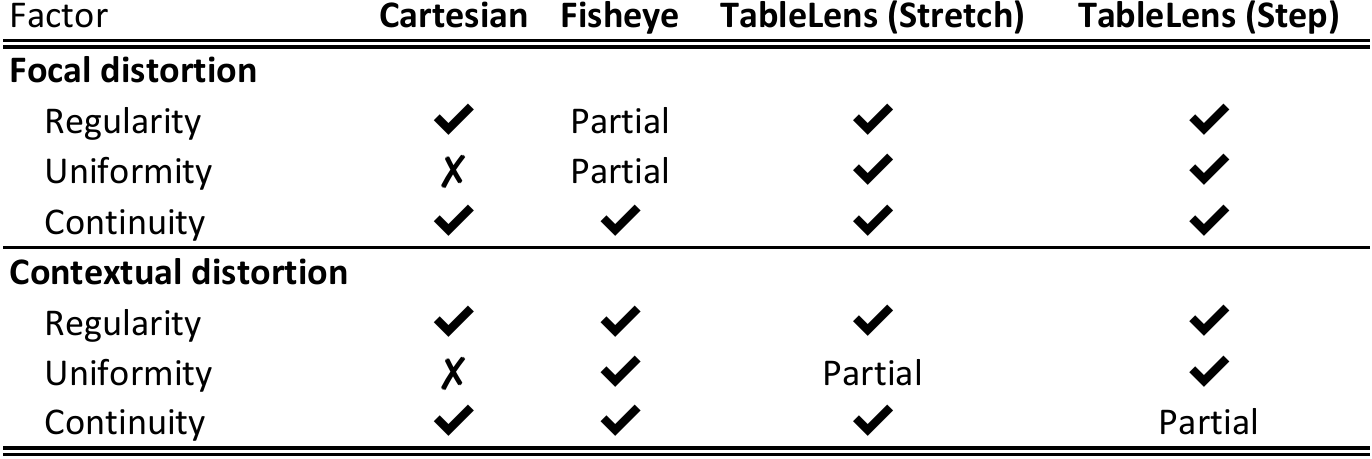}
    \vspace{1.5em}
    \Description{Summary of characteristics of the tested conditions, including focal distortion and contextual distortion. Within these two types of distortion, there are three types of charateristics: regularity, uniformity, and continuity. See Section 3.1 for more details.}
    \caption{\textbf{Characteristic Comparison:} of the four tested lenses.}
    \label{tab:study-1-design-space}
    \vspace{-2em}
\end{table}

\textbf{\lcartesian{}} distorts the entire matrix continuously such that the cells are proportional sized based on their distances to the cursor (\autoref{fig:study-1-lenses-cartesian}).
In \lcartesian{}, cells in focal and contextual regions are all in a regular grid but sized differently.

\textbf{\lfisheye{}} magnifies the center part of the focus and shrinks the surrounding area around the lens's inner boundary. 
The focal cells need to be rendered in a regular grid and sized uniformly. 
To continuously embed the focal region inside the contextual region, distortion must be applied to the focal area's \emph{transition area}.
As a result, cells inside the transition area are rendered irregularly and are sized differently (see \autoref{fig:study-1-lenses-fisheye}). 

\textbf{\ltable{} and \ltabledc{}} are two variations of TableLens that enlarge a fixed number of rows and columns around the focal point and uniformly compress the remaining matrix. 
\ltable{} stretches the enlarged rows and columns on either of the two axes (\autoref{fig:study-1-lenses-stretch});
\ltabledc{} preserves the cells' aspect ratio by adding white space around enlarged rows and columns on either of the two axes, which introduces discontinuities (see the blank space in \autoref{fig:study-1-lenses-step}).
\ltable{} and \ltabledc{} both have regular and uniform cells in the focal region. 

We summarize the characteristics of the four tested lenses in~\autoref{tab:study-1-design-space}.
The four tested lenses cover a variety of characteristics, and the study is to investigate how those characteristics affect perception and interaction performance.

\subsection{Data} \label{sec:study-1-data}

We used time series as the multivariate data in our study, as it is widely used~\cite{behrisch2014visual,bach_small_2015,wood_visualizing_2011,fischer_visual_2021,beck_state_2014,yi_timematrix_2010} and has not been empirically tested in the context of MMV before.
We generated task datasets consisting of three dimensions ($x$, $y$, and $t$): with $x$ and $y$ as the rows and columns in the matrix and $t$ as the number of time instances in the time series.
For a particular value of $t$, the $x$ and $y$ dimensions are shown as a traditional univariate matrix, which we refer to as the \emph{context}. The $t$ dimension is revealed interactively upon placing the lens over a focus region. Enlarging the cells under a lens's focal area provides space for displaying this $t$ dimension as a line chart, which we refer to as the \emph{focus}.
Each data set contains $x \times y \times t$ values, and each cell contains $t$ values as multivariate details.

We included two data sizes: \lsmall{} with $50 \times 50 \times 5$ and \llarge{} with $100 \times 100 \times 5$.
We decided to test large matrices because small matrices have enough space for each cell to show the multivariate details constantly, and interaction is less necessary for them.
We also chose to study the scalability in terms of the matrices' size and keep the detail's size (i.e., the number of time instances) unchanged.

\begin{figure}
  \vspace{2mm}
  \centering
  \includegraphics[width=0.95\columnwidth]{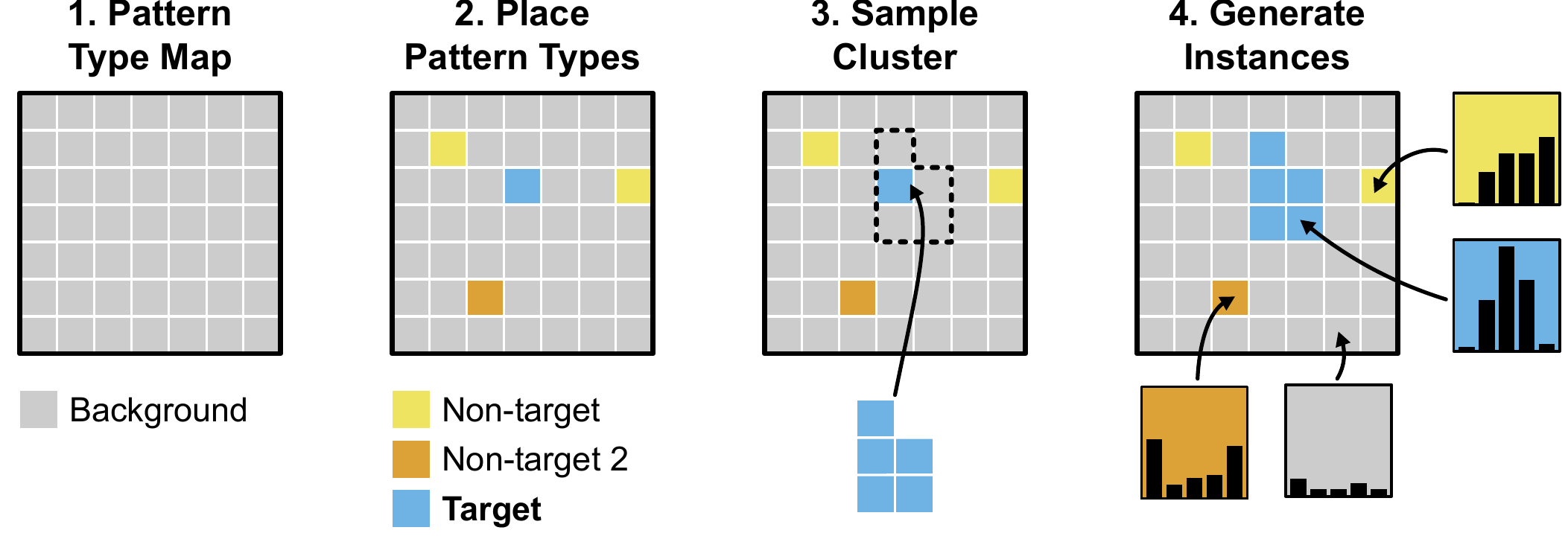}
  \Description{Demonstration of the four steps of data generation: first generate pattern type map, then place pattern types, followed by sample cluster, and finally generate instances. Full details in Section 3.2.}
  \caption{\textbf{Data Generation.} To generate a unique dataset for each trial, we following this 3-step pipeline. First, we sampled the location of target patterns. Second, for each target pattern location we sampled cluster of target pattern types. Finally, for each cell we generate a 5-point temporal pattern instance based on the cell's assigned pattern type.}
  \label{fig:study-1-data-generation}
  \vspace{1mm}
\end{figure}

\begin{figure}[t]
  \centering
  \includegraphics[width=0.95\columnwidth]{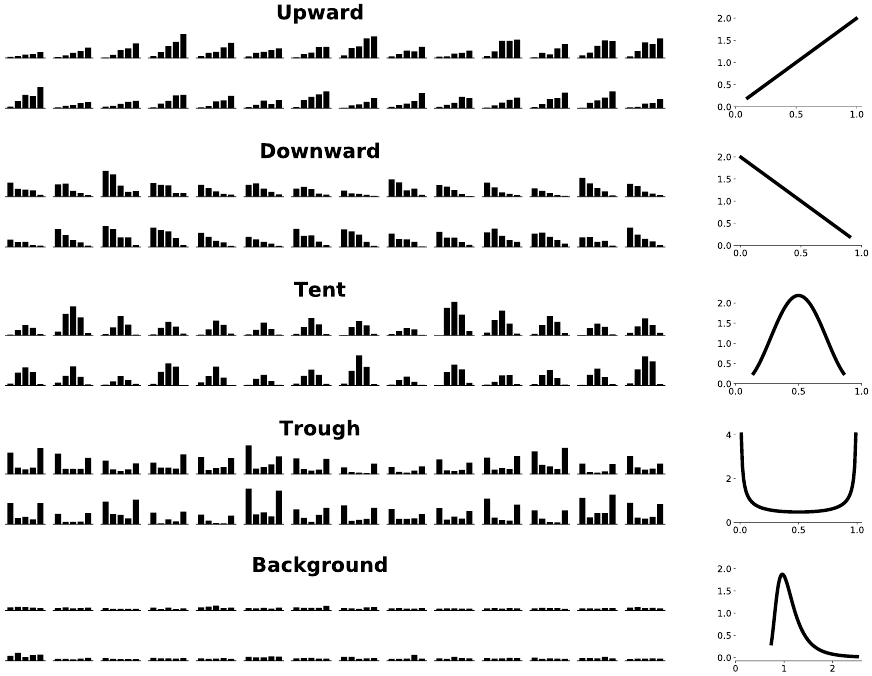}
  \Description{Demonstration of pattern instances of five different types: upward, downward, tent, trough, and background. Full details in Section 3.2.}
  \caption{\textbf{Pattern Types.} Example pattern instances for each of the five pattern types: upward, downward, tent, trough, and background. The instances slightly differ in their shape and magnitude to mimic realistic data. On the right we plot the probability density function of each pattern.}
  \label{fig:study-1-patterns}
  \vspace{-0.5em}
\end{figure}

The goal of the tasks is to evaluate and compare the temporal
patterns that arise along the third dimension (i.e., the \emph{details}). 
As shown in~\autoref{fig:study-1-data-generation}, our data generation consisted of three steps: first we sampled a matrix of different pattern \emph{types}, then we expanded the target pattern types into clusters, and finally sampled the actual pattern \emph{instances} for each cell. To avoid memory effects and ensure that participants would have to inspect the patterns under the focal area, we sampled pattern instances from five distinct distributions (\autoref{fig:study-1-patterns}) inspired by temporal patterns described by Correll and Gleicher~\cite{correll2016semantics}: upward, downward, tent, trough, and background.

In the first step, we created a matrix of pattern types. In the beginning, the matrix contained only background pattern types (\autoref{fig:study-1-data-generation}.1). We then randomly placed non-target pattern types into the matrix (\autoref{fig:study-1-data-generation}.2). We added these non-target patterns as lightweight distractions and made the final dataset more realistic. We then randomly sampled a position for the target pattern type. Further, to make it easier to locate the target cells, we sampled a cluster of target pattern types using a 2D Gaussian distribution centered on the previously determined target pattern type location (\autoref{fig:study-1-data-generation}.3).

Finally, for each cell, we generated a pattern instance by randomly sampling 100 values from different distributions (\autoref{tab:study-1-pattern-distributions}) and aggregated them in a 5-bin histogram. This approach created patterns instances that differ slightly in shape and magnitude while still being distinct enough to avoid ambiguousness (\autoref{fig:study-1-patterns}). This approach strikes a balance between predictability and generality.

\begin{table}[t]
\vspace{1mm}
{\small
\begin{tabularx}{\columnwidth}{ r X }
  \toprule
  Pattern type & Data distribution \\ 
  \midrule
  Upward     & Beta with $\alpha{=}2$ and with $\beta{=}1$ \\  
  Downward   & Beta with $\alpha{=}1$ and with $\beta{=}2$ \\  
  Tent       & Beta with $\alpha{=}4$ and with $\beta{=}4$ \\  
  Trough     & Beta with $\alpha{=}\frac{1}{3}$ and with $\beta{=}\frac{1}{3}$ \\
  Background & Fréchet with $\alpha{=}5$, s=1, m=0 \\
  \bottomrule
\end{tabularx}
}
\vspace{1.5em}
\Description{Parameters used for data generation. Details are in Section 3.2.}
\caption{\textbf{Pattern Distribution Functions:} Pattern instances are generated from the histogram of 100 sampled values from the associated data distribution. The probability density function of each pattern type is shown in~\autoref{fig:study-1-patterns}.\label{tab:study-1-pattern-distributions}}
\vspace{-1.5em}
\end{table}

\subsection{Interactions and Tasks} 
\label{sec:study-1-tasks}

The participants were asked to interact with the MMV, which by default showed a slice of our 3D dataset as a univariate matrix visualization (i.e., a heatmap for one of the five time instances, see \autoref{sec:study-1-data} for details).
We used a continuous color schema from white to black to encode matrix cells.
Darker cells indicate higher values. Such a color scheme is colorblind-friendly. 
Upon moving the mouse cursor over the MMV, the lens enlarges an area to show the details of the time series as a line chart.
In each chart, the line connects five dots representing the five values. The dot representing the value of the background color is additionally highlighted to represent the currently selected time instance. Participants can switch the time instance by clicking on a respective dot in an embedded line chart.
To clearly present interactive line charts while still keeping the context cells legible, we conducted a series of internal tests to find an appropriate combination of the parameters for \emph{the number of cells to be enlarged} and \emph{the magnification factor}. 
We ensure the size of the enlarged area and each line chart to be consistent across different lenses.
For the two data sizes we tested, the enlarged cell is in the same size: four times the side length as the original cell in \lsmall{} data and eight times in \llarge{} data.
\yalong{A line chart in such a size can be reasonably interpreted and interacted by users.}
We kept the ratio of the enlarged cells the same as the matrix, i.e., 1:1, and decided to enlarge $3 \times 3$ matrix cells as the focal area to show their line charts.
Increasing the number of enlarged cells or magnification factor makes it challenging to interpret the color of the surrounding cells, even for screens with a standard resolution. For example, on a Full-HD (or 1080p) screen, we used 800 $\times$ 800 pixels to visualize a 100 $\times$ 100 matrix for \llarge{} data, and the size of a context cell is 5 $\times$ 5 pixels in \lfisheye{}, \ltable{}, and \ltabledc{} when the lens is on top of the matrix. 
Some context cells in \lcartesian{} are even smaller.
According to our internal tests, interpreting colors in context cells smaller than 5 × 5 pixels is difficult.
\autoref{fig:study-1-lenses} demonstrates the tested interaction techniques for enlarging their focal areas.

The most basic interaction for exploring an MMV includes three steps: first finding the cell(s) of interest, then moving the cursor towards them to enlarge them, and finally checking their embedded details. 
Additionally, in some cases, users must inspect both the focal and context areas.
Past research in HCI and visualization proposed taxonomies~\cite{munzner2014visualization,yang_embodied_2021,laviola20173d,nilsson_natural_2018,lam_framework_2008,wang_baldonado_guidelines_2000} for these interactions and conducted studies in various applications (see \autoref{sec:related_work}), but not with MMV.
We analyzed previous work to break down the fundamental MMV interactions into four components. 
First, \textbf{\emph{wayfinding}} is the process of searching targets cells.
Second, \textbf{\emph{travel}} refers to the act of moving the mouse cursor to the target cells. 
Third, \textbf{\emph{interpretation}} is the activity of interpreting the targets' visual encoding.
And finally, \textbf{\emph{context-switching}} refers to re-interpreting a changed view, for example, when updating visualization through interaction or moving the focus to a different part of the view.
We then designed three tasks to cover different aspects of the identified components.
Since we used the same visual encoding for all conditions (i.e., the line chart), we did not expect a noticeable performance difference in \emph{interpretation}. 
Thus, our focus is on evaluating the other three.
In the following, we first describe the study tasks with practical examples in the context of multi-year population data for counties in the United States in an MMV. 
This data is easily accessible and understandable. 
Each matrix cell represents a county and contains multi-year population data. The cells are typically placed according to their relative geographic locations, which is similar to the tile map representation~\cite{mcneill_generating_2017} used by Wood et al.~\cite{wood_visualizing_2011}.
We then discuss the rationale and motivation of our task choices.

\textbf{In the first task (\tlocate{})}, we asked participants to click on a specific cell highlighted with an orange outline. 
Our goal is to test how the distortion influences the participants' perceptual ability to locate a specific cell. 
Thus, we remove the highlighting as soon as the participants move their cursor into the matrix. 
For accessibility, the highlighting reappears once the cursor is moved outside the matrix. Additionally, we followed Blanch et al.'s~\cite{blanch2011benchmarking} approach and added visual distractors, i.e., non-target patterns (\autoref{sec:study-1-data}) in our case.
A frequent operation in analyzing population data is to investigate the temporal trend of a given location, like \emph{``what is the temporal trend of Middlesex County, MA in the last five years?''}

The \tlocate{} task was designed to inspect the \textbf{\emph{travel}} component in interactions. Locating and selecting an element is the most common task in graphical user interfaces and visualizations and is a primitive visualization interaction ~\cite{soukoreff2004towards,munzner2014visualization}. It is also a standard task tested in many user studies (e.g.,~\cite{ronne_jakobsen_sizing_2011,javed_polyzoom_2012}). Fitts' law provides a way to quantify the performance of basic target selection~\cite{soukoreff2004towards}. However, the standard model does not consider the lenses' distortion effects. 
This task aims to investigate how different types of distortion influence the performance of locating target.

\textbf{For the second task (\tcompare{})}, we asked the participants to \emph{search} for the cell with the highest single value among a cluster of cells, which is a 7$\times$7 region for a 3$\times$3 lens in the study. Since we test the ability to locate a cell in the \tlocate{} task, we decided to permanently highlight the search area with an orange outline. To enforce the use of the lenses, we pre-selected a value of the time series that does not reveal the target patterns. Only when the user employs the lens the relevant details of the multivariate pattern will be visible. An example of this task in population analysis can be \emph{``Within New England (a set of counties), which county has the largest number of population of a single year in the last five years?''}

The \tcompare{} task involves both the \textbf{\emph{travel}} and \textbf{\emph{wayfinding}} components. 
\emph{Wayfinding} is an essential step for any high-level visual analytics task~\cite{munzner2014visualization}.
In order to find the cell with the highest value, the participants had to inspect and compare the details of multiple cells.
Similar tasks have been tested in other contexts, for example, by Pietriga et al.~\cite{pietriga_pointing_2007} for multiscale searching and Jakobsen \& Hornbæk for geographic maps~\cite{ronne_jakobsen_sizing_2011}.
We expect that the different lens distortions will influence the performance of \emph{wayfinding}, especially in an interactive scenario.
It is impractical to only test \emph{wayfinding} performance without physically \emph{traveling} to the targets. Thus, we included both these two components in this task.

\textbf{In the third task (\tcontext{})}, we asked the participants to find the largest cluster at the time instance where a given cell reaches its highest value. The participants needed to move the mouse cursor to a cell highlighted with an orange border and click on the dot representing the highest value. 
The representation of MMV (i.e., the heatmap) will then be updated to the time instance corresponding to the clicked dot.
Subsequently, several clusters in the sizes between 5$\times$5 to 7$\times$7 of dark cells appeared in the matrix, and participants were asked to select the largest one. For instance, a practical use case in population analysis can be \emph{``At the year when the population of Orange County, FL reaches its peak value, where is the largest region with high population?''}

The \tcontext{} task includes the \textbf{\emph{travel}}, \textbf{\emph{wayfinding}}, and \textbf{\emph{context-switching}} components. 
Context-switching frequently happens in interactive visualization and multi-scale navigation and has been tested in various scenarios~\cite{yang_embodied_2021,plumlee_zooming_2002,plumlee_zooming_2006,ronne_jakobsen_sizing_2011}, but not with MMV. 
In MMV, users have to \emph{switch their context} in many scenarios, e.g., when enlarging the cells to show the line charts, changing the time instance, and moving their focus between the focal and contextual areas.
We expect different types of distortion will influence the \emph{context-switching} performance.
Again, it is unrealistic to test \emph{context-switching} without \emph{travel} and \emph{wayfinding}. Thus, we include all these three components in this task.

\vspace{-1em}
\subsection{Experimental Design}
\label{sec:study-1-exp-design}
We included two factors in the user study: \flens{} and \fsize{}.
The \flens{} had four different lenses, as described in~\autoref{sec:study-1-exp_conditions}.
The \fsize{} had two data sizes
as described in~\autoref{sec:study-1-data}.
The experiment followed a full-factorial within-subject design. We used a Latin square (4 groups) to balance the visualizations but kept the ordering of tasks consistent: first \tlocate{}, then \tcompare{}, and finally \tcontext{}.
Each participant completed 48 study trials: 4 visualizations $\times$ 2 data sizes $\times$ 3 tasks $\times$ 2 repetitions. 
The entire study has been tested on common resolution settings (FHD, QHD and UHD).

\textbf{Participants.}
We recruited 48 participants on Prolific (\url{https://www.prolific.co}). All participants were located in the US and spoke English natively. To ensure data quality, we restrict participation to workers who had an acceptance rate above 90\%.
Our final participant pool consisted of 19 female, 26 male, and three non-binary participants. Out of those participants, twelve had a master's degree, 
16 had a bachelor's degree, 14 had a high school degree, and six did not specify their education levels. 
Finally, 4 participants were between 18-20, 17 participants were between the age of 21-30, 18 participants were between 31-40, five participants were between 41-50, four participants were above 50.
We compensated each participant with 9 USD, for an hourly rate of 12 USD.

\textbf{Procedures.}
Participants were first presented with the consent form, which provided information about the study's purpose and the procedure.
After signing the consent form electronically, the participants had to watch a short training video (1 minute and 13 seconds) that demonstrated how to read and interact the MMV. 

Participants completed the three tasks one by one based on a Latin square design.
Prior to working with a new lens, we showed a video demonstrating how to interact with the matrix using the current lens.
Each (visualization $\times$ task) block started with two training trials followed by study trials. 
Before each training trial, we encouraged participants to get familiar with the visualization condition and explicitly told them they were not timed for the training.
We also ensured that participants submitted the correct answers in training trials before we allowed them to proceed. 
Before starting the study trials, we asked the participants to complete the trials \emph{``as accurately and as quickly as they can, and accuracy is more important,''} and informed them that these trials were timed.
To start a trial, participants had to click on a ``start'' button placed in the same location above the MMV. This ensured a consistent cursor starting point 
and precisely measured the task duration. The visualization only appeared after clicking the start button.

After each task, participants were asked to rate each visualization’s perceived difficulty and write their justifications. We collected the demographic information as the final step. The average completion time was around 45 minutes.

\textbf{Measurements.}
We collected the following measurements during the user study: 
\textbf{\emph{Time.}}
We measured the time in milliseconds from the moment the user clicked on the start button until they selected an answer. 
\textbf{\emph{Accuracy.}}
We measured the participants accuracy as the ratio of correct over all answers.
\textbf{\emph{Perceived Difficulty Rating.}}
After the user completed a task with all four lenses we asked the participants to rate \emph{``how hard was performing the task with each of the visualizations?''} on a 5-point Likert scale ranging from \emph{easy (1)} to \emph{hard (5)}.
The questionnaire listed visualizations in the same order as presented in the user study with figures.
\textbf{\emph{Qualitative Feedback.}} We also asked participants to optionally justify their perceived difficulty ratings in text. 

\textbf{Statistical Analysis.}
For dependent variables or their transformed values that met the normality assumption (i.e., time), we used \emph{linear mixed modeling} to evaluate the effect of independent variables on the dependent variables~\cite{Bates2015}. 
Compared to repeated measure ANOVA, linear mixed modeling 
does not have the constraint of sphericity~\cite[Ch.\ 13]{field2012discovering}.
We modeled all independent variables (four visualization techniques and two data sizes) and their interactions as fixed effects. A within-subject design with random intercepts was used for all models. 
We evaluated the significance of the inclusion of an independent variable or interaction terms using log-likelihood ratio. 
We then performed Tukey's HSD post-hoc tests for pair-wise comparisons using the least square means~\cite{Lenth2016}. 
We used predicted vs. residual and Q---Q plots to graphically evaluate the homoscedasticity and normality of the Pearson residuals respectively. 
For other dependent variables that cannot meet the normality assumption (i.e., accuracy and perceived difficulty rating), we used the \emph{Friedman} test to evaluate the effect of the independent variable, as well as a Wilcoxon-Nemenyi-McDonald-Thompson test for pair-wise comparisons. Significance values are reported for $p < .05 (*)$, $p < .01 (**)$, and $p < .001 (***)$, abbreviated by the number of stars in parenthesis.
Numbers in parentheses indicate mean values and 95\% confidence intervals (CI). 
We also calculated the Cohen's d as indicators of effect size for significant comparisons.

\begin{figure}
	\centering
	\includegraphics[width=\columnwidth]{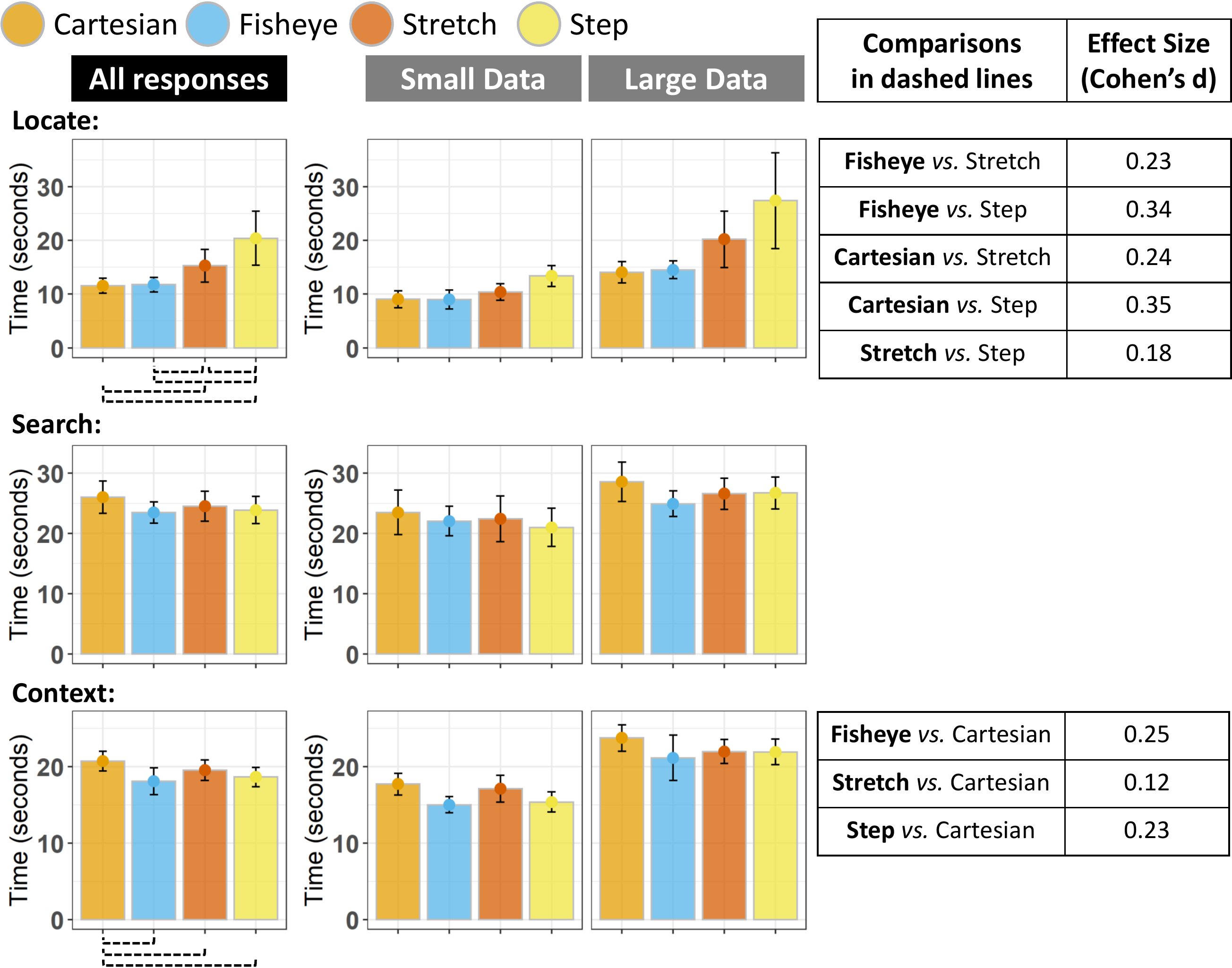}
	\Description{Left are bar charts with error bars showing 95\% confidence intervals. The bar charts are showing the time performance of four tested conditions. Right are tables showing effect sizes of significant comparisons. Details are in Section 3.5.}
	\caption{\emph{Time} by task and in different data sizes. Confidence intervals indicate 95\% confidence for mean values. Dashed lines indicate statistical significance for $p<.05$. Tables are showing their effect sizes.}
	\label{fig:study-1-result-time}
\end{figure} 

\subsection{Results}
\begin{figure}[t]
	\centering
	\includegraphics[width=0.98\columnwidth]{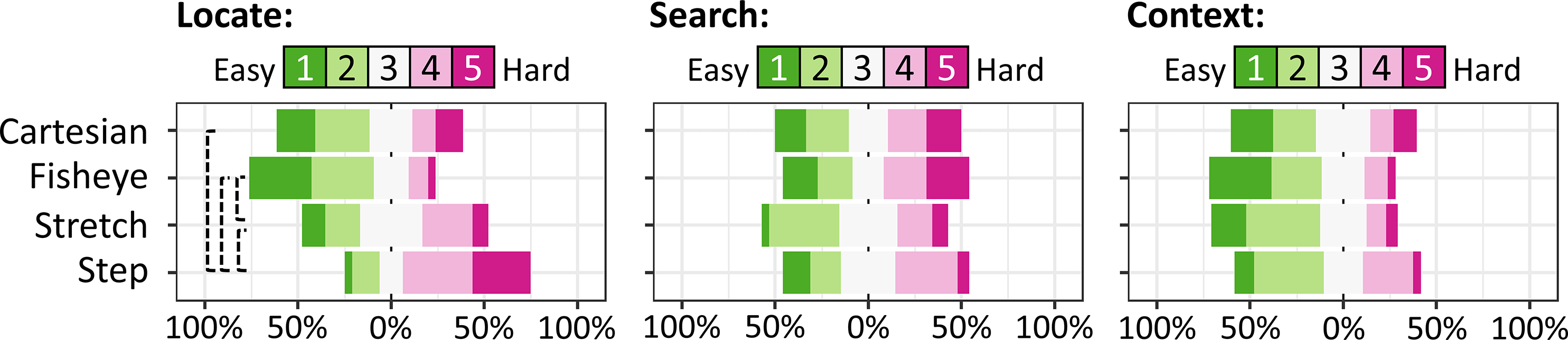}
	\vspace{1mm}
	\Description{Stacked bar charts showing the subjective ratings. Details are in Section 3.5.}
	\caption{\emph{Perceived difficulty ratings} by task. Dashed lines indicate $p<.05$.}
	\label{fig:study-1-result-rating}
	\vspace{-0.3em}
\end{figure}

The accuracy was similarly high across all conditions: on average, 95.3\% for \tlocate{}, 92.3\% for \tcompare{}, and 78.1\% for \tcontext{}. We did not find any significant differences between \flens{} and \fsize{} on accuracy. Therefore, we focus our analysis on the time (\autoref{fig:study-1-result-time}), perceived difficulty (\autoref{fig:study-1-result-rating}), and qualitative feedback.

We found \flens{} had a significant effect on time in both \tlocate{} ($***$) and \tcontext{} ($***$) tasks, but no significant effect in the \tcompare{} task ($p=0.163$).
We also found \fsize{} had a significant effect on time in all tasks (all $***$).
No significant effect has been found in the interaction between \flens{} and \fsize{} for all tasks.
For the perceived difficulty ratings, \flens{} had a significant effect in the \tlocate{} task ($***$), but not in \tcompare{} 
and \tcontext{} 
tasks.
All statistical results are included in the supplementary materials.

\smallskip
\noindent\textbf{Quantitative Key Findings.}

\textbf{\emph{\lfisheye{} was the best performing technique.}}
\lfisheye{} (11.8s, CI=1.4s) and \lcartesian{} (11.7s, CI=1.4s) had a similar  performance in the \tlocate{} task, and they both outperformed \ltable{} (15.3s, CI=3s) and \ltabledc{} (20.4s, CI=5s) (all $***$).
The perceived difficulty ratings mostly aligned with the performance results. I.e., participants rated \lfisheye{} (2.19, CI=0.33) easier than \ltable{} (3, CI=0.33, $*$) and \ltabledc{} (3.77, CI=0.34, $***$). \lcartesian{} (2.71, CI=0.39) was also rated easier than \ltabledc{} ($***$).  
\lfisheye{} (18s, CI=1.8s) also outperformed \lcartesian{} (20.7s, CI=1.3s) in the \tcontext{} task ($***$).
Overall, \lfisheye{} was the best choice for the tested tasks.

\textbf{\emph{\lcartesian{} was not ideal for the \tcontext{} task.}}
\lcartesian{} (20.7s, CI=1.3s) was slower than \lfisheye{} (18s, CI=1.8s, $***$), \ltabledc{} (18.6s, CI=1.3s, $**$), and \ltable{} (19.5s, CI=1.3s, $*$). Participants also tended to consider \lcartesian{} (2.69, CI=0.38) more difficult than \lfisheye{} (2.27, CI=0.34) and \ltable{} (2.46, CI=0.32), but not statistically significant.

\textbf{\emph{\ltable{} had advantage over \ltabledc{} in the \tlocate{} task.}}
The only performance difference between these two condition is that \ltable{} (15.3s, CI=3s) was faster than \ltabledc{} (20.4s, CI=5s) in the \tlocate{} task ($***$). 
Again, the perceived difficulty ratings aligned with the performance, where participants found \ltable{} (3, CI=0.33) easier than \ltabledc{} (3.77, CI=0.34) in the \tlocate{} task ($*$).

\textbf{\emph{All performed similar in the \tcompare{} task.}}
We did not find an effect of visualization on time or on the perceived difficulty rating.

\smallskip
\noindent\textbf{Qualitative Feedback.}

We also asked participants to justify their perceived difficulty rating after each task.
We analyze the collected feedback to get an overview of the pros and cons of each lens.

\textbf{\emph{\lcartesian{}}} was commented to be \emph{``natural''} by six participants. They found it to be intuitive to have cells that are closer to the cursor to be larger.
However, 18 participants complained its distortion. More specifically, five participants found it difficult to know the cursor's current location within the matrix. 
Six found the distortion results in unexpected \emph{``jump''} and made it challenging \emph{``to get to the right cell.''}
One participant also felt \emph{``sea-sick''.}
In the \tcompare{} task, one participant found the distortion made it hard to \emph{``see the boundary of the highlighted region,''} 
In the \tcontext{} task, two participants found it tough to \emph{``see far away clusters.''}

\textbf{\emph{\lfisheye{}}} was commented to be \emph{``easy to use''} by 18 participants. More specifically, nine found it \emph{``easy to follow,''} four found it \emph{``not jump so much''} and \emph{``more in line with the cursor,''} two found it \emph{``pinpoint fast,''} two found it \emph{``easy to locate,''} and one found it \emph{``easy to know the current location.''} 
Four participants also found \emph{``(cells are) the same size outside the fishseye''} and easy to \emph{``compare the clusters at once''} in the \tcontext{} task.
However, nine participants found it \emph{``hard to see the surroundings'' due to irregular shapes in the the transition area} and they sometimes found it hard to precisely identify the highlighted box in the \tcompare{} task.

\textbf{\emph{\ltable{}}} was reported by four participants who found its regularity to be beneficial: \emph{``lined up with the boxes''} and \emph{``(easy) to keep context in my head.''}
Three participants explicitly commented it to be \emph{``better than the step.''}
However, 14 participants found it disorienting, like \emph{``hard to get my bearing''} and \emph{``alignment off.''} 

\textbf{\emph{\ltabledc{}}} was found positive with its regularity by two participants. However, ten people found it \emph{``disorienting.''} Eight also found the empty space in the enlarged row and column confusing, with one specifically pointed out that \emph{``the gap breaks out the clusters''} in the \tcontext{} task. Four reported it challenging for \emph{``precise moves.''}

\vspace{-0.2em}
\subsection{Discussion}
\label{sec:discussion-lenses}

Most lenses performed similarly in the first user study, with a few notable differences.
We discuss the potential reasons for these differences, and provide guidelines for future lens design in MMV.

\textbf{Correspondence facilitates precise locating.}
\tlocate{} is a fundamental part of many high-level tasks in exploring MMV.
In this task, after moving the mouse cursor into the matrix, the context gets distorted. 
Thus, it is important to find a good entry point to facilitate this task. 
A common strategy is to enter at the same row or column as the target cell. However, due to distortion, the cursor may not land on the target row or column.
We define the difference between the expected and actually hovered row or column after entering the matrix as correspondence.
Higher correspondence means less offset and gives the user more predictable interactions.
To find out the lenses' correspondence, we simulated the cursor moving into the matrix from the top and scanning the entire boundary with an incremental one pixel each time.
We found \lfisheye{} and \lcartesian{} have a perfect correspondence.
However, for \ltable{} and \ltabledc{}, the offsets vary from 0 to 3 cells in 50$\times$50 matrices and from 0 to 6 cells in 100$\times$100 matrices (see supplementary material for details).
In summary, the ranking of correspondence is: \lfisheye{} $=$ \lcartesian{} $>$ \ltable{} $=$ \ltabledc{}. 
The performance and perceived difficulty results align with correspondence, where \lfisheye{} and \lcartesian{} were faster and  generally perceived as easier than \ltable{} and \ltabledc{} in the \tlocate{} task. 
Appert et al.~\cite{appert_high-precision_2010} discussed this in pixel-based lenses and proposed interaction techniques to improve correspondence. 
However, it is unclear how to adapt their techniques to MMV. 
On the other hand, their evaluation results partially aligned with our results and confirmed our hypothesis: techniques with higher correspondence have better locating performance.

\textbf{Discontinuity affects the performance for precise locating.}
\ltabledc{} was slower than \ltable{} in the \tlocate{} task.
The perceived difficulty also aligned with the time performance, where \ltabledc{} was considered more difficult than \ltable{} in the \tlocate{} task.
The only difference in these two lenses is the way they visualize the enlarged rows and columns: \ltable{} stretches them, while \ltabledc{} aligns the cells in the center and leaves blank space with discontinuity.
We conjectured that this discontinuity hinders the ability of precise movement in the MMV, thus degrading the performance of \ltabledc{}.
This is also reflected in participants' comments, where  eight specifically found the ``gaps'' confusing.

\textbf{Uniformity facilitates contextualizing patterns.}
\lcartesian{} was the slowest in the \tcontext{} task. 
This task had two components, where the first component is similar to the \tlocate{} task, and the second component required identifying the cluster with most number of cells in the context. 
The second component started right after the first one, which means the visualization was still distorted by the lenses.
For uniform distortion, the participants only need to compare the areas of the clusters.
However, when the context was distorted non-uniformly, comparing the areas of clusters may lead to a wrong answer. As a result, participants had to count the number of cells, which is expected to take longer.
Participants might also first remove the distortion by moving the cursor outside, which would prolong the task. 
From~\autoref{tab:study-1-design-space} and \autoref{sec:study-1-exp_conditions}, we can see that the ranking of contextual uniformity is: 
\lfisheye{} $=$ \ltabledc{} $>$ \ltable{} $>$ \lcartesian{}.
This ranking aligns with the time performance.
In summary, our results suggest that the performance was proportional to the level of contextual uniformity.

\textbf{Small regions with irregular distortion might not affect performance.}
Within the lenses, all conditions had perfect regularity, except for the \lfisheye{}, where the cells in the transition area were not in a regular grid.
Despite this irregularity, \lfisheye{} had the best overall performance.
This does not necessarily mean that regularity is not important for lenses in MMV, since \lfisheye{} only has a limited region that is irregular. 
Further studies are required to confirm the effect of regularity in other regions (i.e., focal and contextual regions) and at different sizes.  

\textbf{Different distortions do not affect coarse locating.}
All lenses had similar performance in the \tcompare{} task.
We believe this is due to having a large target region (7$\times$7) in this task. 
With a large target, participants only needed to coarsely locate a region instead of precisely locating a single cell like in the \tlocate{} task.
As a result, the correspondence, discontinuity and other distortion characteristics do not lead to significant performance difference in coarse locating.

\section{Study 2 --- Focus+Context, Overview+\\Detail and Pan\&Zoom in MMV}

This study is intended to address the literature gaps in terms of identifying which is the best interaction techniques among \lfc{}, \lod{} and \lzoom{} for MMV.
We chose \lfisheye{} as the representative technique for \lfc{} as it was the best performing technique from the first study.
We designed our first study to be generalizable for testing interaction techniques in MMV, i.e., the same experimental setups can be used to test interaction techniques other than just lenses.
Therefore, we reused many materials from the first study in our second study.
Same as the first user study, we designed the second user study as an exploratory study rather than hypothesis testing. 
This is because the literature had mixed results for the comparisons of the three generic interaction techniques and with little empirical knowledge about the comparison of them in the context of MMV. 
As a result, there was not enough guidance to generate reliable hypotheses.
We also pre-registered this study at \url{https://osf.io/q4zp9}.

\begin{figure*}[t]
    \centering
        \subfigure[\lod{}] {
        \includegraphics[height=3.5cm]{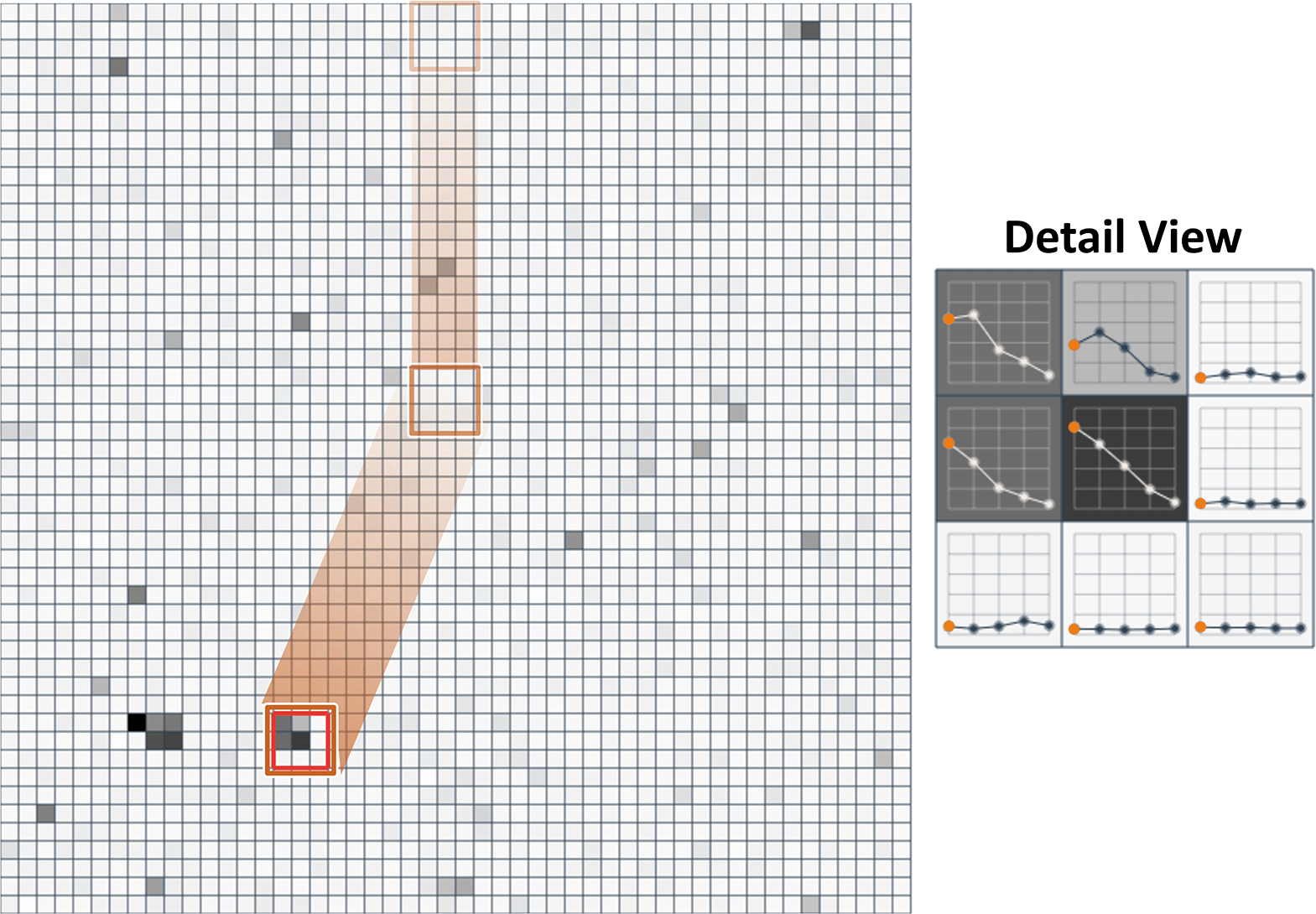}
            \label{fig:study-2-od}
        }
        \rulesep
        \subfigure[\lzoom{}] {
        \includegraphics[height=3.5cm]{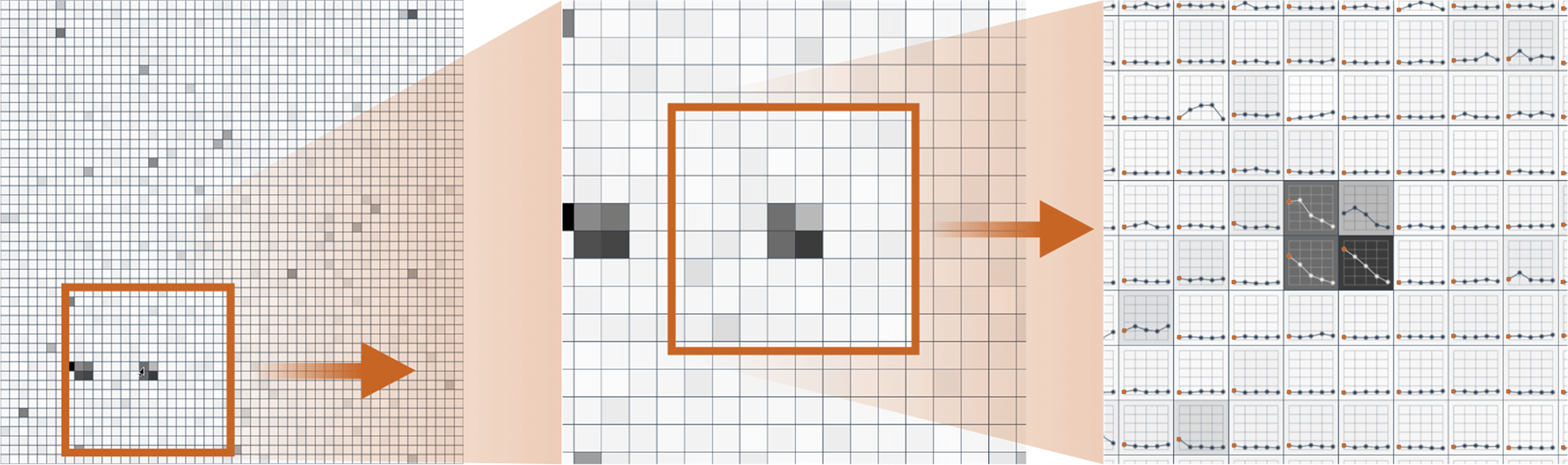}
            \label{fig:study-2-zoom}
        }
    \Description{Demonstration of overview+detail and pan\&zoom techniques. Details in Section 4.1.}
    \caption{\textbf{Study 2 visualization additional conditions (with 50$\times$50 matrices)}: 
    (a) \lod{}, the \emph{detail view} on the right shows the details of the \emph{red box} in left matrix. The user can drag the red box to update the detail view in real-time.
    (b) \lzoom{}, the user can scroll the mouse wheel to zoom in or out a certain region of the matrix. In addition to \lod{} and \lzoom{} conditions, we used the same design of \lfisheye{} from the first user study, as demonstrated in~\autoref{fig:study-1-lenses-fisheye}. An interactive demo is available at
    \url{https://mmvdemo.github.io/},
    and has been tested with Chrome and Edge browsers.}
    \label{fig:study-2-conditions}
\end{figure*}

\subsection{Experimental Conditions}
\label{sec:study-2-exp-conditions}
Same as the first user study, the tested conditions use different ways to selectively enlarge an ROI of the matrix so that the enlarged cells have sufficient display space to show the multivariate details. 
Unlike the first study, where all conditions superimpose the enlarged ROI (or focused view) within the matrix (or the contextual view), different conditions use different strategies to manage the focused and contextual views in the second study.

\textbf{\lfc{}}: we used the same design of \lfisheye{} from the first user study (\autoref{fig:study-1-lenses-fisheye}). \lfc{} displays the focused view inside the contextual view.

\textbf{\lod{}}: we placed a separate view as the \emph{detail view} on the right of the overview (the matrix). 
Some designs place one view at a fixed location (e.g., top right corner) inside the other view (e.g., in~\cite{ronne_jakobsen_sizing_2011}). However, such a design is not suitable in our case, as it will occlude part of the matrix. Thus, we decided to place the two views side-by-side. 
In the detail view, the multivariate details (i.e., lines chart in this study) are rendered for a selected ROI.
A red box is used to indicate the ROI in the matrix. 
The user can drag the red box within the matrix, and the detail view will update in real-time. 
This tightly coupled design between overview and detail view is suggested by Hornbæk et al.~\cite{hornbaek_navigation_2002}.
We set the size of the detail view and the number of line charts to the same as the \lfisheye{}, i.e., the detail view always renders 3$\times$3 line charts in the same size as they are in \lfisheye{}. 
A demonstration of \lod{} is presented in~\autoref{fig:study-2-od}.
\lod{} uses a spatial separation between the focused and contextual views.

\textbf{\lzoom{}}: the participant can scroll the mouse wheel to zoom in or out an ROI of the matrix continuously. The mouse cursor is used as the center of zooming, and the transitions are animated.
When the user zooms into a certain level, where the cells' size is equal to or larger than a threshold, the line charts will be rendered inside the cells. 
We set the threshold as the size of enlarged cells with line charts in \lfisheye{}.
The user can also pan to inspect different parts of the matrix. 
The design of \lzoom{} follows the widespread map interfaces (e.g., Google Maps), and it is a standard design in many user studies (e.g., in~\cite{ronne_jakobsen_sizing_2011,pietriga_pointing_2007,woodburn2019interactive}).
A demonstration of \lzoom{} is presented in~\autoref{fig:study-2-zoom}.
\lzoom{} uses a temporal separation between the focused and contextual views, i.e., only one zoom level can be viewed at a time.

\subsection{Experimental Setups}
\textbf{Experimental Design.}
Similar to the first study, we have two factors: \ftechnique{} (see \autoref{fig:study-2-conditions}) and \fsize{}. 
Each participant completed 36 study trials: 3 visualizations$\times$2 data sizes$\times$3 tasks$\times$2 repetitions.

\textbf{Data and Tasks.}
We reused the data from the first study. To avoid learning effects, we used a screening tool from Prolific to limit participants to people who have not seen our first study.
We slightly modified the \tlocate{} task from the first study to adapt to the new interaction conditions.
The \tlocate{} task in the first study asked participants to click on a highlighted cell as it is important for understanding how different distortion from lenses affects precise selection.
However, in the second study, \lod{} and \lzoom{} do not have any distortion. Thus, the previous task can lead to undesired bias.
Instead, in the second study, we asked participants to interpret the temporal pattern and select an answer from five options (see~\autoref{fig:study-1-patterns}).
With the adapted \tlocate{} task, we can compare the effectiveness of interpreting a given cell's detail in MMV, which involves locating the target cell and navigating to details.
For the \tcompare{} and \tcontext{} tasks, we believe \lod{} and \lzoom{} do not introduce performance bias.
Therefore, we used the same \tcompare{} and \tcontext{} tasks from the first user study.

\textbf{Participants.}
We recruited 45 participants on Prolific. 
As mentioned, to avoid learning effect, we filtered out participants from the first study at screening stage. All participants were located in the US and spoke English natively. To ensure data quality,  we again restrict participation to workers who had an acceptance rate above 90\%.
Our final participant pool consisted of 16 female, and 29 male. 
Out of those participants, one had a PhD degree, one had a master degree, 15 had a bachelor degree, 21 had a high school degree, and seven did not specify their education levels. 
Finally, 7 participants were between the age of 18-20, 
22 participants were between the age of 21-30, 
11 participants were between the age of 31-40,
one participant was between the age of 41-50, and
four participants were above 50.
We compensated each participant with 7 USD.

\textbf{Procedures.}
We used similar procedures as in the first study, except after each task, instead of only rating the perceived difficulty, we asked participants to rate the overall \emph{usability}, \emph{mental demand}, and \emph{physical demand} for each visualization.
This change is intended to obtain a more nuanced understanding of the  perceived effectiveness.
The average completion time was around 35 minutes.

\textbf{Measurements and Statistical Analysis.}
We collected similar measures as in the first study, including \emph{time}, \emph{accuracy}, and \emph{qualitative feedback}.
As described in the \emph{procedures}, we also collected the \emph{subjective ratings} of \emph{usability}, \emph{mental demand}, and \emph{physical demand} for each visualization at each task.
We expected that the additional ratings could help us towards a more nuanced understanding of the perceived performance of different techniques.
We used the same method as in the first study to analyze the collected data.

\subsection{Results}

\begin{figure}[t]
	\centering
	\includegraphics[width=\columnwidth]{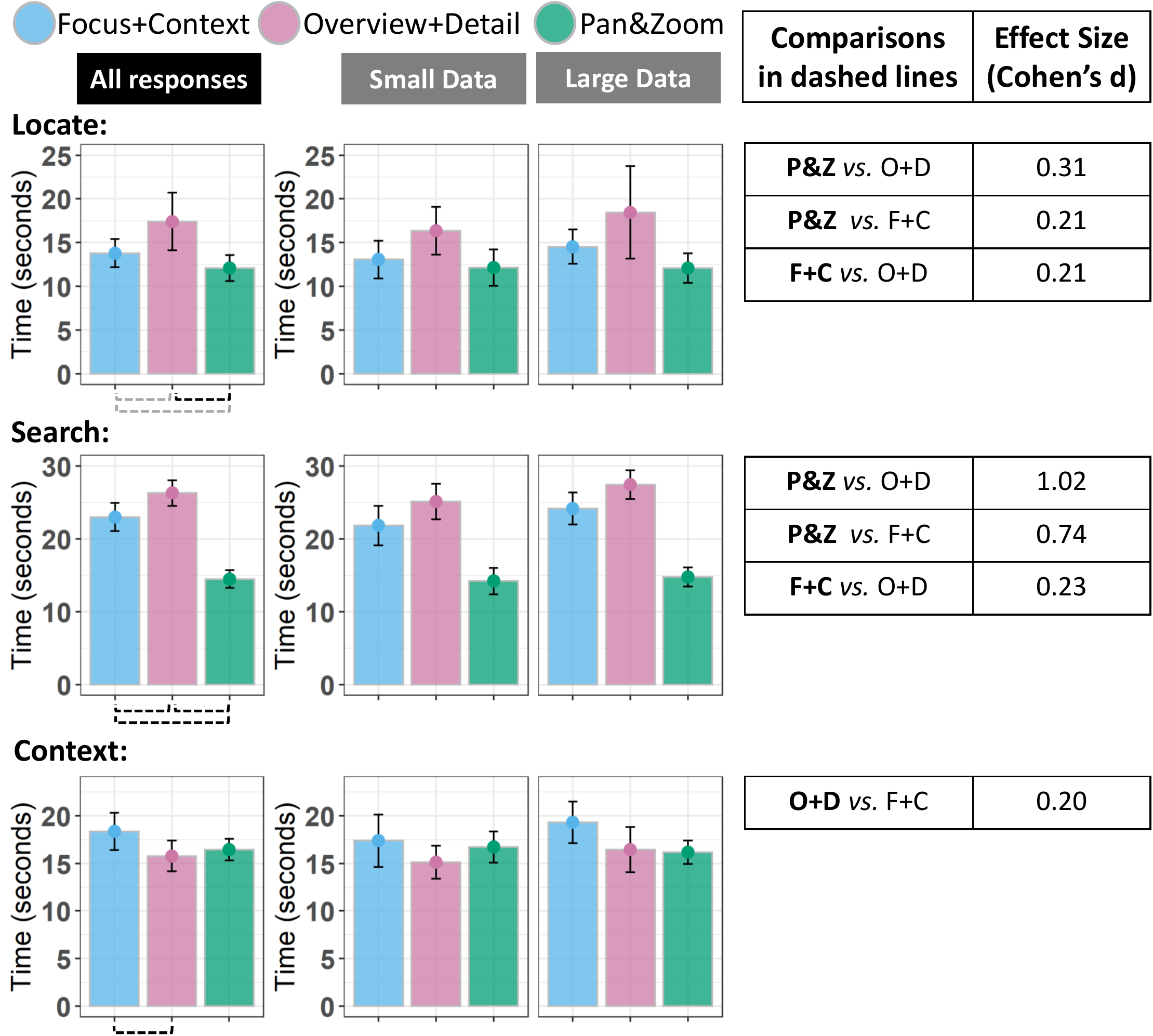}
	\Description{Left are bar charts with error bars showing 95\% confidence intervals. The bar charts are showing the time performance of four tested conditions. Right are tables showing effect sizes of significant comparisons. Details are in Section 4.3.}
	\caption{\emph{Time} by task and in different data sizes. Confidence intervals indicate 95\% confidence for mean values. Dashed lines indicate statistical significance for $p<.05$ (black) and $0.05<p<0.1$ (gray). }
	\label{fig:study-2-result-time}
\end{figure}

\begin{figure}
	\centering
	\includegraphics[width=0.98\columnwidth]{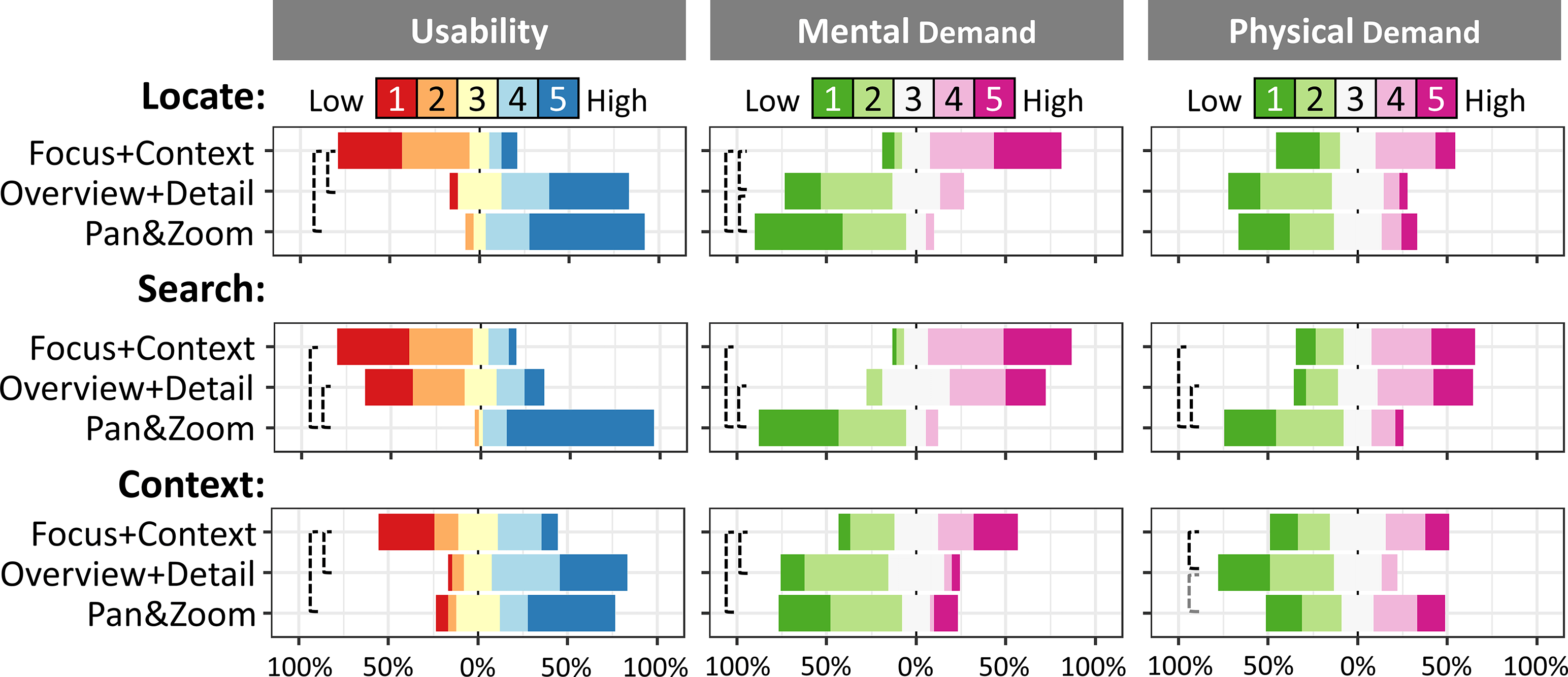}
	\Description{Stacked bar charts showing the subjective ratings. Details are in Section 4.3.}
	\caption{\emph{Usability}, \emph{mental demand}, and \emph{physical demand} ratings by task. Dashed lines indicate $p<.05$ (black) and $0.05<p<0.1$ (gray). }
	\label{fig:study-2-result-rating}
\end{figure}

Same as the first user study, the accuracy was high across all conditions: on average, 98.7\% for \tlocate{}, 93.0\% for \tcompare{}, and 74.8\% for \tcontext{}. 
We did not find any significant differences
on accuracy. 
Therefore, we focus our analysis on the time (\autoref{fig:study-2-result-time}), subjective ratings (\autoref{fig:study-2-result-rating}), and qualitative feedback.

We found \ftechnique{} had a significant effect on time in all tasks: \tlocate{} ($***$), \tcompare{} ($***$), and \tcontext{} ($*$).
We also found \fsize{} had a significant effect on time in \tcompare{} ($*$), and a marginal effect in \tcontext{} ($p=0.092$), but not in \tlocate{} ($p=0.172$).
No significant effect has been found in the interaction between \flens{} and \fsize{} on time for all tasks.
In terms of subjective ratings, we found \ftechnique{} had a significant effect on usability and mental demand in tall tasks (all $***$).
For physical demand, we found significance in the \tcompare{} ($***$) and \tcontext{} ($**$) tasks. 
All statistical results are included in the supplementary materials.

\smallskip
\noindent\textbf{Quantitative Key Findings}
\label{sec:study-2-quantitative-results}

\emph{\textbf{\lzoom{} was the best performing technique.}}
In the \tlocate{} task, \lzoom{} (12.1s, CI=1.5s) was faster than \lfc{} (13.8s, CI=1.5s, $p=0.081$) and \lod{} (17.4s, CI=3.3s, $***$).
In the \tcompare{} task, \lzoom{} (14.5s, CI=1.2s) was faster then \lfc{} (23.0s, CI=1.9s, $***$) and \lod{} (26.3s, CI=1.7s, $***$).
In the \tcontext{} task, \lzoom{} (16.5s, CI=1.1s) had a similar performance as \lod{} (15.8s, CI=1.6s), and tended to be faster than \lfc{} (18.4s, CI=2.0s), but not significant.
The subjective ratings mostly aligned with the performance: participants rated \lzoom{} 
with a higher usability and lower mental demand than \lfc{} 
in all tasks, all $***$.
Participants also found \lzoom{} with a higher usability, lower mental and physical demand than \lod{} in the \tcompare{} task, all $***$.
Overall, \lzoom{} was the best choice for the tested tasks.

\emph{\textbf{\lod{} performed well in the \tcontext{} task.}}
In the \tcontext{} task, \lod{} (15.8s, CI=1.6s) was faster than \lfc{} (18.4s, CI=2.0s, $*$).
It also tended to be slightly faster than \lzoom{} (16.5s, CI=1.1s), but that was not statistically significant.
Again, subjective ratings mostly aligned with the performance results. 
\lod{} was rated to have a higher usability ($***$), lower mental ($**$) and physical demand ($**$) than \lfc{} for the \tcontext{} task.
\lod{} was also rated to be marginally less physical demand than \lod{} for the \tcontext{} task ($p=0.090$).

\emph{\textbf{\lod{} was the slowest technique in the \tlocate{} and \tcompare{} tasks.}}
Despite its good performance in the \tcontext{} task, \lod{} 
was slower than \lzoom{} 
, all $***$.
\lod{} was also slower than \lfc{} (23.0s, CI=1.9s) in the \tcompare{} task ($***$), and was marginally slower than \lfc{} (13.8s, CI=1.5s) in the \tlocate{} task ($p=0.055$).

\emph{\textbf{\lfc{} received the worst subjective ratings.}}
\lfc{} had the second best performance in the \tlocate{} task. However, it was rated as with the lowest usability and highest mental demand (all $***$).
For the \tcompare{} task, it was rated with a lower usability, higher mental and physical demand than \lzoom{} (all $***$).
For the \tcontext{} task, it was again rated with lowest usability and highest mental demand (all $>**$), and with a higher physical demand than \lod{} ($**$).

\smallskip
\noindent\textbf{Qualitative Feedback}

Same as the first study, in addition to quantitative data, we also asked participants to justify their subjective ratings after each task.
We analyze the collected feedback to get an overview of the pros and cons of each interaction technique.

\emph{\textbf{\lfc{}}} was complained to be \emph{``difficult for precise selection''} and \emph{``hard to get where I wanted to be''} by 21 participants. 
17 participants also considered it to be \emph{``disorienting''} as \emph{``hard to tell where I was.''} 
11 participants did not feel confident with it, and \emph{``had to double check.''}
Five participants found it \emph{``difficult to anticipate the mapping.''}
In the \tcompare{} task, eight participants also found using it to \emph{``scan a large region is difficult,''} which \emph{``requires high working memory,''} as they need to keep the context in mind.
In the \tcontext{} task, four participants also commented that the distortion makes it difficult to identify and inspect close clusters.

\emph{\textbf{\lod{}}} was reported to be beneficial for \emph{``clearly knowing where you are''} by four participants.
However, 12 participants also pointed out it \emph{``requires a bit of working memory to translate the position in the matrix to the detail view.''}
Five reported that they \emph{``had to double check.''}
Two participants found it \emph{``becomes more difficult for large matrices.''}
In the \tcompare{} task, 13 participants found using it to \emph{``scan a large region is difficult.''} Potentially also because participants need to keep switching between the overview and the detail view all the time.
In the \tcontext{} task, 11 participants found \emph{``having two view at once''} helps complete the task.

\emph{\textbf{\lzoom{}}} was found to be \emph{``intuitive''} and \emph{``familiar''} by 15 participants.
One participant also took advantage of the large number of cells it can enlarge: \emph{``there is no need to precisely zoom in.''}
In the \tcompare{} task, 18 participants reported it can show a large number of enlarged cells, and \emph{``having all at once''} makes this task easy.
In the \tcontext{} task, 12 participants complained about \emph{``the extra physical movements required to zoom in and out.''}

\section{Discussion}
The overarching goal of our studies is to answer the question \emph{``Which is the best interaction technique for exploring MMV?''}.
Our results show that \lzoom{}, was as fast as or faster than the overall best performing \lfc{} (i.e., the \lfisheye{}) and \lod{}.
Participants also rated \lzoom{} the overall best option in terms of usability, and mental and physical demand.

\subsection{What leads to different performance for focus+context, overview+detail, and pan\&zoom?}
\label{sec:discussion-three-techniques}

\textbf{Spatial separation of views requires extra time.}
\lod{} was the slowest in the \tlocate{} and \tcompare{} tasks.
We believe a potential reason is the spatial separation of two views in \lod{}.
In \lod{}, the participants had to interact with the overview, and then inspect the ``far away'' detail view.
In \lfc{} and \lzoom{}, the participants have all the information in just one display space.
As a result, \lod{} was likely to require more eye movements, and potentially introduce extra context-switching cost. 
Our findings partially align with previous studies~\cite{cockburn_review_2009}, where they also found \lod{} required more time in some applications.

\textbf{Spatial separation of views is beneficial for contextualizing details.}
\lod{} was faster than \lfc{} in the \tcontext{} task, and tended to be slightly faster than \lzoom{}, but not significantly.
As mentioned earlier, the \tcontext{} task has two components, with the first one similar to the \tlocate{} task, and the second one in identifying the largest cluster.
\lod{} was the slowest in the \tlocate{} task, which means its good performance in the \tcontext{} tasks was mainly from identifying the clusters.
With \lod{}, there is no further interaction needed to finish the second component after the first one.
While, for \lzoom{}, participants had to zoom out to complete the second component. This is also confirmed by the reported physical demand ratings, where 33 out of 45 participants found \lzoom{} required equal (nine participants) or more physical (24 participants) movements than \lod{} in the \tcontext{} task.
On the other hand, compared to \lod{}, the distortion in \lfc{} 
was likely to affect the contextualizing performance, as  
participants might require extra effort to interpret the distortion. This is also confirmed by the usability, mental demand, and physical demand ratings.
In summary, in contextualizing details, the gain of having spatial separation of views outweighs its loss.

\textbf{More cells showing details lead to better search performance.}
In the \tcompare{} task, \lzoom{} can show more enlarged cells with line charts.
\lzoom{} can treat the entire space of MMV as the focal area, as a result, more enlarged cells showing details can fit in the space. In our user study, a maximum of roughly 10$\times$10 cells can be presented with details.
On the other hand, \lfc{} and \lod{} only represent 3$\times$3 cells with details in the study, and more travels (or scanning) was required to complete this task.
This is also confirmed by the subjective rating of usability, mental demand, and physical demand, where \lzoom{} clearly received the best ratings.
Increasing the number of cells in detail for \lfc{} and \lod{} can potentially improve their performance.
For \lfc{}, however, a larger focal area also introduces more distortion. Moreover, it is not possible to have a focal area as large as the entire matrix like in \lzoom{}.
On the other hand, it can be straightforward to increase the size of detail view in \lod{}. However, more screen space will be required, which may not be an option for scenarios with limited screen estate.
Another potential reason might be that the majority of users are more familiar with \lzoom{} compared to other techniques, as it is a standard interaction technique in many web applications (e.g., Google Maps, photo viewers).
Our results partially aligned with Yang et al~\cite{yang_embodied_2021}, which found that \lzoom{} had better performance than \lod{}.
Our results differ from the study by Pietriga et al~\cite{pietriga_pointing_2007}, where they found \lfc{} and \lod{} outperformed \lzoom{}.
One possible reason is that they only consider one navigation task, and did not consider multivariate details. 
Interpreting multivariate details and switching between the focal and contextual areas can introduce additional effort for different conditions.

\textbf{Distortion results in bad user experience.}
\lfc{} was rated lowest on usability and highest in mental demand for almost all tasks, despite its generally good performance in the \tlocate{} and \tcompare{} tasks.
An interesting fact is that the \lfc{} technique used in the second study (i.e., \lfisheye{}) was rated as the easiest technique in the first study. 
However, when compared to the regularity and uniformity in \lod{} and \lzoom{}, participants clearly disliked the \lfisheye{}.  
These results were not found in previous studies~\cite{baudisch_keeping_2002,gutwin_fisheyes_2003,shoemaker_supporting_2007}. 
A possible explanation is that prior studies employed applications, where the regularity and uniformity are not important. However, keeping rows and columns in a regular grid is critical for matrix visualization and should be considered for the design of MMV.

\vspace{-0.5em}
\subsection{Generalization, Limitations and Future Work}

\textbf{Interaction Techniques.}
In our first study, we followed Carpendale et al's taxonomy~\cite{carpendale_extending_1997}, and tested four representative lenses.
In the second study, we compared the best performing lens (focus+context) to overview+detail, pan\&zoom. 
Those are the most widely-used techniques, and are likely to be among the first choices when designing interactions for MMV.
Thus, we believe our selected techniques cover a wide range of interaction techniques for MMV, and provide practical guidance on selecting the most effective and applicable technique. 
There are other interaction techniques that can be adapted to MMV, like insets~\cite{lekschas_pattern-driven_2020}, editing values, aggregating values across cells and adapting visualizations to the aggregated data~\cite{horak_responsive_2021} and re-ordering the matrix~\cite{behrisch_matrix_2016}.
Our study is meant as a first assessment of the fundamental interactions for MMV.
Including these techniques in a future study can obtain a more comprehensive understanding of the effectiveness of interaction techniques in MMV but beyond the scope of this paper.

Our results and discussion can inspire improvements to existing approaches and generate potential new techniques.
In the first study, we found correspondence necessary for precise locating. TableLenses particularly suffered from its low correspondence. 
One possible way to increase the correspondence in TableLenses is to dynamically move the entire matrix based on the mouse cursor to compensate for its row or column offsets. 
However, such a design requires extra screen space and may confuse the users.
Future design can also consider adapting 3D distortion and techniques, like the Perspective Wall~\cite{mackinlay_perspective_1991} and Mélange~\cite{elmqvist_melange_2010} to MMV.
In the second user study, we found \lzoom{} was overall a good option, but \lod{} had a similar performance as \lzoom{} and was rated lowest on physical demand in the \tcontext{} task.
To have the benefits of both, the two techniques could be combined. 
However, adding an overview to a zooming interface leads to mixed results in the literature~\cite{hornbaek_navigation_2002}: some found it useful for navigation, and some found it unnecessary~\cite{nekrasovski_evaluation_2006,yang_embodied_2021}.

The performance of \lfc{} was not ideal in our studies. One potential way to improve lenses is to allow the users to select multiple focal areas, which has been explored in some applications~\cite{elmqvist_melange_2010,lekschas_pattern-driven_2020,horak_responsive_2021}. 
However, as the first controlled study for MMV, we decided to have only one focal area to focus on the basic MMV interactions.
Further tests are required to understand the effectiveness of these techniques.
Meanwhile, we believe our results can be partially generalized to multi-focal interactions. The tested \tlocate{} and \tcompare{} tasks investigated the \emph{wayfinding} and  \emph{travel} components. 
We conjectured that adding the multi-focal feature to our tested conditions would not significantly change our results of these two tasks, as they are basic interactions and do not require participants to investigate multiple areas of interest.
Having multiple focus areas is likely to severely change the distortion for the \tcontext{} task, and additional investigation is required for checking the  \emph{context-switching} component.
Moreover, one key motivation of having multiple focus areas is to reduce \emph{the number of travels}~\cite{lekschas_pattern-driven_2020,elmqvist_melange_2010,yang_embodied_2021}. 
Our tested tasks did not explicitly test \emph{the number of travels} component, and it should be systematically explored in future studies.

\textbf{Technique Configurations.} We discuss the rationale and limitation of the chosen parameters for the tested techniques:

\emph{The size of the focal area.} 
We chose the parameters for lenses to ensure that the users can interact with the enlarged cells while the contextual cells are still legible on screens with standard resolutions (\autoref{sec:study-1-tasks}). 
With higher resolution, other settings could be tested, and we expect larger focal areas to be beneficial for some tasks (e.g., the \tcompare{} task).
The focal area for \lod{} is not as constrained as \lfc{}, but a larger focal area will require more screen space, and we intentionally kept their sizes consistent to reduce confounding factors. 
One future direction is to investigate the effect of focal area size on different interaction techniques.

\emph{Lens on-demand.}
Providing the ability to switch on and off the lenses is likely to improve their correspondence. 
However, one key motivation for using lenses is interactive exploration~\cite{tominski_survey_2014,tominski_interactive_2017}, where the location of the targets is not known upfront. Our studies were designed to investigate the performance of interactive exploration and simulate different interaction components of the exploration scenario in the tasks. 
Additionally, allowing the participants to turn on and off the lenses might bring extra complexities to the interactions, which could potentially affect their performance.
Providing extra training might reduce this side effect but significantly increases the user study time.
However, experts can get familiar with enabling/disabling lenses with less time constraint in real-world applications, and its effectiveness should be evaluated.

\emph{Dragging interaction in \lod{}.} 
There are two ways to select the focal area in \lod{}: point-and-click and drag-and-drop~\cite{yang_embodied_2021}. Point-and-click requires fewer steps, while drag-and-drop provides a better estimation of the interaction~\cite{kumar_browsing_1997}.
It is unclear which is a better choice for MMV. 
We chose drag-and-drop in our study because the point-and-click method conflicts with our target selection interaction. 
A future study is desired to compare the effectiveness of these two methods for MMV.

\textbf{Embedded Visualization and Tasks.}
In our studies, we tested time series data, one type of widely used multivariate data.  
Our tested tasks focus on the \emph{interactions} to locate, search, and contextualize multivariate details.
These tasks were chosen to investigate the tested conditions' \emph{wayfinding}, \emph{travel}, and \emph{context-switching} performance.
We intentionally lower the difficulty of \emph{interpreting} the embedded visualization, so that the participants do not need deep knowledge about a particular type of visualization and can focus on the interactions.
Changing the embedded visualization is likely to affect the interpretation performance but will likely bring minimum influence on the \emph{wayfinding}, \emph{travel}, and \emph{context-switching} performance.
Thus, we expect our findings on the effectiveness of different interaction techniques to be partially generalized to MMV with other embedded visualizations.
Future studies are required to confirm our hypothesis.
Horak et al~\cite{horak_responsive_2021} demonstrated embedding different types of visualizations for different cells. 
Such an adaptive design can facilitate complex data analysis process and should be tested in the future.
We also plan to study more specific MMV applications with more sophisticated and high-level tasks.

\textbf{Scalability.} We identified three potential effects related to scalability that were not fully investigated in our studies:

\emph{The number of data points and pattern types in the line chart.} Inspired by Correll \& Gleicher~\cite{correll_semantics_2016}, we used five primitive temporal patterns in our study. We did not include more complicated patterns, as we want to focus on studying the performance of different interactions. 
We also chose to have five data points in each cell, as this is enough for representing all selected temporal patterns while still allowing interaction within the line chart. 
Interpreting more complicated patterns or increasing the number of points in the line chart is likely to increase the difficulty constantly for all testing conditions. 
Thus, our findings can still provide helpful guidelines for selecting the appropriate interaction technique. However, further investigations are required to confirm this conjecture.

\emph{Size of the matrix.} We tested two different sizes of matrices. 
We believe the tested sizes are representative as they can be reasonably rendered and interacted on a standard screen.
We found that the performance of almost all conditions decreased in the larger data set. However, we did not find significant evidence that one specific condition resists the increasing data size better than others. Future studies are required to investigate MMV in larger data sets.

\emph{Size of target regions.}
In the \tcompare{} task, we used 7$\times$7 as the size of the target regions, which was larger than the size of the lenses, so that participants had to move the lenses to fully explore it. 
In the \tcontext{} task, we controlled the range of cluster sizes (from 5$\times$5 to 7$\times$7) to make the task less obvious and more challenging for participants.
We cannot find any literature indicating a significant effect of cluster size, and it should be tested in the future.

\section{Conclusion}

We have presented two studies comparing interaction techniques for exploring MMV. The findings extend our understanding of the different interaction techniques' effectiveness for exploring MMV. 
Our results suggest that pan\&zoom was the overall best performing technique, while for contextualizing details, overview+detail can also be a good choice. 
We also believe there is potential to improve the design of lenses in MMV, for example, reducing the influence of distortion through lensing on demand. 
To provide structured guidelines for future research and design, we discussed the effect of correspondence, uniformity, irregularity, and continuity of lenses. 
Our results indicate that high correspondence, uniformity, and continuity led to better performance for lenses. Future lens design should take these metrics into account.
Another potential future direction is to investigate hybrid techniques, such as adding an overview to a zooming interface or providing interactive zooming inside the lenses. 
In summary, we believe there is much unexplored space in MMV, and our study results and discussion can potentially lead to improved and novel interaction designs in MMV.

\begin{acks}
This work was partially supported by NSF grants III-2107328 and IIS-1901030, NIH grant 5U54CA225088-03, the Harvard Data Science Initiative, and a Harvard Physical Sciences and Engineering Accelerator Award.
\end{acks}

\bibliographystyle{ACM-Reference-Format}
\balance
\bibliography{zotero,other_references}


\begin{thebibliography}{91}


\ifx \showCODEN    \undefined \def \showCODEN     #1{\unskip}     \fi
\ifx \showDOI      \undefined \def \showDOI       #1{#1}\fi
\ifx \showISBNx    \undefined \def \showISBNx     #1{\unskip}     \fi
\ifx \showISBNxiii \undefined \def \showISBNxiii  #1{\unskip}     \fi
\ifx \showISSN     \undefined \def \showISSN      #1{\unskip}     \fi
\ifx \showLCCN     \undefined \def \showLCCN      #1{\unskip}     \fi
\ifx \shownote     \undefined \def \shownote      #1{#1}          \fi
\ifx \showarticletitle \undefined \def \showarticletitle #1{#1}   \fi
\ifx \showURL      \undefined \def \showURL       {\relax}        \fi
\providecommand\bibfield[2]{#2}
\providecommand\bibinfo[2]{#2}
\providecommand\natexlab[1]{#1}
\providecommand\showeprint[2][]{arXiv:#2}

\bibitem[\protect\citeauthoryear{Anders}{Anders}{2009}]%
        {anders2009visualization}
\bibfield{author}{\bibinfo{person}{Simon Anders}.}
  \bibinfo{year}{2009}\natexlab{}.
\newblock \showarticletitle{Visualization of genomic data with the Hilbert
  curve}.
\newblock \bibinfo{journal}{\emph{Bioinformatics}} \bibinfo{volume}{25},
  \bibinfo{number}{10} (\bibinfo{year}{2009}), \bibinfo{pages}{1231--1235}.
\newblock
\urldef\tempurl%
\url{https://doi.org/10.1093/bioinformatics/btp152}
\showDOI{\tempurl}


\bibitem[\protect\citeauthoryear{Andrienko, Andrienko, Bak, Keim, Kisilevich,
  and Wrobel}{Andrienko et~al\mbox{.}}{2011}]%
        {andrienko_conceptual_2011}
\bibfield{author}{\bibinfo{person}{Gennady Andrienko}, \bibinfo{person}{Natalia
  Andrienko}, \bibinfo{person}{Peter Bak}, \bibinfo{person}{Daniel Keim},
  \bibinfo{person}{Slava Kisilevich}, {and} \bibinfo{person}{Stefan Wrobel}.}
  \bibinfo{year}{2011}\natexlab{}.
\newblock \showarticletitle{A conceptual framework and taxonomy of techniques
  for analyzing movement}.
\newblock \bibinfo{journal}{\emph{Journal of Visual Languages \& Computing}}
  \bibinfo{volume}{22}, \bibinfo{number}{3} (\bibinfo{date}{June}
  \bibinfo{year}{2011}), \bibinfo{pages}{213--232}.
\newblock
\showISSN{1045926X}
\urldef\tempurl%
\url{https://doi.org/10.1016/j.jvlc.2011.02.003}
\showDOI{\tempurl}


\bibitem[\protect\citeauthoryear{Apperley, Tzavaras, and Spence}{Apperley
  et~al\mbox{.}}{1982}]%
        {apperley_bifocal_1982}
\bibfield{author}{\bibinfo{person}{Mark~D Apperley}, \bibinfo{person}{I
  Tzavaras}, {and} \bibinfo{person}{Robert Spence}.}
  \bibinfo{year}{1982}\natexlab{}.
\newblock \showarticletitle{A bifocal display technique for data presentation}.
\newblock  (\bibinfo{year}{1982}).
\newblock
\showISSN{1017-4656}
\urldef\tempurl%
\url{https://doi.org/10.2312/eg.19821002}
\showDOI{\tempurl}


\bibitem[\protect\citeauthoryear{Appert, Chapuis, and Pietriga}{Appert
  et~al\mbox{.}}{2010}]%
        {appert_high-precision_2010}
\bibfield{author}{\bibinfo{person}{Caroline Appert}, \bibinfo{person}{Olivier
  Chapuis}, {and} \bibinfo{person}{Emmanuel Pietriga}.}
  \bibinfo{year}{2010}\natexlab{}.
\newblock \showarticletitle{High-precision magnification lenses}. In
  \bibinfo{booktitle}{\emph{Proceedings of the 28th international conference on
  {Human} factors in computing systems - {CHI} '10}}. \bibinfo{publisher}{ACM
  Press}, \bibinfo{address}{Atlanta, Georgia, USA}, \bibinfo{pages}{273}.
\newblock
\showISBNx{978-1-60558-929-9}
\urldef\tempurl%
\url{https://doi.org/10.1145/1753326.1753366}
\showDOI{\tempurl}


\bibitem[\protect\citeauthoryear{Bach, Dragicevic, Archambault, Hurter, and
  Carpendale}{Bach et~al\mbox{.}}{2017}]%
        {bach_descriptive_2017}
\bibfield{author}{\bibinfo{person}{Benjamin Bach}, \bibinfo{person}{Pierre
  Dragicevic}, \bibinfo{person}{Daniel Archambault},
  \bibinfo{person}{Christophe Hurter}, {and} \bibinfo{person}{Sheelagh
  Carpendale}.} \bibinfo{year}{2017}\natexlab{}.
\newblock \showarticletitle{A {Descriptive} {Framework} for {Temporal} {Data}
  {Visualizations} {Based} on {Generalized} {Space}-{Time} {Cubes}:
  {Generalized} {Space}-{Time} {Cube}}.
\newblock \bibinfo{journal}{\emph{Computer Graphics Forum}}
  \bibinfo{volume}{36}, \bibinfo{number}{6} (\bibinfo{date}{Sept.}
  \bibinfo{year}{2017}), \bibinfo{pages}{36--61}.
\newblock
\showISSN{01677055}
\urldef\tempurl%
\url{https://doi.org/10.1111/cgf.12804}
\showDOI{\tempurl}


\bibitem[\protect\citeauthoryear{Bach, Henry-Riche, Dwyer, Madhyastha, Fekete,
  and Grabowski}{Bach et~al\mbox{.}}{2015}]%
        {bach_small_2015}
\bibfield{author}{\bibinfo{person}{Benjamin Bach}, \bibinfo{person}{Nathalie
  Henry-Riche}, \bibinfo{person}{Tim Dwyer}, \bibinfo{person}{Tara Madhyastha},
  \bibinfo{person}{J-D Fekete}, {and} \bibinfo{person}{Thomas Grabowski}.}
  \bibinfo{year}{2015}\natexlab{}.
\newblock \showarticletitle{Small {MultiPiles}: {Piling} {Time} to {Explore}
  {Temporal} {Patterns} in {Dynamic} {Networks}}.
\newblock \bibinfo{journal}{\emph{Computer Graphics Forum}}
  \bibinfo{volume}{34}, \bibinfo{number}{3} (\bibinfo{year}{2015}),
  \bibinfo{pages}{31--40}.
\newblock
\showISSN{1467-8659}
\urldef\tempurl%
\url{https://doi.org/10.1111/cgf.12615}
\showDOI{\tempurl}


\bibitem[\protect\citeauthoryear{Bach, Pietriga, and Fekete}{Bach
  et~al\mbox{.}}{2014}]%
        {bach2014visualizing}
\bibfield{author}{\bibinfo{person}{Benjamin Bach}, \bibinfo{person}{Emmanuel
  Pietriga}, {and} \bibinfo{person}{Jean-Daniel Fekete}.}
  \bibinfo{year}{2014}\natexlab{}.
\newblock \showarticletitle{Visualizing dynamic networks with matrix cubes}. In
  \bibinfo{booktitle}{\emph{Proceedings of the {SIGCHI} {Conference} on {Human}
  {Factors} in {Computing} {Systems}}}. \bibinfo{publisher}{ACM},
  \bibinfo{address}{Toronto Ontario Canada}, \bibinfo{pages}{877--886}.
\newblock
\showISBNx{978-1-4503-2473-1}
\urldef\tempurl%
\url{https://doi.org/10.1145/2556288.2557010}
\showDOI{\tempurl}


\bibitem[\protect\citeauthoryear{Bates, Mächler, Bolker, and Walker}{Bates
  et~al\mbox{.}}{2015}]%
        {Bates2015}
\bibfield{author}{\bibinfo{person}{Douglas Bates}, \bibinfo{person}{Martin
  Mächler}, \bibinfo{person}{Ben Bolker}, {and} \bibinfo{person}{Steve
  Walker}.} \bibinfo{year}{2015}\natexlab{}.
\newblock \showarticletitle{Fitting {Linear} {Mixed}-{Effects} {Models} {Using}
  \textbf{lme4}}.
\newblock \bibinfo{journal}{\emph{Journal of Statistical Software}}
  \bibinfo{volume}{67}, \bibinfo{number}{1} (\bibinfo{year}{2015}),
  \bibinfo{numpages}{47}~pages.
\newblock
\showISSN{1548-7660}
\urldef\tempurl%
\url{https://doi.org/10.18637/jss.v067.i01}
\showDOI{\tempurl}


\bibitem[\protect\citeauthoryear{Baudisch, Good, Bellotti, and
  Schraedley}{Baudisch et~al\mbox{.}}{2002}]%
        {baudisch_keeping_2002}
\bibfield{author}{\bibinfo{person}{Patrick Baudisch},
  \bibinfo{person}{Nathaniel Good}, \bibinfo{person}{Victoria Bellotti}, {and}
  \bibinfo{person}{Pamela Schraedley}.} \bibinfo{year}{2002}\natexlab{}.
\newblock \showarticletitle{Keeping things in context: a comparative evaluation
  of focus plus context screens, overviews, and zooming}. In
  \bibinfo{booktitle}{\emph{Proceedings of the {SIGCHI} {Conference} on {Human}
  {Factors} in {Computing} {Systems}}} \emph{(\bibinfo{series}{{CHI} '02})}.
  \bibinfo{publisher}{Association for Computing Machinery},
  \bibinfo{address}{New York, NY, USA}, \bibinfo{pages}{259--266}.
\newblock
\showISBNx{978-1-58113-453-7}
\urldef\tempurl%
\url{https://doi.org/10.1145/503376.503423}
\showDOI{\tempurl}


\bibitem[\protect\citeauthoryear{Beck, Burch, Diehl, and Weiskopf}{Beck
  et~al\mbox{.}}{2014}]%
        {beck_state_2014}
\bibfield{author}{\bibinfo{person}{Fabian Beck}, \bibinfo{person}{Michael
  Burch}, \bibinfo{person}{Stephan Diehl}, {and} \bibinfo{person}{Daniel
  Weiskopf}.} \bibinfo{year}{2014}\natexlab{}.
\newblock \showarticletitle{The {State} of the {Art} in {Visualizing} {Dynamic}
  {Graphs}}.
\newblock \bibinfo{journal}{\emph{EuroVis - STARs}} (\bibinfo{year}{2014}),
  \bibinfo{pages}{21 pages}.
\newblock
\urldef\tempurl%
\url{https://doi.org/10.2312/EUROVISSTAR.20141174}
\showDOI{\tempurl}


\bibitem[\protect\citeauthoryear{Behrisch, Bach, Riche, Schreck, and
  Fekete}{Behrisch et~al\mbox{.}}{2016}]%
        {behrisch_matrix_2016}
\bibfield{author}{\bibinfo{person}{Michael Behrisch}, \bibinfo{person}{Benjamin
  Bach}, \bibinfo{person}{Nathalie~Henry Riche}, \bibinfo{person}{Tobias
  Schreck}, {and} \bibinfo{person}{Jean-Daniel Fekete}.}
  \bibinfo{year}{2016}\natexlab{}.
\newblock \showarticletitle{Matrix {Reordering} {Methods} for {Table} and
  {Network} {Visualization}}.
\newblock \bibinfo{journal}{\emph{Computer Graphics Forum}}
  \bibinfo{volume}{35}, \bibinfo{number}{3} (\bibinfo{year}{2016}),
  \bibinfo{pages}{693--716}.
\newblock
\showISSN{1467-8659}
\urldef\tempurl%
\url{https://doi.org/10.1111/cgf.12935}
\showDOI{\tempurl}


\bibitem[\protect\citeauthoryear{Behrisch, Davey, Fischer, Thonnard, Schreck,
  Keim, and Kohlhammer}{Behrisch et~al\mbox{.}}{2014}]%
        {behrisch2014visual}
\bibfield{author}{\bibinfo{person}{Michael Behrisch}, \bibinfo{person}{James
  Davey}, \bibinfo{person}{Fabian Fischer}, \bibinfo{person}{Olivier Thonnard},
  \bibinfo{person}{Tobias Schreck}, \bibinfo{person}{Daniel Keim}, {and}
  \bibinfo{person}{J{\"o}rn Kohlhammer}.} \bibinfo{year}{2014}\natexlab{}.
\newblock \showarticletitle{Visual Analysis of Sets of Heterogeneous Matrices
  Using Projection-Based Distance Functions and Semantic Zoom}.
\newblock \bibinfo{journal}{\emph{Computer Graphics Forum}}
  \bibinfo{volume}{33}, \bibinfo{number}{3} (\bibinfo{year}{2014}),
  \bibinfo{pages}{411--420}.
\newblock


\bibitem[\protect\citeauthoryear{Bier, Stone, Pier, Buxton, and DeRose}{Bier
  et~al\mbox{.}}{1993}]%
        {bier_toolglass_1993}
\bibfield{author}{\bibinfo{person}{Eric~A. Bier}, \bibinfo{person}{Maureen~C.
  Stone}, \bibinfo{person}{Ken Pier}, \bibinfo{person}{William Buxton}, {and}
  \bibinfo{person}{Tony~D. DeRose}.} \bibinfo{year}{1993}\natexlab{}.
\newblock \showarticletitle{Toolglass and magic lenses: the see-through
  interface}. In \bibinfo{booktitle}{\emph{Proceedings of the 20th annual
  conference on {Computer} graphics and interactive techniques}}.
  \bibinfo{pages}{73--80}.
\newblock


\bibitem[\protect\citeauthoryear{Blanch and Ortega}{Blanch and Ortega}{2011}]%
        {blanch2011benchmarking}
\bibfield{author}{\bibinfo{person}{Renaud Blanch} {and}
  \bibinfo{person}{Michael Ortega}.} \bibinfo{year}{2011}\natexlab{}.
\newblock \showarticletitle{Benchmarking pointing techniques with distractors:
  adding a density factor to {Fitts}' pointing paradigm}. In
  \bibinfo{booktitle}{\emph{Proceedings of the {SIGCHI} {Conference} on {Human}
  {Factors} in {Computing} {Systems}}}. \bibinfo{publisher}{ACM},
  \bibinfo{address}{Vancouver BC Canada}, \bibinfo{pages}{1629--1638}.
\newblock
\showISBNx{978-1-4503-0228-9}
\urldef\tempurl%
\url{https://doi.org/10.1145/1978942.1979180}
\showDOI{\tempurl}


\bibitem[\protect\citeauthoryear{Boix, James, Park, Meuleman, and Kellis}{Boix
  et~al\mbox{.}}{2021}]%
        {boix2021regulatory}
\bibfield{author}{\bibinfo{person}{Carles~A. Boix},
  \bibinfo{person}{Benjamin~T. James}, \bibinfo{person}{Yongjin~P. Park},
  \bibinfo{person}{Wouter Meuleman}, {and} \bibinfo{person}{Manolis Kellis}.}
  \bibinfo{year}{2021}\natexlab{}.
\newblock \showarticletitle{Regulatory genomic circuitry of human disease loci
  by integrative epigenomics}.
\newblock \bibinfo{journal}{\emph{Nature}} \bibinfo{volume}{590},
  \bibinfo{number}{7845} (\bibinfo{date}{Feb.} \bibinfo{year}{2021}),
  \bibinfo{pages}{300--307}.
\newblock
\showISSN{0028-0836, 1476-4687}
\urldef\tempurl%
\url{https://doi.org/10.1038/s41586-020-03145-z}
\showDOI{\tempurl}


\bibitem[\protect\citeauthoryear{Boulos}{Boulos}{2003}]%
        {boulos_use_2003}
\bibfield{author}{\bibinfo{person}{Maged N.~Kamel Boulos}.}
  \bibinfo{year}{2003}\natexlab{}.
\newblock \showarticletitle{The use of interactive graphical maps for browsing
  medical/health {Internet} information resources}.
\newblock \bibinfo{journal}{\emph{International Journal of Health Geographics}}
  \bibinfo{volume}{2}, \bibinfo{number}{1} (\bibinfo{date}{Jan.}
  \bibinfo{year}{2003}), \bibinfo{pages}{1}.
\newblock
\showISSN{1476-072X}
\urldef\tempurl%
\url{https://doi.org/10.1186/1476-072X-2-1}
\showDOI{\tempurl}


\bibitem[\protect\citeauthoryear{Burch, Schmidt, and Weiskopf}{Burch
  et~al\mbox{.}}{2013}]%
        {burch_matrix-based_2013}
\bibfield{author}{\bibinfo{person}{Michael Burch}, \bibinfo{person}{Benjamin
  Schmidt}, {and} \bibinfo{person}{Daniel Weiskopf}.}
  \bibinfo{year}{2013}\natexlab{}.
\newblock \showarticletitle{A {Matrix}-{Based} {Visualization} for {Exploring}
  {Dynamic} {Compound} {Digraphs}}. In \bibinfo{booktitle}{\emph{2013 17th
  {International} {Conference} on {Information} {Visualisation}}}.
  \bibinfo{publisher}{IEEE}, \bibinfo{address}{London, United Kingdom},
  \bibinfo{pages}{66--73}.
\newblock
\showISBNx{978-0-7695-5049-7}
\urldef\tempurl%
\url{https://doi.org/10.1109/IV.2013.8}
\showDOI{\tempurl}


\bibitem[\protect\citeauthoryear{Burigat, Chittaro, and Parlato}{Burigat
  et~al\mbox{.}}{2008}]%
        {burigat_map_2008}
\bibfield{author}{\bibinfo{person}{Stefano Burigat}, \bibinfo{person}{Luca
  Chittaro}, {and} \bibinfo{person}{Edoardo Parlato}.}
  \bibinfo{year}{2008}\natexlab{}.
\newblock \showarticletitle{Map, diagram, and web page navigation on mobile
  devices: the effectiveness of zoomable user interfaces with overviews}. In
  \bibinfo{booktitle}{\emph{Proceedings of the 10th international conference on
  {Human} computer interaction with mobile devices and services - {MobileHCI}
  '08}}. \bibinfo{publisher}{ACM Press}, \bibinfo{pages}{147}.
\newblock
\showISBNx{978-1-59593-952-4}
\urldef\tempurl%
\url{https://doi.org/10.1145/1409240.1409257}
\showDOI{\tempurl}


\bibitem[\protect\citeauthoryear{Carpendale, Cowperthwaite, and
  Fracchia}{Carpendale et~al\mbox{.}}{1997}]%
        {carpendale_extending_1997}
\bibfield{author}{\bibinfo{person}{Sheelagh Carpendale},
  \bibinfo{person}{David~J Cowperthwaite}, {and} \bibinfo{person}{F~David
  Fracchia}.} \bibinfo{year}{1997}\natexlab{}.
\newblock \showarticletitle{Extending distortion viewing from {2D} to {3D}}.
\newblock \bibinfo{journal}{\emph{IEEE Computer Graphics and Applications}}
  \bibinfo{volume}{17}, \bibinfo{number}{4} (\bibinfo{date}{Aug.}
  \bibinfo{year}{1997}), \bibinfo{pages}{42--51}.
\newblock
\showISSN{02721716}
\urldef\tempurl%
\url{https://doi.org/10.1109/38.595268}
\showDOI{\tempurl}


\bibitem[\protect\citeauthoryear{Chimera}{Chimera}{1998}]%
        {chimera_value_1998}
\bibfield{author}{\bibinfo{person}{Richard Chimera}.}
  \bibinfo{year}{1998}\natexlab{}.
\newblock \showarticletitle{Value {Bars}: an information visualization and
  navigation tool for multi-attribute listings and tables}.
\newblock  (\bibinfo{date}{Oct.} \bibinfo{year}{1998}).
\newblock
\urldef\tempurl%
\url{https://drum.lib.umd.edu/handle/1903/376}
\showURL{%
\tempurl}


\bibitem[\protect\citeauthoryear{Cockburn, Karlson, and Bederson}{Cockburn
  et~al\mbox{.}}{2009}]%
        {cockburn_review_2009}
\bibfield{author}{\bibinfo{person}{Andy Cockburn}, \bibinfo{person}{Amy
  Karlson}, {and} \bibinfo{person}{Benjamin~B. Bederson}.}
  \bibinfo{year}{2009}\natexlab{}.
\newblock \showarticletitle{A review of overview+detail, zooming, and
  focus+context interfaces}.
\newblock \bibinfo{journal}{\emph{Comput. Surveys}} \bibinfo{volume}{41},
  \bibinfo{number}{1} (\bibinfo{date}{Jan.} \bibinfo{year}{2009}),
  \bibinfo{pages}{2:1--2:31}.
\newblock
\showISSN{0360-0300}
\urldef\tempurl%
\url{https://doi.org/10.1145/1456650.1456652}
\showDOI{\tempurl}


\bibitem[\protect\citeauthoryear{Correll and Gleicher}{Correll and
  Gleicher}{2016a}]%
        {correll2016semantics}
\bibfield{author}{\bibinfo{person}{Michael Correll} {and}
  \bibinfo{person}{Michael Gleicher}.} \bibinfo{year}{2016}\natexlab{a}.
\newblock \showarticletitle{The semantics of sketch: Flexibility in visual
  query systems for time series data}. In \bibinfo{booktitle}{\emph{2016 IEEE
  Conference on Visual Analytics Science and Technology (VAST)}}. IEEE,
  \bibinfo{pages}{131--140}.
\newblock


\bibitem[\protect\citeauthoryear{Correll and Gleicher}{Correll and
  Gleicher}{2016b}]%
        {correll_semantics_2016}
\bibfield{author}{\bibinfo{person}{Michael Correll} {and}
  \bibinfo{person}{Michael Gleicher}.} \bibinfo{year}{2016}\natexlab{b}.
\newblock \showarticletitle{The semantics of sketch: {Flexibility} in visual
  query systems for time series data}. In \bibinfo{booktitle}{\emph{2016 {IEEE}
  {Conference} on {Visual} {Analytics} {Science} and {Technology} ({VAST})}}.
  \bibinfo{pages}{131--140}.
\newblock
\urldef\tempurl%
\url{https://doi.org/10.1109/VAST.2016.7883519}
\showDOI{\tempurl}


\bibitem[\protect\citeauthoryear{Dang, Cui, and Forbes}{Dang
  et~al\mbox{.}}{2016}]%
        {dang2016multilayermatrix}
\bibfield{author}{\bibinfo{person}{Tuan~Nhon Dang}, \bibinfo{person}{Hong Cui},
  {and} \bibinfo{person}{Angus~G Forbes}.} \bibinfo{year}{2016}\natexlab{}.
\newblock \showarticletitle{MultiLayerMatrix: visualizing large taxonomic
  datasets}. In \bibinfo{booktitle}{\emph{EuroVis Workshop on Visual Analytics
  (EuroVA). The Eurographics Association}}. \bibinfo{numpages}{6}~pages.
\newblock


\bibitem[\protect\citeauthoryear{Ellis, Bertini, and Dix}{Ellis
  et~al\mbox{.}}{2005}]%
        {ellis_sampling_2005}
\bibfield{author}{\bibinfo{person}{Geoffrey Ellis}, \bibinfo{person}{Enrico
  Bertini}, {and} \bibinfo{person}{Alan Dix}.} \bibinfo{year}{2005}\natexlab{}.
\newblock \showarticletitle{The sampling lens: making sense of saturated
  visualisations}. In \bibinfo{booktitle}{\emph{{CHI} '05 {Extended}
  {Abstracts} on {Human} {Factors} in {Computing} {Systems}}}
  \emph{(\bibinfo{series}{{CHI} {EA} '05})}. \bibinfo{publisher}{Association
  for Computing Machinery}, \bibinfo{address}{New York, NY, USA},
  \bibinfo{pages}{1351--1354}.
\newblock
\showISBNx{978-1-59593-002-6}
\urldef\tempurl%
\url{https://doi.org/10.1145/1056808.1056914}
\showDOI{\tempurl}


\bibitem[\protect\citeauthoryear{Elmqvist, Do, Goodell, Henry, and
  Fekete}{Elmqvist et~al\mbox{.}}{2008a}]%
        {elmqvist_zame_2008}
\bibfield{author}{\bibinfo{person}{Niklas Elmqvist},
  \bibinfo{person}{Thanh-Nghi Do}, \bibinfo{person}{Howard Goodell},
  \bibinfo{person}{Nathalie Henry}, {and} \bibinfo{person}{Jean-Daniel
  Fekete}.} \bibinfo{year}{2008}\natexlab{a}.
\newblock \showarticletitle{{ZAME}: {Interactive} {Large}-{Scale} {Graph}
  {Visualization}}. In \bibinfo{booktitle}{\emph{2008 {IEEE} {Pacific}
  {Visualization} {Symposium}}}. \bibinfo{publisher}{IEEE},
  \bibinfo{address}{Kyoto}, \bibinfo{pages}{215--222}.
\newblock
\showISBNx{978-1-4244-1966-1}
\urldef\tempurl%
\url{https://doi.org/10.1109/PACIFICVIS.2008.4475479}
\showDOI{\tempurl}


\bibitem[\protect\citeauthoryear{Elmqvist, Henry, Ri~he, and Fekete}{Elmqvist
  et~al\mbox{.}}{2008b}]%
        {elmqvist_melange_2008}
\bibfield{author}{\bibinfo{person}{Niklas Elmqvist}, \bibinfo{person}{Nathalie
  Henry}, \bibinfo{person}{Yann Ri~he}, {and} \bibinfo{person}{Jean-Daniel
  Fekete}.} \bibinfo{year}{2008}\natexlab{b}.
\newblock \showarticletitle{Melange: space folding for multi-focus
  interaction}. In \bibinfo{booktitle}{\emph{Proceeding of the twenty-sixth
  annual {CHI} conference on {Human} factors in computing systems - {CHI}
  '08}}. \bibinfo{publisher}{ACM Press}, \bibinfo{address}{Florence, Italy},
  \bibinfo{pages}{1333}.
\newblock
\showISBNx{978-1-60558-011-1}
\urldef\tempurl%
\url{https://doi.org/10.1145/1357054.1357263}
\showDOI{\tempurl}


\bibitem[\protect\citeauthoryear{Elmqvist, Riche, Henry-Riche, and
  Fekete}{Elmqvist et~al\mbox{.}}{2010}]%
        {elmqvist_melange_2010}
\bibfield{author}{\bibinfo{person}{Niklas Elmqvist}, \bibinfo{person}{Yann
  Riche}, \bibinfo{person}{Nathalie Henry-Riche}, {and}
  \bibinfo{person}{Jean-Daniel Fekete}.} \bibinfo{year}{2010}\natexlab{}.
\newblock \showarticletitle{Mélange: {Space} {Folding} for {Visual}
  {Exploration}}.
\newblock \bibinfo{journal}{\emph{IEEE Transactions on Visualization and
  Computer Graphics}} \bibinfo{volume}{16}, \bibinfo{number}{3}
  (\bibinfo{date}{May} \bibinfo{year}{2010}), \bibinfo{pages}{468--483}.
\newblock
\showISSN{1941-0506}
\urldef\tempurl%
\url{https://doi.org/10.1109/TVCG.2009.86}
\showDOI{\tempurl}


\bibitem[\protect\citeauthoryear{Field, Miles, and Field}{Field
  et~al\mbox{.}}{2012}]%
        {field2012discovering}
\bibfield{author}{\bibinfo{person}{Andy Field}, \bibinfo{person}{Jeremy Miles},
  {and} \bibinfo{person}{Zo{\"e} Field}.} \bibinfo{year}{2012}\natexlab{}.
\newblock \bibinfo{booktitle}{\emph{Discovering statistics using R}}.
\newblock \bibinfo{publisher}{Sage publications}.
\newblock


\bibitem[\protect\citeauthoryear{Fischer, Arya, Streeb, Seebacher, Keim, and
  Worring}{Fischer et~al\mbox{.}}{2021}]%
        {fischer_visual_2021}
\bibfield{author}{\bibinfo{person}{Maximilian~T Fischer},
  \bibinfo{person}{Devanshu Arya}, \bibinfo{person}{Dirk Streeb},
  \bibinfo{person}{Daniel Seebacher}, \bibinfo{person}{Daniel~A Keim}, {and}
  \bibinfo{person}{Marcel Worring}.} \bibinfo{year}{2021}\natexlab{}.
\newblock \showarticletitle{Visual {Analytics} for {Temporal} {Hypergraph}
  {Model} {Exploration}}.
\newblock \bibinfo{journal}{\emph{IEEE Transactions on Visualization and
  Computer Graphics}} \bibinfo{volume}{27}, \bibinfo{number}{2}
  (\bibinfo{date}{Feb.} \bibinfo{year}{2021}), \bibinfo{pages}{550--560}.
\newblock
\showISSN{1941-0506}
\urldef\tempurl%
\url{https://doi.org/10.1109/TVCG.2020.3030408}
\showDOI{\tempurl}


\bibitem[\protect\citeauthoryear{Fitts}{Fitts}{1954}]%
        {fitts_information_1954}
\bibfield{author}{\bibinfo{person}{Paul~M. Fitts}.}
  \bibinfo{year}{1954}\natexlab{}.
\newblock \showarticletitle{The information capacity of the human motor system
  in controlling the amplitude of movement}.
\newblock \bibinfo{journal}{\emph{Journal of Experimental Psychology}}
  \bibinfo{volume}{47}, \bibinfo{number}{6} (\bibinfo{year}{1954}),
  \bibinfo{pages}{381--391}.
\newblock
\showISSN{0022-1015(Print)}
\urldef\tempurl%
\url{https://doi.org/10.1037/h0055392}
\showDOI{\tempurl}


\bibitem[\protect\citeauthoryear{Ghoniem, Fekete, and Castagliola}{Ghoniem
  et~al\mbox{.}}{2005}]%
        {ghoniem2005readability}
\bibfield{author}{\bibinfo{person}{Mohammad Ghoniem},
  \bibinfo{person}{Jean-Daniel Fekete}, {and} \bibinfo{person}{Philippe
  Castagliola}.} \bibinfo{year}{2005}\natexlab{}.
\newblock \showarticletitle{On the readability of graphs using node-link and
  matrix-based representations: a controlled experiment and statistical
  analysis}.
\newblock \bibinfo{journal}{\emph{Information Visualization}}
  \bibinfo{volume}{4}, \bibinfo{number}{2} (\bibinfo{year}{2005}),
  \bibinfo{pages}{114--135}.
\newblock


\bibitem[\protect\citeauthoryear{Goodwin, Dykes, Slingsby, and Turkay}{Goodwin
  et~al\mbox{.}}{2016}]%
        {goodwin_visualizing_2016}
\bibfield{author}{\bibinfo{person}{Sarah Goodwin}, \bibinfo{person}{Jason
  Dykes}, \bibinfo{person}{Aidan Slingsby}, {and} \bibinfo{person}{Cagatay
  Turkay}.} \bibinfo{year}{2016}\natexlab{}.
\newblock \showarticletitle{Visualizing {Multiple} {Variables} {Across} {Scale}
  and {Geography}}.
\newblock \bibinfo{journal}{\emph{IEEE Transactions on Visualization and
  Computer Graphics}} \bibinfo{volume}{22}, \bibinfo{number}{1}
  (\bibinfo{date}{Jan.} \bibinfo{year}{2016}), \bibinfo{pages}{599--608}.
\newblock
\showISSN{1077-2626}
\urldef\tempurl%
\url{https://doi.org/10.1109/TVCG.2015.2467199}
\showDOI{\tempurl}


\bibitem[\protect\citeauthoryear{Gutwin}{Gutwin}{2002}]%
        {gutwin_improving_2002}
\bibfield{author}{\bibinfo{person}{Carl Gutwin}.}
  \bibinfo{year}{2002}\natexlab{}.
\newblock \showarticletitle{Improving focus targeting in interactive fisheye
  views}. In \bibinfo{booktitle}{\emph{Proceedings of the {SIGCHI} {Conference}
  on {Human} {Factors} in {Computing} {Systems}}} \emph{(\bibinfo{series}{{CHI}
  '02})}. \bibinfo{publisher}{Association for Computing Machinery},
  \bibinfo{address}{New York, NY, USA}, \bibinfo{pages}{267--274}.
\newblock
\showISBNx{978-1-58113-453-7}
\urldef\tempurl%
\url{https://doi.org/10.1145/503376.503424}
\showDOI{\tempurl}


\bibitem[\protect\citeauthoryear{Gutwin and Skopik}{Gutwin and Skopik}{2003}]%
        {gutwin_fisheyes_2003}
\bibfield{author}{\bibinfo{person}{Carl Gutwin} {and} \bibinfo{person}{Amy
  Skopik}.} \bibinfo{year}{2003}\natexlab{}.
\newblock \showarticletitle{Fisheyes are good for large steering tasks}. In
  \bibinfo{booktitle}{\emph{Proceedings of the {SIGCHI} {Conference} on {Human}
  {Factors} in {Computing} {Systems}}} \emph{(\bibinfo{series}{{CHI} '03})}.
  \bibinfo{publisher}{Association for Computing Machinery},
  \bibinfo{address}{New York, NY, USA}, \bibinfo{pages}{201--208}.
\newblock
\showISBNx{978-1-58113-630-2}
\urldef\tempurl%
\url{https://doi.org/10.1145/642611.642648}
\showDOI{\tempurl}


\bibitem[\protect\citeauthoryear{Henry and Fekete}{Henry and Fekete}{2007}]%
        {henry_matlink_2007}
\bibfield{author}{\bibinfo{person}{Nathalie Henry} {and}
  \bibinfo{person}{Jean-Daniel Fekete}.} \bibinfo{year}{2007}\natexlab{}.
\newblock \showarticletitle{{MatLink}: {Enhanced} {Matrix} {Visualization} for
  {Analyzing} {Social} {Networks}}. In
  \bibinfo{booktitle}{\emph{Human-{Computer} {Interaction} – {INTERACT}
  2007}} \emph{(\bibinfo{series}{Lecture {Notes} in {Computer} {Science}})},
  \bibfield{editor}{\bibinfo{person}{Cécilia Baranauskas},
  \bibinfo{person}{Philippe Palanque}, \bibinfo{person}{Julio Abascal}, {and}
  \bibinfo{person}{Simone Diniz~Junqueira Barbosa}} (Eds.).
  \bibinfo{publisher}{Springer}, \bibinfo{address}{Berlin, Heidelberg},
  \bibinfo{pages}{288--302}.
\newblock
\showISBNx{978-3-540-74800-7}
\urldef\tempurl%
\url{https://doi.org/10.1007/978-3-540-74800-7_24}
\showDOI{\tempurl}


\bibitem[\protect\citeauthoryear{Henry, Fekete, and McGuffin}{Henry
  et~al\mbox{.}}{2007}]%
        {henry_nodetrix_2007}
\bibfield{author}{\bibinfo{person}{Nathalie Henry},
  \bibinfo{person}{Jean-Daniel Fekete}, {and} \bibinfo{person}{Michael~J.
  McGuffin}.} \bibinfo{year}{2007}\natexlab{}.
\newblock \showarticletitle{{NodeTrix}: a {Hybrid} {Visualization} of {Social}
  {Networks}}.
\newblock \bibinfo{journal}{\emph{IEEE Transactions on Visualization and
  Computer Graphics}} \bibinfo{volume}{13}, \bibinfo{number}{6}
  (\bibinfo{date}{Nov.} \bibinfo{year}{2007}).
\newblock
\showISSN{1941-0506}
\urldef\tempurl%
\url{https://doi.org/10.1109/TVCG.2007.70582}
\showDOI{\tempurl}


\bibitem[\protect\citeauthoryear{Horak, Berger, Schumann, Dachselt, and
  Tominski}{Horak et~al\mbox{.}}{2021}]%
        {horak_responsive_2021}
\bibfield{author}{\bibinfo{person}{Tom Horak}, \bibinfo{person}{Philip Berger},
  \bibinfo{person}{Heidrun Schumann}, \bibinfo{person}{Raimund Dachselt}, {and}
  \bibinfo{person}{Christian Tominski}.} \bibinfo{year}{2021}\natexlab{}.
\newblock \showarticletitle{Responsive {Matrix} {Cells}: {A} {Focus}+{Context}
  {Approach} for {Exploring} and {Editing} {Multivariate} {Graphs}}.
\newblock \bibinfo{journal}{\emph{IEEE Transactions on Visualization and
  Computer Graphics}} \bibinfo{volume}{27}, \bibinfo{number}{2}
  (\bibinfo{date}{Feb.} \bibinfo{year}{2021}), \bibinfo{pages}{1644--1654}.
\newblock
\showISSN{1077-2626, 1941-0506, 2160-9306}
\urldef\tempurl%
\url{https://doi.org/10.1109/TVCG.2020.3030371}
\showDOI{\tempurl}


\bibitem[\protect\citeauthoryear{Hornbæk, Bederson, and Plaisant}{Hornbæk
  et~al\mbox{.}}{2002}]%
        {hornbaek_navigation_2002}
\bibfield{author}{\bibinfo{person}{Kasper Hornbæk},
  \bibinfo{person}{Benjamin~B. Bederson}, {and} \bibinfo{person}{Catherine
  Plaisant}.} \bibinfo{year}{2002}\natexlab{}.
\newblock \showarticletitle{Navigation patterns and usability of zoomable user
  interfaces with and without an overview}.
\newblock \bibinfo{journal}{\emph{ACM Transactions on Computer-Human
  Interaction}} \bibinfo{volume}{9}, \bibinfo{number}{4} (\bibinfo{date}{Dec.}
  \bibinfo{year}{2002}), \bibinfo{pages}{362--389}.
\newblock
\showISSN{1073-0516}
\urldef\tempurl%
\url{https://doi.org/10.1145/586081.586086}
\showDOI{\tempurl}


\bibitem[\protect\citeauthoryear{Hornbæk and Frøkjær}{Hornbæk and
  Frøkjær}{2001}]%
        {hornbaek_reading_2001}
\bibfield{author}{\bibinfo{person}{Kasper Hornbæk} {and} \bibinfo{person}{Erik
  Frøkjær}.} \bibinfo{year}{2001}\natexlab{}.
\newblock \showarticletitle{Reading of electronic documents: the usability of
  linear, fisheye, and overview+detail interfaces}. In
  \bibinfo{booktitle}{\emph{Proceedings of the {SIGCHI} {Conference} on {Human}
  {Factors} in {Computing} {Systems}}} \emph{(\bibinfo{series}{{CHI} '01})}.
  \bibinfo{publisher}{Association for Computing Machinery},
  \bibinfo{address}{New York, NY, USA}, \bibinfo{pages}{293--300}.
\newblock
\showISBNx{978-1-58113-327-1}
\urldef\tempurl%
\url{https://doi.org/10.1145/365024.365118}
\showDOI{\tempurl}


\bibitem[\protect\citeauthoryear{Isenberg, Carpendale, Bezerianos, Henry, and
  Fekete}{Isenberg et~al\mbox{.}}{2009}]%
        {isenberg_coconuttrix_2009}
\bibfield{author}{\bibinfo{person}{Petra Isenberg}, \bibinfo{person}{Sheelegh
  Carpendale}, \bibinfo{person}{Anastasia Bezerianos},
  \bibinfo{person}{Nathalie Henry}, {and} \bibinfo{person}{Jean-Daniel
  Fekete}.} \bibinfo{year}{2009}\natexlab{}.
\newblock \showarticletitle{{CoCoNutTrix}: {Collaborative} {Retrofitting} for
  {Information} {Visualization}}.
\newblock \bibinfo{journal}{\emph{IEEE Computer Graphics and Applications}}
  \bibinfo{volume}{29}, \bibinfo{number}{5} (\bibinfo{date}{Sept.}
  \bibinfo{year}{2009}).
\newblock
\showISSN{1558-1756}
\urldef\tempurl%
\url{https://doi.org/10.1109/MCG.2009.78}
\showDOI{\tempurl}


\bibitem[\protect\citeauthoryear{Javed, Ghani, and Elmqvist}{Javed
  et~al\mbox{.}}{2012}]%
        {javed_polyzoom_2012}
\bibfield{author}{\bibinfo{person}{Waqas Javed}, \bibinfo{person}{Sohaib
  Ghani}, {and} \bibinfo{person}{Niklas Elmqvist}.}
  \bibinfo{year}{2012}\natexlab{}.
\newblock \showarticletitle{Polyzoom: multiscale and multifocus exploration in
  2d visual spaces}. In \bibinfo{booktitle}{\emph{Proceedings of the {SIGCHI}
  {Conference} on {Human} {Factors} in {Computing} {Systems}}}
  \emph{(\bibinfo{series}{{CHI} '12})}. \bibinfo{publisher}{Association for
  Computing Machinery}, \bibinfo{address}{New York, NY, USA},
  \bibinfo{pages}{287--296}.
\newblock
\showISBNx{978-1-4503-1015-4}
\urldef\tempurl%
\url{https://doi.org/10.1145/2207676.2207716}
\showDOI{\tempurl}


\bibitem[\protect\citeauthoryear{Kastner, Erb, and Haberl}{Kastner
  et~al\mbox{.}}{2014}]%
        {kastner_rapid_2014}
\bibfield{author}{\bibinfo{person}{Thomas Kastner}, \bibinfo{person}{Karl-Heinz
  Erb}, {and} \bibinfo{person}{Helmut Haberl}.}
  \bibinfo{year}{2014}\natexlab{}.
\newblock \showarticletitle{Rapid growth in agricultural trade: effects on
  global area efficiency and the role of management}.
\newblock \bibinfo{journal}{\emph{Environmental Research Letters}}
  \bibinfo{volume}{9}, \bibinfo{number}{3} (\bibinfo{date}{March}
  \bibinfo{year}{2014}), \bibinfo{pages}{034015}.
\newblock
\showISSN{1748-9326}
\urldef\tempurl%
\url{https://doi.org/10.1088/1748-9326/9/3/034015}
\showDOI{\tempurl}


\bibitem[\protect\citeauthoryear{Kerpedjiev, Abdennur, Lekschas, McCallum,
  Dinkla, Strobelt, Luber, Ouellette, Azhir, Kumar, Hwang, Lee, Alver, Pfister,
  Mirny, Park, and Gehlenborg}{Kerpedjiev et~al\mbox{.}}{2018}]%
        {kerpedjiev_higlass_2018}
\bibfield{author}{\bibinfo{person}{Peter Kerpedjiev}, \bibinfo{person}{Nezar
  Abdennur}, \bibinfo{person}{Fritz Lekschas}, \bibinfo{person}{Chuck
  McCallum}, \bibinfo{person}{Kasper Dinkla}, \bibinfo{person}{Hendrik
  Strobelt}, \bibinfo{person}{Jacob~M. Luber}, \bibinfo{person}{Scott~B.
  Ouellette}, \bibinfo{person}{Alaleh Azhir}, \bibinfo{person}{Nikhil Kumar},
  \bibinfo{person}{Jeewon Hwang}, \bibinfo{person}{Soohyun Lee},
  \bibinfo{person}{Burak~H. Alver}, \bibinfo{person}{Hanspeter Pfister},
  \bibinfo{person}{Leonid~A. Mirny}, \bibinfo{person}{Peter~J. Park}, {and}
  \bibinfo{person}{Nils Gehlenborg}.} \bibinfo{year}{2018}\natexlab{}.
\newblock \showarticletitle{{HiGlass}: web-based visual exploration and
  analysis of genome interaction maps}.
\newblock \bibinfo{journal}{\emph{Genome Biology}} \bibinfo{volume}{19},
  \bibinfo{number}{1} (\bibinfo{date}{Aug.} \bibinfo{year}{2018}),
  \bibinfo{pages}{125}.
\newblock
\showISSN{1474-760X}
\urldef\tempurl%
\url{https://doi.org/10.1186/s13059-018-1486-1}
\showDOI{\tempurl}


\bibitem[\protect\citeauthoryear{Krüger, Thom, Wörner, Bosch, and
  Ertl}{Krüger et~al\mbox{.}}{2013}]%
        {kruger_trajectorylensesset-based_2013}
\bibfield{author}{\bibinfo{person}{Robert Krüger}, \bibinfo{person}{Dennis
  Thom}, \bibinfo{person}{Michael Wörner}, \bibinfo{person}{Harald Bosch},
  {and} \bibinfo{person}{Thomas Ertl}.} \bibinfo{year}{2013}\natexlab{}.
\newblock \showarticletitle{{TrajectoryLenses}–{A} {Set}-based {Filtering}
  and {Exploration} {Technique} for {Long}-term {Trajectory} {Data}}. In
  \bibinfo{booktitle}{\emph{Computer {Graphics} {Forum}}},
  Vol.~\bibinfo{volume}{32}. \bibinfo{publisher}{Wiley Online Library},
  \bibinfo{pages}{451--460}.
\newblock


\bibitem[\protect\citeauthoryear{Kumar, Plaisant, and Shneiderman}{Kumar
  et~al\mbox{.}}{1997}]%
        {kumar_browsing_1997}
\bibfield{author}{\bibinfo{person}{Harsha~P. Kumar}, \bibinfo{person}{Catherine
  Plaisant}, {and} \bibinfo{person}{Ben Shneiderman}.}
  \bibinfo{year}{1997}\natexlab{}.
\newblock \showarticletitle{Browsing hierarchical data with multi-level dynamic
  queries and pruning}.
\newblock \bibinfo{journal}{\emph{International Journal of Human-Computer
  Studies}} \bibinfo{volume}{46}, \bibinfo{number}{1} (\bibinfo{date}{Jan.}
  \bibinfo{year}{1997}), \bibinfo{pages}{103--124}.
\newblock
\showISSN{10715819}
\urldef\tempurl%
\url{https://doi.org/10.1006/ijhc.1996.0085}
\showDOI{\tempurl}


\bibitem[\protect\citeauthoryear{Lam}{Lam}{2008}]%
        {lam_framework_2008}
\bibfield{author}{\bibinfo{person}{Heidi Lam}.}
  \bibinfo{year}{2008}\natexlab{}.
\newblock \showarticletitle{A {Framework} of {Interaction} {Costs} in
  {Information} {Visualization}}.
\newblock \bibinfo{journal}{\emph{IEEE Transactions on Visualization and
  Computer Graphics}} \bibinfo{volume}{14}, \bibinfo{number}{6}
  (\bibinfo{date}{Nov.} \bibinfo{year}{2008}), \bibinfo{pages}{1149--1156}.
\newblock
\showISSN{1077-2626}
\urldef\tempurl%
\url{https://doi.org/10.1109/TVCG.2008.109}
\showDOI{\tempurl}


\bibitem[\protect\citeauthoryear{LaViola~Jr, Kruijff, McMahan, Bowman, and
  Poupyrev}{LaViola~Jr et~al\mbox{.}}{2017}]%
        {laviola20173d}
\bibfield{author}{\bibinfo{person}{Joseph~J LaViola~Jr}, \bibinfo{person}{Ernst
  Kruijff}, \bibinfo{person}{Ryan~P McMahan}, \bibinfo{person}{Doug Bowman},
  {and} \bibinfo{person}{Ivan~P Poupyrev}.} \bibinfo{year}{2017}\natexlab{}.
\newblock \bibinfo{booktitle}{\emph{3D user interfaces: theory and practice}}.
\newblock \bibinfo{publisher}{Addison-Wesley Professional}.
\newblock


\bibitem[\protect\citeauthoryear{Lekschas, Bach, Kerpedjiev, Gehlenborg, and
  Pfister}{Lekschas et~al\mbox{.}}{2018}]%
        {lekschas_hipiler_2018}
\bibfield{author}{\bibinfo{person}{Fritz Lekschas}, \bibinfo{person}{Benjamin
  Bach}, \bibinfo{person}{Peter Kerpedjiev}, \bibinfo{person}{Nils Gehlenborg},
  {and} \bibinfo{person}{Hanspeter Pfister}.} \bibinfo{year}{2018}\natexlab{}.
\newblock \showarticletitle{{HiPiler}: {Visual} {Exploration} of {Large}
  {Genome} {Interaction} {Matrices} with {Interactive} {Small} {Multiples}}.
\newblock \bibinfo{journal}{\emph{IEEE Transactions on Visualization and
  Computer Graphics}} \bibinfo{volume}{24}, \bibinfo{number}{1}
  (\bibinfo{date}{Jan.} \bibinfo{year}{2018}), \bibinfo{pages}{522--531}.
\newblock
\showISSN{1077-2626}
\urldef\tempurl%
\url{https://doi.org/10.1109/TVCG.2017.2745978}
\showDOI{\tempurl}


\bibitem[\protect\citeauthoryear{Lekschas, Behrisch, Bach, Kerpedjiev,
  Gehlenborg, and Pfister}{Lekschas et~al\mbox{.}}{2020}]%
        {lekschas_pattern-driven_2020}
\bibfield{author}{\bibinfo{person}{Fritz Lekschas}, \bibinfo{person}{Michael
  Behrisch}, \bibinfo{person}{Benjamin Bach}, \bibinfo{person}{Peter
  Kerpedjiev}, \bibinfo{person}{Nils Gehlenborg}, {and}
  \bibinfo{person}{Hanspeter Pfister}.} \bibinfo{year}{2020}\natexlab{}.
\newblock \showarticletitle{Pattern-{Driven} {Navigation} in {2D} {Multiscale}
  {Visualizations} with {Scalable} {Insets}}.
\newblock \bibinfo{journal}{\emph{IEEE Transactions on Visualization and
  Computer Graphics}} \bibinfo{volume}{26}, \bibinfo{number}{1}
  (\bibinfo{date}{Jan.} \bibinfo{year}{2020}), \bibinfo{pages}{611--621}.
\newblock
\showISSN{1941-0506}
\urldef\tempurl%
\url{https://doi.org/10.1109/TVCG.2019.2934555}
\showDOI{\tempurl}


\bibitem[\protect\citeauthoryear{Lekschas, Zhou, Chen, Gehlenborg, Bach, and
  Pfister}{Lekschas et~al\mbox{.}}{2021}]%
        {lekschas_generic_2021}
\bibfield{author}{\bibinfo{person}{Fritz Lekschas}, \bibinfo{person}{Xinyi
  Zhou}, \bibinfo{person}{Wei Chen}, \bibinfo{person}{Nils Gehlenborg},
  \bibinfo{person}{Benjamin Bach}, {and} \bibinfo{person}{Hanspeter Pfister}.}
  \bibinfo{year}{2021}\natexlab{}.
\newblock \showarticletitle{A {Generic} {Framework} and {Library} for
  {Exploration} of {Small} {Multiples} through {Interactive} {Piling}}.
\newblock \bibinfo{journal}{\emph{IEEE Transactions on Visualization and
  Computer Graphics}} \bibinfo{volume}{27}, \bibinfo{number}{2}
  (\bibinfo{date}{Feb.} \bibinfo{year}{2021}), \bibinfo{pages}{358--368}.
\newblock
\showISSN{1941-0506}
\urldef\tempurl%
\url{https://doi.org/10.1109/TVCG.2020.3028948}
\showDOI{\tempurl}


\bibitem[\protect\citeauthoryear{Lenth}{Lenth}{2016}]%
        {Lenth2016}
\bibfield{author}{\bibinfo{person}{Russell~V. Lenth}.}
  \bibinfo{year}{2016}\natexlab{}.
\newblock \showarticletitle{Least-{Squares} {Means}: {The} \textit{{R}}
  {Package} \textbf{lsmeans}}.
\newblock \bibinfo{journal}{\emph{Journal of Statistical Software}}
  \bibinfo{volume}{69}, \bibinfo{number}{1} (\bibinfo{year}{2016}),
  \bibinfo{numpages}{33}~pages.
\newblock
\showISSN{1548-7660}
\urldef\tempurl%
\url{https://doi.org/10.18637/jss.v069.i01}
\showDOI{\tempurl}


\bibitem[\protect\citeauthoryear{Mackinlay, Robertson, and Card}{Mackinlay
  et~al\mbox{.}}{1991}]%
        {mackinlay_perspective_1991}
\bibfield{author}{\bibinfo{person}{Jock~D. Mackinlay},
  \bibinfo{person}{George~G. Robertson}, {and} \bibinfo{person}{Stuart~K.
  Card}.} \bibinfo{year}{1991}\natexlab{}.
\newblock \showarticletitle{The perspective wall: detail and context smoothly
  integrated {Mackinlay} color plates}. In
  \bibinfo{booktitle}{\emph{Proceedings of the {SIGCHI} conference on {Human}
  factors in computing systems {Reaching} through technology - {CHI} '91}}.
  \bibinfo{publisher}{ACM Press}, \bibinfo{address}{New Orleans, Louisiana,
  United States}, \bibinfo{pages}{173--176}.
\newblock
\showISBNx{978-0-89791-383-6}
\urldef\tempurl%
\url{https://doi.org/10.1145/108844.108870}
\showDOI{\tempurl}


\bibitem[\protect\citeauthoryear{McGuffin and Balakrishnan}{McGuffin and
  Balakrishnan}{2002}]%
        {mcguffin_acquisition_2002}
\bibfield{author}{\bibinfo{person}{Michael McGuffin} {and}
  \bibinfo{person}{Ravin Balakrishnan}.} \bibinfo{year}{2002}\natexlab{}.
\newblock \showarticletitle{Acquisition of expanding targets}. In
  \bibinfo{booktitle}{\emph{Proceedings of the {SIGCHI} {Conference} on {Human}
  {Factors} in {Computing} {Systems}}} \emph{(\bibinfo{series}{{CHI} '02})}.
  \bibinfo{publisher}{Association for Computing Machinery},
  \bibinfo{address}{New York, NY, USA}, \bibinfo{pages}{57--64}.
\newblock
\showISBNx{978-1-58113-453-7}
\urldef\tempurl%
\url{https://doi.org/10.1145/503376.503388}
\showDOI{\tempurl}


\bibitem[\protect\citeauthoryear{McLachlan, Munzner, Koutsofios, and
  North}{McLachlan et~al\mbox{.}}{2008}]%
        {mclachlan_liverac_2008}
\bibfield{author}{\bibinfo{person}{Peter McLachlan}, \bibinfo{person}{Tamara
  Munzner}, \bibinfo{person}{Eleftherios Koutsofios}, {and}
  \bibinfo{person}{Stephen North}.} \bibinfo{year}{2008}\natexlab{}.
\newblock \showarticletitle{{LiveRAC}: interactive visual exploration of system
  management time-series data}. In \bibinfo{booktitle}{\emph{Proceedings of the
  {SIGCHI} {Conference} on {Human} {Factors} in {Computing} {Systems}}}
  \emph{(\bibinfo{series}{{CHI} '08})}. \bibinfo{publisher}{Association for
  Computing Machinery}, \bibinfo{address}{New York, NY, USA},
  \bibinfo{pages}{1483--1492}.
\newblock
\showISBNx{978-1-60558-011-1}
\urldef\tempurl%
\url{https://doi.org/10.1145/1357054.1357286}
\showDOI{\tempurl}


\bibitem[\protect\citeauthoryear{McNeill and Hale}{McNeill and Hale}{2017}]%
        {mcneill_generating_2017}
\bibfield{author}{\bibinfo{person}{Graham McNeill} {and}
  \bibinfo{person}{Scott~A. Hale}.} \bibinfo{year}{2017}\natexlab{}.
\newblock \showarticletitle{Generating {Tile} {Maps}}.
\newblock \bibinfo{journal}{\emph{Computer Graphics Forum}}
  \bibinfo{volume}{36}, \bibinfo{number}{3} (\bibinfo{date}{June}
  \bibinfo{year}{2017}), \bibinfo{pages}{435--445}.
\newblock
\showISSN{0167-7055, 1467-8659}
\urldef\tempurl%
\url{https://doi.org/10.1111/cgf.13200}
\showDOI{\tempurl}


\bibitem[\protect\citeauthoryear{Munzner}{Munzner}{2014}]%
        {munzner2014visualization}
\bibfield{author}{\bibinfo{person}{Tamara Munzner}.}
  \bibinfo{year}{2014}\natexlab{}.
\newblock \bibinfo{booktitle}{\emph{Visualization analysis and design}}.
\newblock \bibinfo{publisher}{CRC press}.
\newblock


\bibitem[\protect\citeauthoryear{Nekrasovski, Bodnar, McGrenere, Guimbretière,
  and Munzner}{Nekrasovski et~al\mbox{.}}{2006}]%
        {nekrasovski_evaluation_2006}
\bibfield{author}{\bibinfo{person}{Dmitry Nekrasovski}, \bibinfo{person}{Adam
  Bodnar}, \bibinfo{person}{Joanna McGrenere}, \bibinfo{person}{François
  Guimbretière}, {and} \bibinfo{person}{Tamara Munzner}.}
  \bibinfo{year}{2006}\natexlab{}.
\newblock \showarticletitle{An evaluation of pan \&amp; zoom and rubber sheet
  navigation with and without an overview}. In
  \bibinfo{booktitle}{\emph{Proceedings of the {SIGCHI} {Conference} on {Human}
  {Factors} in {Computing} {Systems}}} \emph{(\bibinfo{series}{{CHI} '06})}.
  \bibinfo{publisher}{Association for Computing Machinery},
  \bibinfo{address}{New York, NY, USA}, \bibinfo{pages}{11--20}.
\newblock
\showISBNx{978-1-59593-372-0}
\urldef\tempurl%
\url{https://doi.org/10.1145/1124772.1124775}
\showDOI{\tempurl}


\bibitem[\protect\citeauthoryear{Neto and Paulovich}{Neto and
  Paulovich}{2021}]%
        {neto2020explainable}
\bibfield{author}{\bibinfo{person}{Mario~Popolin Neto} {and}
  \bibinfo{person}{Fernando~V. Paulovich}.} \bibinfo{year}{2021}\natexlab{}.
\newblock \showarticletitle{Explainable {Matrix} - {Visualization} for {Global}
  and {Local} {Interpretability} of {Random} {Forest} {Classification}
  {Ensembles}}.
\newblock \bibinfo{journal}{\emph{IEEE Transactions on Visualization and
  Computer Graphics}} \bibinfo{volume}{27}, \bibinfo{number}{2}
  (\bibinfo{date}{Feb.} \bibinfo{year}{2021}), \bibinfo{pages}{1427--1437}.
\newblock
\showISSN{1077-2626, 1941-0506, 2160-9306}
\urldef\tempurl%
\url{https://doi.org/10.1109/TVCG.2020.3030354}
\showDOI{\tempurl}


\bibitem[\protect\citeauthoryear{Niederer, Stitz, Hourieh, Grassinger, Aigner,
  and Streit}{Niederer et~al\mbox{.}}{2017}]%
        {niederer2017taco}
\bibfield{author}{\bibinfo{person}{Christina Niederer}, \bibinfo{person}{Holger
  Stitz}, \bibinfo{person}{Reem Hourieh}, \bibinfo{person}{Florian Grassinger},
  \bibinfo{person}{Wolfgang Aigner}, {and} \bibinfo{person}{Marc Streit}.}
  \bibinfo{year}{2017}\natexlab{}.
\newblock \showarticletitle{TACO: visualizing changes in tables over time}.
\newblock \bibinfo{journal}{\emph{IEEE transactions on visualization and
  computer graphics}} \bibinfo{volume}{24}, \bibinfo{number}{1}
  (\bibinfo{year}{2017}), \bibinfo{pages}{677--686}.
\newblock


\bibitem[\protect\citeauthoryear{Nilsson, Serafin, Steinicke, and
  Nordahl}{Nilsson et~al\mbox{.}}{2018}]%
        {nilsson_natural_2018}
\bibfield{author}{\bibinfo{person}{Niels~Christian Nilsson},
  \bibinfo{person}{Stefania Serafin}, \bibinfo{person}{Frank Steinicke}, {and}
  \bibinfo{person}{Rolf Nordahl}.} \bibinfo{year}{2018}\natexlab{}.
\newblock \showarticletitle{Natural {Walking} in {Virtual} {Reality}: {A}
  {Review}}.
\newblock \bibinfo{journal}{\emph{Computers in Entertainment}}
  \bibinfo{volume}{16}, \bibinfo{number}{2} (\bibinfo{date}{April}
  \bibinfo{year}{2018}), \bibinfo{pages}{1--22}.
\newblock
\showISSN{15443574}
\urldef\tempurl%
\url{https://doi.org/10.1145/3180658}
\showDOI{\tempurl}


\bibitem[\protect\citeauthoryear{Nobre, Meyer, Streit, and Lex}{Nobre
  et~al\mbox{.}}{2019}]%
        {nobre_state_2019}
\bibfield{author}{\bibinfo{person}{Carolina Nobre}, \bibinfo{person}{Miriah
  Meyer}, \bibinfo{person}{Marc Streit}, {and} \bibinfo{person}{Alexander
  Lex}.} \bibinfo{year}{2019}\natexlab{}.
\newblock \showarticletitle{The {State} of the {Art} in {Visualizing}
  {Multivariate} {Networks}}.
\newblock \bibinfo{journal}{\emph{Computer Graphics Forum}}
  \bibinfo{volume}{38}, \bibinfo{number}{3} (\bibinfo{year}{2019}),
  \bibinfo{pages}{807--832}.
\newblock
\showISSN{1467-8659}
\urldef\tempurl%
\url{https://doi.org/10.1111/cgf.13728}
\showDOI{\tempurl}


\bibitem[\protect\citeauthoryear{Pearce}{Pearce}{2020}]%
        {Pearce2020}
\bibfield{author}{\bibinfo{person}{Adam Pearce}.}
  \bibinfo{year}{2020}\natexlab{}.
\newblock \bibinfo{title}{Communicating Model Uncertainty Over Space,
  https://pair-code.github.io/interpretability/uncertainty-over-space/}.
\newblock
\newblock
\urldef\tempurl%
\url{https://pair-code.github.io/interpretability/uncertainty-over-space/}
\showURL{%
\tempurl}


\bibitem[\protect\citeauthoryear{Pietriga and Appert}{Pietriga and
  Appert}{2008}]%
        {pietriga_sigma_2008}
\bibfield{author}{\bibinfo{person}{Emmanuel Pietriga} {and}
  \bibinfo{person}{Caroline Appert}.} \bibinfo{year}{2008}\natexlab{}.
\newblock \showarticletitle{Sigma lenses: focus-context transitions combining
  space, time and translucence}. In \bibinfo{booktitle}{\emph{Proceeding of the
  twenty-sixth annual {CHI} conference on {Human} factors in computing systems
  - {CHI} '08}}. \bibinfo{publisher}{ACM Press}, \bibinfo{address}{Florence,
  Italy}, \bibinfo{pages}{1343}.
\newblock
\showISBNx{978-1-60558-011-1}
\urldef\tempurl%
\url{https://doi.org/10.1145/1357054.1357264}
\showDOI{\tempurl}


\bibitem[\protect\citeauthoryear{Pietriga, Appert, and
  Beaudouin-Lafon}{Pietriga et~al\mbox{.}}{2007}]%
        {pietriga_pointing_2007}
\bibfield{author}{\bibinfo{person}{Emmanuel Pietriga},
  \bibinfo{person}{Caroline Appert}, {and} \bibinfo{person}{Michel
  Beaudouin-Lafon}.} \bibinfo{year}{2007}\natexlab{}.
\newblock \showarticletitle{Pointing and beyond: an operationalization and
  preliminary evaluation of multi-scale searching}. In
  \bibinfo{booktitle}{\emph{Proceedings of the {SIGCHI} {Conference} on {Human}
  {Factors} in {Computing} {Systems}}} \emph{(\bibinfo{series}{{CHI} '07})}.
  \bibinfo{publisher}{Association for Computing Machinery},
  \bibinfo{address}{New York, NY, USA}, \bibinfo{pages}{1215--1224}.
\newblock
\showISBNx{978-1-59593-593-9}
\urldef\tempurl%
\url{https://doi.org/10.1145/1240624.1240808}
\showDOI{\tempurl}


\bibitem[\protect\citeauthoryear{Plumlee and Ware}{Plumlee and Ware}{2002}]%
        {plumlee_zooming_2002}
\bibfield{author}{\bibinfo{person}{Matthew Plumlee} {and}
  \bibinfo{person}{Colin Ware}.} \bibinfo{year}{2002}\natexlab{}.
\newblock \showarticletitle{Zooming, multiple windows, and visual working
  memory}. In \bibinfo{booktitle}{\emph{Proceedings of the {Working}
  {Conference} on {Advanced} {Visual} {Interfaces} - {AVI} '02}}.
  \bibinfo{publisher}{ACM Press}, \bibinfo{address}{Trento, Italy},
  \bibinfo{pages}{59}.
\newblock
\showISBNx{978-1-58113-537-4}
\urldef\tempurl%
\url{https://doi.org/10.1145/1556262.1556270}
\showDOI{\tempurl}


\bibitem[\protect\citeauthoryear{Plumlee and Ware}{Plumlee and Ware}{2006}]%
        {plumlee_zooming_2006}
\bibfield{author}{\bibinfo{person}{Matthew~D. Plumlee} {and}
  \bibinfo{person}{Colin Ware}.} \bibinfo{year}{2006}\natexlab{}.
\newblock \showarticletitle{Zooming versus multiple window interfaces:
  {Cognitive} costs of visual comparisons}.
\newblock \bibinfo{journal}{\emph{ACM Transactions on Computer-Human
  Interaction (TOCHI)}} \bibinfo{volume}{13}, \bibinfo{number}{2}
  (\bibinfo{date}{June} \bibinfo{year}{2006}), \bibinfo{pages}{179--209}.
\newblock
\showISSN{1073-0516, 1557-7325}
\urldef\tempurl%
\url{https://doi.org/10.1145/1165734.1165736}
\showDOI{\tempurl}


\bibitem[\protect\citeauthoryear{Rao and Card}{Rao and Card}{1994}]%
        {rao_table_1994}
\bibfield{author}{\bibinfo{person}{Ramana Rao} {and} \bibinfo{person}{Stuart~K.
  Card}.} \bibinfo{year}{1994}\natexlab{}.
\newblock \showarticletitle{The table lens: merging graphical and symbolic
  representations in an interactive focus + context visualization for tabular
  information}. In \bibinfo{booktitle}{\emph{Proceedings of the {SIGCHI}
  conference on {Human} factors in computing systems celebrating
  interdependence - {CHI} '94}}. \bibinfo{publisher}{ACM Press},
  \bibinfo{address}{Boston, Massachusetts, United States},
  \bibinfo{pages}{318--322}.
\newblock
\showISBNx{978-0-89791-650-9}
\urldef\tempurl%
\url{https://doi.org/10.1145/191666.191776}
\showDOI{\tempurl}


\bibitem[\protect\citeauthoryear{Roberts}{Roberts}{2007}]%
        {roberts_state_2007}
\bibfield{author}{\bibinfo{person}{Jonathan~C Roberts}.}
  \bibinfo{year}{2007}\natexlab{}.
\newblock \showarticletitle{State of the {Art}: {Coordinated} {Multiple}
  {Views} in {Exploratory} {Visualization}}. In \bibinfo{booktitle}{\emph{Fifth
  {International} {Conference} on {Coordinated} and {Multiple} {Views} in
  {Exploratory} {Visualization} ({CMV} 2007)}}. \bibinfo{pages}{61--71}.
\newblock
\urldef\tempurl%
\url{https://doi.org/10.1109/CMV.2007.20}
\showDOI{\tempurl}


\bibitem[\protect\citeauthoryear{Robertson and Mackinlay}{Robertson and
  Mackinlay}{1993}]%
        {robertson_document_1993}
\bibfield{author}{\bibinfo{person}{George~G. Robertson} {and}
  \bibinfo{person}{Jock~D. Mackinlay}.} \bibinfo{year}{1993}\natexlab{}.
\newblock \showarticletitle{The document lens}. In
  \bibinfo{booktitle}{\emph{Proceedings of the 6th annual {ACM} symposium on
  {User} interface software and technology - {UIST} '93}}.
  \bibinfo{publisher}{ACM Press}, \bibinfo{address}{Atlanta, Georgia, United
  States}, \bibinfo{pages}{101--108}.
\newblock
\showISBNx{978-0-89791-628-8}
\urldef\tempurl%
\url{https://doi.org/10.1145/168642.168652}
\showDOI{\tempurl}


\bibitem[\protect\citeauthoryear{Rønne~Jakobsen and Hornbæk}{Rønne~Jakobsen
  and Hornbæk}{2011}]%
        {ronne_jakobsen_sizing_2011}
\bibfield{author}{\bibinfo{person}{Mikkel Rønne~Jakobsen} {and}
  \bibinfo{person}{Kasper Hornbæk}.} \bibinfo{year}{2011}\natexlab{}.
\newblock \showarticletitle{Sizing up visualizations: effects of display size
  in focus+context, overview+detail, and zooming interfaces}. In
  \bibinfo{booktitle}{\emph{Proceedings of the 2011 annual conference on
  {Human} factors in computing systems - {CHI} '11}}. \bibinfo{publisher}{ACM
  Press}, \bibinfo{address}{Vancouver, BC, Canada}, \bibinfo{pages}{1451}.
\newblock
\showISBNx{978-1-4503-0228-9}
\urldef\tempurl%
\url{https://doi.org/10.1145/1978942.1979156}
\showDOI{\tempurl}


\bibitem[\protect\citeauthoryear{Sadana, Major, Dove, and Stasko}{Sadana
  et~al\mbox{.}}{2014}]%
        {sadana2014onset}
\bibfield{author}{\bibinfo{person}{Ramik Sadana}, \bibinfo{person}{Timothy
  Major}, \bibinfo{person}{Alistair Dove}, {and} \bibinfo{person}{John
  Stasko}.} \bibinfo{year}{2014}\natexlab{}.
\newblock \showarticletitle{Onset: A visualization technique for large-scale
  binary set data}.
\newblock \bibinfo{journal}{\emph{IEEE transactions on visualization and
  computer graphics}} \bibinfo{volume}{20}, \bibinfo{number}{12}
  (\bibinfo{year}{2014}), \bibinfo{pages}{1993--2002}.
\newblock


\bibitem[\protect\citeauthoryear{Sarkar and Brown}{Sarkar and Brown}{1992}]%
        {sarkar_graphical_1992}
\bibfield{author}{\bibinfo{person}{Manojit Sarkar} {and}
  \bibinfo{person}{Marc~H. Brown}.} \bibinfo{year}{1992}\natexlab{}.
\newblock \showarticletitle{Graphical fisheye views of graphs}. In
  \bibinfo{booktitle}{\emph{Proceedings of the {SIGCHI} conference on {Human}
  factors in computing systems - {CHI} '92}}. \bibinfo{publisher}{ACM Press},
  \bibinfo{address}{Monterey, California, United States},
  \bibinfo{pages}{83--91}.
\newblock
\showISBNx{978-0-89791-513-7}
\urldef\tempurl%
\url{https://doi.org/10.1145/142750.142763}
\showDOI{\tempurl}


\bibitem[\protect\citeauthoryear{Shoemaker and Gutwin}{Shoemaker and
  Gutwin}{2007}]%
        {shoemaker_supporting_2007}
\bibfield{author}{\bibinfo{person}{Garth Shoemaker} {and} \bibinfo{person}{Carl
  Gutwin}.} \bibinfo{year}{2007}\natexlab{}.
\newblock \showarticletitle{Supporting multi-point interaction in visual
  workspaces}. In \bibinfo{booktitle}{\emph{Proceedings of the {SIGCHI}
  {Conference} on {Human} {Factors} in {Computing} {Systems}}}
  \emph{(\bibinfo{series}{{CHI} '07})}. \bibinfo{publisher}{Association for
  Computing Machinery}, \bibinfo{address}{New York, NY, USA},
  \bibinfo{pages}{999--1008}.
\newblock
\showISBNx{978-1-59593-593-9}
\urldef\tempurl%
\url{https://doi.org/10.1145/1240624.1240777}
\showDOI{\tempurl}


\bibitem[\protect\citeauthoryear{Siirtola}{Siirtola}{1999}]%
        {siirtola_interaction_1999}
\bibfield{author}{\bibinfo{person}{Harri Siirtola}.}
  \bibinfo{year}{1999}\natexlab{}.
\newblock \bibinfo{booktitle}{\emph{Interaction with the {Reorderable}
  {Matrix}}}.
\newblock


\bibitem[\protect\citeauthoryear{Soukoreff and MacKenzie}{Soukoreff and
  MacKenzie}{2004}]%
        {soukoreff2004towards}
\bibfield{author}{\bibinfo{person}{R~William Soukoreff} {and}
  \bibinfo{person}{I~Scott MacKenzie}.} \bibinfo{year}{2004}\natexlab{}.
\newblock \showarticletitle{Towards a standard for pointing device evaluation,
  perspectives on 27 years of Fitts’ law research in HCI}.
\newblock \bibinfo{journal}{\emph{International journal of human-computer
  studies}} \bibinfo{volume}{61}, \bibinfo{number}{6} (\bibinfo{year}{2004}),
  \bibinfo{pages}{751--789}.
\newblock


\bibitem[\protect\citeauthoryear{{Stefano Burigat} and {Luca
  Chittaro}}{{Stefano Burigat} and {Luca Chittaro}}{2013}]%
        {stefano_burigat_effectiveness_2013}
\bibfield{author}{\bibinfo{person}{{Stefano Burigat}} {and}
  \bibinfo{person}{{Luca Chittaro}}.} \bibinfo{year}{2013}\natexlab{}.
\newblock \showarticletitle{On the effectiveness of {Overview}+{Detail}
  visualization on mobile devices}.
\newblock \bibinfo{journal}{\emph{Personal and Ubiquitous Computing}}
  \bibinfo{volume}{17}, \bibinfo{number}{2} (\bibinfo{year}{2013}),
  \bibinfo{pages}{371--385}.
\newblock
\showISSN{1617-4909}
\urldef\tempurl%
\url{https://doi.org/10.1007/s00779-011-0500-3}
\showDOI{\tempurl}


\bibitem[\protect\citeauthoryear{Tominski, Gladisch, Kister, Dachselt, and
  Schumann}{Tominski et~al\mbox{.}}{2014}]%
        {tominski_survey_2014}
\bibfield{author}{\bibinfo{person}{Christian Tominski}, \bibinfo{person}{Stefan
  Gladisch}, \bibinfo{person}{Ulrike Kister}, \bibinfo{person}{Raimund
  Dachselt}, {and} \bibinfo{person}{Heidrun Schumann}.}
  \bibinfo{year}{2014}\natexlab{}.
\newblock \showarticletitle{A {Survey} on {Interactive} {Lenses} in
  {Visualization}}.
\newblock \bibinfo{journal}{\emph{EuroVis - STARs}} (\bibinfo{year}{2014}),
  \bibinfo{pages}{20 pages}.
\newblock
\urldef\tempurl%
\url{https://doi.org/10.2312/EUROVISSTAR.20141172}
\showDOI{\tempurl}


\bibitem[\protect\citeauthoryear{Tominski, Gladisch, Kister, Dachselt, and
  Schumann}{Tominski et~al\mbox{.}}{2017}]%
        {tominski_interactive_2017}
\bibfield{author}{\bibinfo{person}{Christian Tominski}, \bibinfo{person}{Stefan
  Gladisch}, \bibinfo{person}{Ulrike Kister}, \bibinfo{person}{Raimund
  Dachselt}, {and} \bibinfo{person}{Heidrun Schumann}.}
  \bibinfo{year}{2017}\natexlab{}.
\newblock \showarticletitle{Interactive {Lenses} for {Visualization}: {An}
  {Extended} {Survey}}.
\newblock \bibinfo{journal}{\emph{Computer Graphics Forum}}
  \bibinfo{volume}{36}, \bibinfo{number}{6} (\bibinfo{year}{2017}),
  \bibinfo{pages}{173--200}.
\newblock
\showISSN{1467-8659}
\urldef\tempurl%
\url{https://doi.org/10.1111/cgf.12871}
\showDOI{\tempurl}


\bibitem[\protect\citeauthoryear{Van~Wijk and Nuij}{Van~Wijk and Nuij}{2003}]%
        {wijk_smooth_2003}
\bibfield{author}{\bibinfo{person}{Jarke~J Van~Wijk} {and}
  \bibinfo{person}{Wim~AA Nuij}.} \bibinfo{year}{2003}\natexlab{}.
\newblock \showarticletitle{Smooth and efficient zooming and panning}. In
  \bibinfo{booktitle}{\emph{{IEEE} {Symposium} on {Information} {Visualization}
  2003 ({IEEE} {Cat}. {No}.{03TH8714})}}. \bibinfo{pages}{15--23}.
\newblock
\urldef\tempurl%
\url{https://doi.org/10.1109/INFVIS.2003.1249004}
\showDOI{\tempurl}


\bibitem[\protect\citeauthoryear{Vogogias, Archambault, Bach, and
  Kennedy}{Vogogias et~al\mbox{.}}{2020}]%
        {vogogias_visual_2020}
\bibfield{author}{\bibinfo{person}{Athanasios Vogogias},
  \bibinfo{person}{Daniel Archambault}, \bibinfo{person}{Benjamin Bach}, {and}
  \bibinfo{person}{Jessie Kennedy}.} \bibinfo{year}{2020}\natexlab{}.
\newblock \showarticletitle{Visual {Encodings} for {Networks} with {Multiple}
  {Edge} {Types}}. In \bibinfo{booktitle}{\emph{International {Conference} on
  {Advanced} {Visual} {Interfaces} 2020}}. \bibinfo{pages}{9}.
\newblock


\bibitem[\protect\citeauthoryear{Wang~Baldonado, Woodruff, and
  Kuchinsky}{Wang~Baldonado et~al\mbox{.}}{2000}]%
        {wang_baldonado_guidelines_2000}
\bibfield{author}{\bibinfo{person}{Michelle~Q. Wang~Baldonado},
  \bibinfo{person}{Allison Woodruff}, {and} \bibinfo{person}{Allan Kuchinsky}.}
  \bibinfo{year}{2000}\natexlab{}.
\newblock \showarticletitle{Guidelines for using multiple views in information
  visualization}. In \bibinfo{booktitle}{\emph{Proceedings of the working
  conference on {Advanced} visual interfaces - {AVI} '00}}.
  \bibinfo{publisher}{ACM Press}, \bibinfo{address}{Palermo, Italy},
  \bibinfo{pages}{110--119}.
\newblock
\showISBNx{978-1-58113-252-6}
\urldef\tempurl%
\url{https://doi.org/10.1145/345513.345271}
\showDOI{\tempurl}


\bibitem[\protect\citeauthoryear{Wood, Slingsby, and Dykes}{Wood
  et~al\mbox{.}}{2011}]%
        {wood_visualizing_2011}
\bibfield{author}{\bibinfo{person}{Jo Wood}, \bibinfo{person}{Aidan Slingsby},
  {and} \bibinfo{person}{Jason Dykes}.} \bibinfo{year}{2011}\natexlab{}.
\newblock \showarticletitle{Visualizing the {Dynamics} of {London}'s
  {Bicycle}-{Hire} {Scheme}}.
\newblock \bibinfo{journal}{\emph{Cartographica: The International Journal for
  Geographic Information and Geovisualization}} \bibinfo{volume}{46},
  \bibinfo{number}{4} (\bibinfo{date}{Nov.} \bibinfo{year}{2011}),
  \bibinfo{pages}{239--251}.
\newblock
\showISSN{0317-7173}
\urldef\tempurl%
\url{https://doi.org/10.3138/carto.46.4.239}
\showDOI{\tempurl}


\bibitem[\protect\citeauthoryear{Woodburn, Yang, and Marriott}{Woodburn
  et~al\mbox{.}}{2019}]%
        {woodburn2019interactive}
\bibfield{author}{\bibinfo{person}{Linda Woodburn}, \bibinfo{person}{Yalong
  Yang}, {and} \bibinfo{person}{Kim Marriott}.}
  \bibinfo{year}{2019}\natexlab{}.
\newblock \showarticletitle{Interactive visualisation of hierarchical
  quantitative data: an evaluation}. In \bibinfo{booktitle}{\emph{2019 IEEE
  Visualization Conference (VIS)}}. IEEE, \bibinfo{pages}{96--100}.
\newblock
\urldef\tempurl%
\url{https://doi.org/10.1109/VISUAL.2019.8933545}
\showDOI{\tempurl}


\bibitem[\protect\citeauthoryear{Xu, Jia, Wang, Ai, Zhang, Lai, Eric, and
  Chang}{Xu et~al\mbox{.}}{2017}]%
        {xu2017large}
\bibfield{author}{\bibinfo{person}{Yan Xu}, \bibinfo{person}{Zhipeng Jia},
  \bibinfo{person}{Liang-Bo Wang}, \bibinfo{person}{Yuqing Ai},
  \bibinfo{person}{Fang Zhang}, \bibinfo{person}{Maode Lai}, \bibinfo{person}{I
  Eric}, {and} \bibinfo{person}{Chao Chang}.} \bibinfo{year}{2017}\natexlab{}.
\newblock \showarticletitle{Large scale tissue histopathology image
  classification, segmentation, and visualization via deep convolutional
  activation features}.
\newblock \bibinfo{journal}{\emph{BMC bioinformatics}} \bibinfo{volume}{18},
  \bibinfo{number}{1} (\bibinfo{year}{2017}), \bibinfo{pages}{1--17}.
\newblock


\bibitem[\protect\citeauthoryear{{Yalong Yang}, {Tim Dwyer}, {Sarah Goodwin},
  and {Kim Marriott}}{{Yalong Yang} et~al\mbox{.}}{2017}]%
        {yalong_yang_many--many_2017}
\bibfield{author}{\bibinfo{person}{{Yalong Yang}}, \bibinfo{person}{{Tim
  Dwyer}}, \bibinfo{person}{{Sarah Goodwin}}, {and} \bibinfo{person}{{Kim
  Marriott}}.} \bibinfo{year}{2017}\natexlab{}.
\newblock \showarticletitle{Many-to-{Many} {Geographically}-{Embedded} {Flow}
  {Visualisation}: {An} {Evaluation}}.
\newblock \bibinfo{journal}{\emph{IEEE Transactions on Visualization and
  Computer Graphics}} \bibinfo{volume}{23}, \bibinfo{number}{1}
  (\bibinfo{year}{2017}), \bibinfo{pages}{411--420}.
\newblock
\showISSN{1077-2626}
\urldef\tempurl%
\url{https://doi.org/10.1109/tvcg.2016.2598885}
\showDOI{\tempurl}


\bibitem[\protect\citeauthoryear{Yang, Cordeil, Beyer, Dwyer, Marriott, and
  Pfister}{Yang et~al\mbox{.}}{2021}]%
        {yang_embodied_2021}
\bibfield{author}{\bibinfo{person}{Yalong Yang}, \bibinfo{person}{Maxime
  Cordeil}, \bibinfo{person}{Johanna Beyer}, \bibinfo{person}{Tim Dwyer},
  \bibinfo{person}{Kim Marriott}, {and} \bibinfo{person}{Hanspeter Pfister}.}
  \bibinfo{year}{2021}\natexlab{}.
\newblock \showarticletitle{Embodied {Navigation} in {Immersive} {Abstract}
  {Data} {Visualization}: {Is} {Overview}+{Detail} or {Zooming} {Better} for
  {3D} {Scatterplots}?}
\newblock \bibinfo{journal}{\emph{IEEE Transactions on Visualization and
  Computer Graphics}} \bibinfo{volume}{27}, \bibinfo{number}{2}
  (\bibinfo{date}{Feb.} \bibinfo{year}{2021}), \bibinfo{pages}{1214--1224}.
\newblock
\showISSN{1941-0506}
\urldef\tempurl%
\url{https://doi.org/10.1109/TVCG.2020.3030427}
\showDOI{\tempurl}


\bibitem[\protect\citeauthoryear{Yates, Webb, Sharpnack, Chamberlin, Huang, and
  Machiraju}{Yates et~al\mbox{.}}{2014}]%
        {yates2014visualizing}
\bibfield{author}{\bibinfo{person}{Andrew Yates}, \bibinfo{person}{Amy Webb},
  \bibinfo{person}{Michael Sharpnack}, \bibinfo{person}{Helen Chamberlin},
  \bibinfo{person}{Kun Huang}, {and} \bibinfo{person}{Raghu Machiraju}.}
  \bibinfo{year}{2014}\natexlab{}.
\newblock \showarticletitle{Visualizing {Multidimensional} {Data} with {Glyph}
  {SPLOMs}: {Visualizing} {Multidimensional} {Data} with {Glyph} {SPLOMs}}.
\newblock \bibinfo{journal}{\emph{Computer Graphics Forum}}
  \bibinfo{volume}{33}, \bibinfo{number}{3} (\bibinfo{date}{June}
  \bibinfo{year}{2014}), \bibinfo{pages}{301--310}.
\newblock
\showISSN{01677055}
\urldef\tempurl%
\url{https://doi.org/10.1111/cgf.12386}
\showDOI{\tempurl}


\bibitem[\protect\citeauthoryear{Yi, Elmqvist, and Lee}{Yi
  et~al\mbox{.}}{2010}]%
        {yi_timematrix_2010}
\bibfield{author}{\bibinfo{person}{Ji~Soo Yi}, \bibinfo{person}{Niklas
  Elmqvist}, {and} \bibinfo{person}{Seungyoon Lee}.}
  \bibinfo{year}{2010}\natexlab{}.
\newblock \showarticletitle{{TimeMatrix}: {Analyzing} {Temporal} {Social}
  {Networks} {Using} {Interactive} {Matrix}-{Based} {Visualizations}}.
\newblock \bibinfo{journal}{\emph{International Journal of Human-Computer
  Interaction}} \bibinfo{volume}{26}, \bibinfo{number}{11-12}
  (\bibinfo{date}{Nov.} \bibinfo{year}{2010}), \bibinfo{pages}{1031--1051}.
\newblock
\showISSN{1044-7318, 1532-7590}
\urldef\tempurl%
\url{https://doi.org/10.1080/10447318.2010.516722}
\showDOI{\tempurl}


\bibitem[\protect\citeauthoryear{Zammitto}{Zammitto}{2008}]%
        {zammitto_visualization_2008}
\bibfield{author}{\bibinfo{person}{Veronica Zammitto}.}
  \bibinfo{year}{2008}\natexlab{}.
\newblock \showarticletitle{Visualization Techniques In Video Games}. In
  \bibinfo{booktitle}{\emph{Electronic Visualisation and the Arts}}.
  \bibinfo{pages}{267--276}.
\newblock
\urldef\tempurl%
\url{https://doi.org/10.14236/ewic/EVA2008.30}
\showDOI{\tempurl}


\bibitem[\protect\citeauthoryear{Zanella, Carpendale, and Rounding}{Zanella
  et~al\mbox{.}}{2002}]%
        {zanella_effects_2002}
\bibfield{author}{\bibinfo{person}{Ana Zanella}, \bibinfo{person}{Sheelagh
  Carpendale}, {and} \bibinfo{person}{Michael Rounding}.}
  \bibinfo{year}{2002}\natexlab{}.
\newblock \showarticletitle{On the effects of viewing cues in comprehending
  distortions}. In \bibinfo{booktitle}{\emph{Proceedings of the second {Nordic}
  conference on {Human}-computer interaction}}
  \emph{(\bibinfo{series}{{NordiCHI} '02})}. \bibinfo{publisher}{Association
  for Computing Machinery}, \bibinfo{address}{New York, NY, USA},
  \bibinfo{pages}{119--128}.
\newblock
\showISBNx{978-1-58113-616-6}
\urldef\tempurl%
\url{https://doi.org/10.1145/572020.572035}
\showDOI{\tempurl}


\end{thebibliography}

\end{document}